\documentclass[10pt,english,twoside]{article}
\usepackage{authblk}

\pdfoutput=1
\usepackage{graphicx}
\usepackage{epstopdf, epsfig}
\usepackage{xcolor}
\usepackage{pgfplots}
\pgfplotsset{compat=1.3}
\usetikzlibrary{external}
\usepgfplotslibrary{groupplots}
\tikzset{external/mode=graphics if exists}
\usepackage{relsize}
\def\solidthick{\protect\rule[2pt]{10.pt}{1pt}}

\def\solidshort{\protect\rule[2pt]{3.pt}{1pt}}
\def\dashed{\solidshort$\,$\solidshort$\,$\solidshort}
\def\dotted{$\cdot \hspace{0.05ex} \cdot \hspace{0.05ex} \cdot$}

\definecolor{mygray}{rgb}{0.5,0.5,0.5}
\definecolor{my_dark_blue}{rgb}{0,0,0.5}
\definecolor{col_gray}{rgb}{0.63,0.63,0.63}
\definecolor{col_gray2}{rgb}{0.3,0.3,0.3}
\definecolor{my_dark_blue}{rgb}{0,0,0.5}
\definecolor{my_orange}{rgb}{1.0,0.5,0.0}

\definecolor{col_230}{rgb}{0.0,0.4,0.8}
\definecolor{col_240}{rgb}{0.0,1.0,1.0}
\definecolor{col_250}{rgb}{0.05,0.45,0.37}
\definecolor{col_270}{rgb}{0.9,0.55,0.0}
\definecolor{col_280}{rgb}{1.0,0.0,0.0}
\definecolor{col_290}{rgb}{1.0,0.0,1.0}
\definecolor{col_330}{rgb}{0.0,0.4,0.8}
\definecolor{col_340}{rgb}{0.0,1.0,1.0}
\definecolor{col_350}{rgb}{0.05,0.45,0.37}
\definecolor{col_370}{rgb}{0.9,0.55,0.0}

\definecolor{col_260}{rgb}{0.0,0.4,0.8}
\definecolor{col_221}{rgb}{0.0 0.6 0.51}
\definecolor{col_222}{rgb}{0.0 0.6 0.51}
\definecolor{col_21}{rgb}{0.9,0.55,0.0}
\definecolor{col_511}{rgb}{1.0,0.0,0.0}
\definecolor{col_5}{rgb}{0.13,0.13,0.13}
\definecolor{col_1}{rgb}{0.63,0.63,0.63}

\definecolor{col1_matlab}{rgb}{0.000,0.447,0.741}
\definecolor{col2_matlab}{rgb}{0.850,0.325,0.098}
\definecolor{col3_matlab}{rgb}{0.929,0.694,0.125}
\definecolor{col4_matlab}{rgb}{0.494,0.184,0.556}
\definecolor{col5_matlab}{rgb}{0.466,0.674,0.188}
\definecolor{col6_matlab}{rgb}{0.301,0.745,0.933}
\definecolor{col7_matlab}{rgb}{0.635,0.078,0.184}

\pgfplotsset{tikzPlotsDefault/.style=
    {hide axis,scale only axis,
     height=0pt,width=0pt,
     colorbar right,
     colorbar style={
             anchor=west,
             major tick length=0.05cm,
             ylabel near ticks,
             label style={font=\footnotesize},
             ylabel shift={-0.05cm}
             }
     }
}

\pgfplotsset{colorbar style={
        anchor=west,
        major tick length=0.05cm,
        ylabel near ticks,
        label style={font=\small}
        }
}

\pgfplotsset{
    colormap={bluered}{
        rgb255=(0,39,93)
        rgb255=(222,222,222)
        rgb255=(222,100,25)
    },
}
\pgfplotsset{
    colormap={whiteblue}{
        rgb255=(0,0,255)
        rgb255=(255,255,255)
    },
}
\pgfplotsset{
    colormap={whitered}{
        rgb255=(255,0,0)
        rgb255=(255,255,255)
    },
}
\pgfplotsset{
    colormap={mygraymap}{
    [1pt]
    rgb(0pt)=(0.2422,0.1504,0.6603); rgb(1pt)=(0.2444,0.1534,0.6728); rgb(2pt)=(0.2464,0.1569,0.6847); rgb(3pt)=(0.2484,0.1607,0.6961); rgb(4pt)=(0.2503,0.1648,0.7071); rgb(5pt)=(0.2522,0.1689,0.7179); rgb(6pt)=(0.254,0.1732,0.7286); rgb(7pt)=(0.2558,0.1773,0.7393); rgb(8pt)=(0.2576,0.1814,0.7501); rgb(9pt)=(0.2594,0.1854,0.761); rgb(11pt)=(0.2628,0.1932,0.7828); rgb(12pt)=(0.2645,0.1972,0.7937); rgb(13pt)=(0.2661,0.2011,0.8043); rgb(14pt)=(0.2676,0.2052,0.8148); rgb(15pt)=(0.2691,0.2094,0.8249); rgb(16pt)=(0.2704,0.2138,0.8346); rgb(17pt)=(0.2717,0.2184,0.8439); rgb(18pt)=(0.2729,0.2231,0.8528); rgb(19pt)=(0.274,0.228,0.8612); rgb(20pt)=(0.2749,0.233,0.8692); rgb(21pt)=(0.2758,0.2382,0.8767); rgb(22pt)=(0.2766,0.2435,0.884); rgb(23pt)=(0.2774,0.2489,0.8908); rgb(24pt)=(0.2781,0.2543,0.8973); rgb(25pt)=(0.2788,0.2598,0.9035); rgb(26pt)=(0.2794,0.2653,0.9094); rgb(27pt)=(0.2798,0.2708,0.915); rgb(28pt)=(0.2802,0.2764,0.9204); rgb(29pt)=(0.2806,0.2819,0.9255); rgb(30pt)=(0.2809,0.2875,0.9305); rgb(31pt)=(0.2811,0.293,0.9352); rgb(32pt)=(0.2813,0.2985,0.9397); rgb(33pt)=(0.2814,0.304,0.9441); rgb(34pt)=(0.2814,0.3095,0.9483); rgb(35pt)=(0.2813,0.315,0.9524); rgb(36pt)=(0.2811,0.3204,0.9563); rgb(37pt)=(0.2809,0.3259,0.96); rgb(38pt)=(0.2807,0.3313,0.9636); rgb(39pt)=(0.2803,0.3367,0.967); rgb(40pt)=(0.2798,0.3421,0.9702); rgb(41pt)=(0.2791,0.3475,0.9733); rgb(42pt)=(0.2784,0.3529,0.9763); rgb(43pt)=(0.2776,0.3583,0.9791); rgb(44pt)=(0.2766,0.3638,0.9817); rgb(45pt)=(0.2754,0.3693,0.984); rgb(46pt)=(0.2741,0.3748,0.9862); rgb(47pt)=(0.2726,0.3804,0.9881); rgb(48pt)=(0.271,0.386,0.9898); rgb(49pt)=(0.2691,0.3916,0.9912); rgb(50pt)=(0.267,0.3973,0.9924); rgb(51pt)=(0.2647,0.403,0.9935); rgb(52pt)=(0.2621,0.4088,0.9946); rgb(53pt)=(0.2591,0.4145,0.9955); rgb(54pt)=(0.2556,0.4203,0.9965); rgb(55pt)=(0.2517,0.4261,0.9974); rgb(56pt)=(0.2473,0.4319,0.9983); rgb(57pt)=(0.2424,0.4378,0.9991); rgb(58pt)=(0.2369,0.4437,0.9996); rgb(59pt)=(0.2311,0.4497,0.9995); rgb(60pt)=(0.225,0.4559,0.9985); rgb(61pt)=(0.2189,0.462,0.9968); rgb(62pt)=(0.2128,0.4682,0.9948); rgb(63pt)=(0.2066,0.4743,0.9926); rgb(64pt)=(0.2006,0.4803,0.9906); rgb(65pt)=(0.195,0.4861,0.9887); rgb(66pt)=(0.1903,0.4919,0.9867); rgb(67pt)=(0.1869,0.4975,0.9844); rgb(68pt)=(0.1847,0.503,0.9819); rgb(69pt)=(0.1831,0.5084,0.9793); rgb(70pt)=(0.1818,0.5138,0.9766); rgb(71pt)=(0.1806,0.5191,0.9738); rgb(72pt)=(0.1795,0.5244,0.9709); rgb(73pt)=(0.1785,0.5296,0.9677); rgb(74pt)=(0.1778,0.5349,0.9641); rgb(75pt)=(0.1773,0.5401,0.9602); rgb(76pt)=(0.1768,0.5452,0.956); rgb(77pt)=(0.1764,0.5504,0.9516); rgb(78pt)=(0.1755,0.5554,0.9473); rgb(79pt)=(0.174,0.5605,0.9432); rgb(80pt)=(0.1716,0.5655,0.9393); rgb(81pt)=(0.1686,0.5705,0.9357); rgb(82pt)=(0.1649,0.5755,0.9323); rgb(83pt)=(0.161,0.5805,0.9289); rgb(84pt)=(0.1573,0.5854,0.9254); rgb(85pt)=(0.154,0.5902,0.9218); rgb(86pt)=(0.1513,0.595,0.9182); rgb(87pt)=(0.1492,0.5997,0.9147); rgb(88pt)=(0.1475,0.6043,0.9113); rgb(89pt)=(0.1461,0.6089,0.908); rgb(90pt)=(0.1446,0.6135,0.905); rgb(91pt)=(0.1429,0.618,0.9022); rgb(92pt)=(0.1408,0.6226,0.8998); rgb(93pt)=(0.1383,0.6272,0.8975); rgb(94pt)=(0.1354,0.6317,0.8953); rgb(95pt)=(0.1321,0.6363,0.8932); rgb(96pt)=(0.1288,0.6408,0.891); rgb(97pt)=(0.1253,0.6453,0.8887); rgb(98pt)=(0.1219,0.6497,0.8862); rgb(99pt)=(0.1185,0.6541,0.8834); rgb(100pt)=(0.1152,0.6584,0.8804); rgb(101pt)=(0.1119,0.6627,0.877); rgb(102pt)=(0.1085,0.6669,0.8734); rgb(103pt)=(0.1048,0.671,0.8695); rgb(104pt)=(0.1009,0.675,0.8653); rgb(105pt)=(0.0964,0.6789,0.8609); rgb(106pt)=(0.0914,0.6828,0.8562); rgb(107pt)=(0.0855,0.6865,0.8513); rgb(108pt)=(0.0789,0.6902,0.8462); rgb(109pt)=(0.0713,0.6938,0.8409); rgb(110pt)=(0.0628,0.6972,0.8355); rgb(111pt)=(0.0535,0.7006,0.8299); rgb(112pt)=(0.0433,0.7039,0.8242); rgb(113pt)=(0.0328,0.7071,0.8183); rgb(114pt)=(0.0234,0.7103,0.8124); rgb(115pt)=(0.0155,0.7133,0.8064); rgb(116pt)=(0.0091,0.7163,0.8003); rgb(117pt)=(0.0046,0.7192,0.7941); rgb(118pt)=(0.0019,0.722,0.7878); rgb(119pt)=(0.0009,0.7248,0.7815); rgb(120pt)=(0.0018,0.7275,0.7752); rgb(121pt)=(0.0046,0.7301,0.7688); rgb(122pt)=(0.0094,0.7327,0.7623); rgb(123pt)=(0.0162,0.7352,0.7558); rgb(124pt)=(0.0253,0.7376,0.7492); rgb(125pt)=(0.0369,0.74,0.7426); rgb(126pt)=(0.0504,0.7423,0.7359); rgb(127pt)=(0.0638,0.7446,0.7292); rgb(128pt)=(0.077,0.7468,0.7224); rgb(129pt)=(0.0899,0.7489,0.7156); rgb(130pt)=(0.1023,0.751,0.7088); rgb(131pt)=(0.1141,0.7531,0.7019); rgb(132pt)=(0.1252,0.7552,0.695); rgb(133pt)=(0.1354,0.7572,0.6881); rgb(134pt)=(0.1448,0.7593,0.6812); rgb(135pt)=(0.1532,0.7614,0.6741); rgb(136pt)=(0.1609,0.7635,0.6671); rgb(137pt)=(0.1678,0.7656,0.6599); rgb(138pt)=(0.1741,0.7678,0.6527); rgb(139pt)=(0.1799,0.7699,0.6454); rgb(140pt)=(0.1853,0.7721,0.6379); rgb(141pt)=(0.1905,0.7743,0.6303); rgb(142pt)=(0.1954,0.7765,0.6225); rgb(143pt)=(0.2003,0.7787,0.6146); rgb(144pt)=(0.2061,0.7808,0.6065); rgb(145pt)=(0.2118,0.7828,0.5983); rgb(146pt)=(0.2178,0.7849,0.5899); rgb(147pt)=(0.2244,0.7869,0.5813); rgb(148pt)=(0.2318,0.7887,0.5725); rgb(149pt)=(0.2401,0.7905,0.5636); rgb(150pt)=(0.2491,0.7922,0.5546); rgb(151pt)=(0.2589,0.7937,0.5454); rgb(152pt)=(0.2695,0.7951,0.536); rgb(153pt)=(0.2809,0.7964,0.5266); rgb(154pt)=(0.2929,0.7975,0.517); rgb(155pt)=(0.3052,0.7985,0.5074); rgb(156pt)=(0.3176,0.7994,0.4975); rgb(157pt)=(0.3301,0.8002,0.4876); rgb(158pt)=(0.3424,0.8009,0.4774); rgb(159pt)=(0.3548,0.8016,0.4669); rgb(160pt)=(0.3671,0.8021,0.4563); rgb(161pt)=(0.3795,0.8026,0.4454); rgb(162pt)=(0.3921,0.8029,0.4344); rgb(163pt)=(0.405,0.8031,0.4233); rgb(164pt)=(0.4184,0.803,0.4122); rgb(165pt)=(0.4322,0.8028,0.4013); rgb(166pt)=(0.4463,0.8024,0.3904); rgb(167pt)=(0.4608,0.8018,0.3797); rgb(168pt)=(0.4753,0.8011,0.3691); rgb(169pt)=(0.4899,0.8002,0.3586); rgb(170pt)=(0.5044,0.7993,0.348); rgb(171pt)=(0.5187,0.7982,0.3374); rgb(172pt)=(0.5329,0.797,0.3267); rgb(173pt)=(0.547,0.7957,0.3159); rgb(175pt)=(0.5748,0.7929,0.2941); rgb(176pt)=(0.5886,0.7913,0.2833); rgb(177pt)=(0.6024,0.7896,0.2726); rgb(178pt)=(0.6161,0.7878,0.2622); rgb(179pt)=(0.6297,0.7859,0.2521); rgb(180pt)=(0.6433,0.7839,0.2423); rgb(181pt)=(0.6567,0.7818,0.2329); rgb(182pt)=(0.6701,0.7796,0.2239); rgb(183pt)=(0.6833,0.7773,0.2155); rgb(184pt)=(0.6963,0.775,0.2075); rgb(185pt)=(0.7091,0.7727,0.1998); rgb(186pt)=(0.7218,0.7703,0.1924); rgb(187pt)=(0.7344,0.7679,0.1852); rgb(188pt)=(0.7468,0.7654,0.1782); rgb(189pt)=(0.759,0.7629,0.1717); rgb(190pt)=(0.771,0.7604,0.1658); rgb(191pt)=(0.7829,0.7579,0.1608); rgb(192pt)=(0.7945,0.7554,0.157); rgb(193pt)=(0.806,0.7529,0.1546); rgb(194pt)=(0.8172,0.7505,0.1535); rgb(195pt)=(0.8281,0.7481,0.1536); rgb(196pt)=(0.8389,0.7457,0.1546); rgb(197pt)=(0.8495,0.7435,0.1564); rgb(198pt)=(0.86,0.7413,0.1587); rgb(199pt)=(0.8703,0.7392,0.1615); rgb(200pt)=(0.8804,0.7372,0.165); rgb(201pt)=(0.8903,0.7353,0.1695); rgb(202pt)=(0.9,0.7336,0.1749); rgb(203pt)=(0.9093,0.7321,0.1815); rgb(204pt)=(0.9184,0.7308,0.189); rgb(205pt)=(0.9272,0.7298,0.1973); rgb(206pt)=(0.9357,0.729,0.2061); rgb(207pt)=(0.944,0.7285,0.2151); rgb(208pt)=(0.9523,0.7284,0.2237); rgb(209pt)=(0.9606,0.7285,0.2312); rgb(210pt)=(0.9689,0.7292,0.2373); rgb(211pt)=(0.977,0.7304,0.2418); rgb(212pt)=(0.9842,0.733,0.2446); rgb(213pt)=(0.99,0.7365,0.2429); rgb(214pt)=(0.9946,0.7407,0.2394); rgb(215pt)=(0.9966,0.7458,0.2351); rgb(216pt)=(0.9971,0.7513,0.2309); rgb(217pt)=(0.9972,0.7569,0.2267); rgb(218pt)=(0.9971,0.7626,0.2224); rgb(219pt)=(0.9969,0.7683,0.2181); rgb(220pt)=(0.9966,0.774,0.2138); rgb(221pt)=(0.9962,0.7798,0.2095); rgb(222pt)=(0.9957,0.7856,0.2053); rgb(223pt)=(0.9949,0.7915,0.2012); rgb(224pt)=(0.9938,0.7974,0.1974); rgb(225pt)=(0.9923,0.8034,0.1939); rgb(226pt)=(0.9906,0.8095,0.1906); rgb(227pt)=(0.9885,0.8156,0.1875); rgb(228pt)=(0.9861,0.8218,0.1846); rgb(229pt)=(0.9835,0.828,0.1817); rgb(230pt)=(0.9807,0.8342,0.1787); rgb(231pt)=(0.9778,0.8404,0.1757); rgb(232pt)=(0.9748,0.8467,0.1726); rgb(233pt)=(0.972,0.8529,0.1695); rgb(234pt)=(0.9694,0.8591,0.1665); rgb(235pt)=(0.9671,0.8654,0.1636); rgb(236pt)=(0.9651,0.8716,0.1608); rgb(237pt)=(0.9634,0.8778,0.1582); rgb(238pt)=(0.9619,0.884,0.1557); rgb(239pt)=(0.9608,0.8902,0.1532); rgb(240pt)=(0.9601,0.8963,0.1507); rgb(241pt)=(0.9596,0.9023,0.148); rgb(242pt)=(0.9595,0.9084,0.145); rgb(243pt)=(0.9597,0.9143,0.1418); rgb(244pt)=(0.9601,0.9203,0.1382); rgb(245pt)=(0.9608,0.9262,0.1344); rgb(246pt)=(0.9618,0.932,0.1304); rgb(247pt)=(0.9629,0.9379,0.1261); rgb(248pt)=(0.9642,0.9437,0.1216); rgb(249pt)=(0.9657,0.9494,0.1168); rgb(250pt)=(0.9674,0.9552,0.1116); rgb(251pt)=(0.9692,0.9609,0.1061); rgb(252pt)=(0.9711,0.9667,0.1001); rgb(253pt)=(0.973,0.9724,0.0938); rgb(254pt)=(0.9749,0.9782,0.0872); rgb(255pt)=(0.9769,0.9839,0.0805)
    }
}

\usepgfplotslibrary{colormaps}

\usepackage{amsmath,amssymb,amsfonts}
\usepackage{mathtools}
\usepackage{accents}
\usepackage{gensymb}
\usepackage{textcomp}
\usepackage{siunitx}

\usepackage[colorlinks,linkcolor=blue,citecolor=blue,urlcolor=blue]{hyperref}

\usepackage[round,sort&compress]{natbib}
\let\doi\undefined
\usepackage{doi}

\usepackage[
a4paper,
headheight=1.5\baselineskip,
top=25mm,
bottom=25mm,
inner=25mm,
outer=25mm,
heightrounded=true]
{geometry}

\usepackage{titlesec}
\titleformat{\section}{\large\bfseries}{\thesection}{1em}{}
\titleformat{\subsection}{\normalsize\bfseries}{\thesubsection}{1em}{}

\usepackage[titletoc,title]{appendix}

\title{On the role of turbulent large-scale streaks in generating sediment ridges}

\author[1]{Markus Scherer}
\author[1]{Markus Uhlmann \footnote{Email address for correspondence: \href{markus.uhlmann@kit.edu}{markus.uhlmann@kit.edu}}}
\author[2]{Aman G. Kidanemariam}
\author[1]{Michael Krayer}
\affil[1]{\small Institute for Hydromechanics, Karlsruhe Institute of Technology, 76131 Karlsruhe, Germany}
\affil[2]{\small Department of Mechanical Engineering, The University of Melbourne, Victoria 3010, Australia}
\vspace{-1.3cm}
\date{\small (Dated: \today \ -- original submission, revised version has been accepted \\ for publication in \textit{J.\ Fluid Mech.} 2021)}

\usepackage{fancyhdr}
\newcommand\xavg[1]{\ensuremath{\langle {#1} \rangle_{x}}}
\newcommand\zavg[1]{\ensuremath{\langle {#1} \rangle_{z}}}
\newcommand\xzavg[1]{\ensuremath{\langle {#1} \rangle_{xz}}}
\newcommand\tavg[1]{\ensuremath{\langle {#1} \rangle_{t}}}
\newcommand\xtavg[1]{\ensuremath{\langle {#1} \rangle_{xt}}}
\newcommand\xztavg[1]{\ensuremath{\langle {#1} \rangle_{xzt}}}
\newcommand\gfilter[1]{\ensuremath{\widetilde{#1}}}

\newcommand{\ichanD}{\ensuremath{S250}}
\newcommand{\ichanLRZinit}{\ensuremath{L250}}

\newcommand{\ispecLRZinit}{\ensuremath{L250_{smooth}}}
\newcommand{\ichanKUinit}{\ensuremath{M250}}
\newcommand{\ichanHLRSlonginit}{\ensuremath{M850}}
\newcommand{\ispecHLRSlonginit}{\ensuremath{M650_{smooth}}}

\newcommand\xvec{\ensuremath{\boldsymbol{x}}}

\newcommand\xp{\ensuremath{x_p}}
\newcommand\yp{\ensuremath{y_p}}
\newcommand\zp{\ensuremath{z_p}}
\newcommand\tbulk{\ensuremath{T_b}}

\newcommand\tobs{\ensuremath{T_{obs}}}

\newcommand\rhof{\ensuremath{\rho_f}} 
\newcommand\fnu{\ensuremath{\nu_f}} 
\newcommand\absgrav{\ensuremath{|\boldsymbol{g}|}} 
\newcommand\hf{\ensuremath{h_f}}

\newcommand\hfflucx{\ensuremath{h_f^{\prime\prime}}}
\newcommand\hb{\ensuremath{h_b}}
\newcommand\hbx{\ensuremath{\xavg{h_b}}}
\newcommand\hbz{\ensuremath{\zavg{h_b}}}

\newcommand\hbflucx{\ensuremath{h_b^{\prime\prime}}}
\newcommand\hmean{\ensuremath{H_f}}
\newcommand\hbedmean{\ensuremath{H_b}}
\newcommand\HftoD{\ensuremath{H_f/D}}
\newcommand\HbtoD{\ensuremath{H_b/D}}
\newcommand\sigmax{\ensuremath{\sigma_{h,x}}}
\newcommand\sigmaz{\ensuremath{\sigma_{h,z}}}

\newcommand\lambdahz{\ensuremath{\lambda_{h,z}}}
\newcommand\yrel{\ensuremath{\tilde{y}}}
\newcommand\yrelh{\ensuremath{\tilde{y}/\hmean}}
\newcommand\yrelplus{\ensuremath{\tilde{y}^+}}
\newcommand\kx{\ensuremath{k_x}}
\newcommand\kz{\ensuremath{k_z}}
\newcommand\lamx{\ensuremath{\lambda_x}}
\newcommand\lamz{\ensuremath{\lambda_z}}
\newcommand\Lx{\ensuremath{L_x}}
\newcommand\Ly{\ensuremath{L_y}}
\newcommand\Lz{\ensuremath{L_z}}

\newcommand\qf{\ensuremath{q_f}} 
\newcommand\uvec{\ensuremath{\boldsymbol{u}_f}}
\newcommand\ufi{\ensuremath{u_{f,i}}}
\newcommand\uf{\ensuremath{u_f}}
\newcommand\vf{\ensuremath{v_f}}
\newcommand\wf{\ensuremath{w_f}}

\newcommand\uvect{\ensuremath{\tavg{\boldsymbol{u}_f}}}
\newcommand\uvecx{\ensuremath{\xavg{\boldsymbol{u}_f}}}
\newcommand\uvecxz{\ensuremath{\xzavg{\boldsymbol{u}_f}}}

\newcommand\uvecxzt{\ensuremath{\xztavg{\boldsymbol{u}_f}}}
\newcommand\uvecflucxz{\ensuremath{\boldsymbol{u}_f^\prime}}
\newcommand\uvecflucx{\ensuremath{\boldsymbol{u}_f^{\prime\prime}}}
\newcommand\uvecfluct{\ensuremath{\overline{\boldsymbol{u}}}_f}

\newcommand\uxt{\ensuremath{\xtavg{u_f}}}
\newcommand\uxzt{\ensuremath{\xztavg{u_f}}}
\newcommand\uflucxz{\ensuremath{u_f^\prime}}
\newcommand\uflucx{\ensuremath{u_f^{\prime\prime}}}
\newcommand\ufluct{\ensuremath{\overline{u}}_f}

\newcommand\vxt{\ensuremath{\xtavg{v_f}}}

\newcommand\vflucxz{\ensuremath{v_f^\prime}}

\newcommand\vfluct{\ensuremath{\overline{v}}_f}
\newcommand\wxt{\ensuremath{\xtavg{w_f}}}

\newcommand\wfluct{\ensuremath{\overline{w}}_f}
\newcommand\deltaxfilt{\ensuremath{\Delta_x}}
\newcommand\deltayfilt{\ensuremath{\Delta_y}}
\newcommand\deltazfilt{\ensuremath{\Delta_z}}

\newcommand\ufifilt{\ensuremath{\gfilter{u}_{f,i}}}

\newcommand\ufour{\ensuremath{\hat{u}_f}}
\newcommand\uconj{\ensuremath{\hat{u}_f^*}}
\newcommand\uufvec{\ensuremath{\tavg{\phi_{\alpha\alpha}}}}

\newcommand\uuf{\ensuremath{\phi_{uu}}}
\newcommand\uuft{\ensuremath{\tavg{\phi_{uu}}}}

\newcommand\vvft{\ensuremath{\tavg{\phi_{vv}}}}

\newcommand\uuprime{\ensuremath{\xztavg{\ufluct\ufluct}}}
\newcommand\vvprime{\ensuremath{\xztavg{\vfluct\vfluct}}}
\newcommand\wwprime{\ensuremath{\xztavg{\wfluct\wfluct}}}

\newcommand\usec{\ensuremath{u_{\perp}}}
\newcommand\ubulk{\ensuremath{u_b}}
\newcommand\utau{\ensuremath{u_\tau}}
\newcommand\ugrav{\ensuremath{u_{g}}}

\newcommand\uroughshift{\ensuremath{\Delta U^+}}

\newcommand\tauw{\ensuremath{\tau_w}}
\newcommand\tauf{\ensuremath{\tau_f}}
\newcommand\taub{\ensuremath{\tau_b}}
\newcommand\taubflucx{\ensuremath{\tau^{\prime\prime}_b}}
\newcommand\taubflucxz{\ensuremath{\tau^{\prime}_b}}

\newcommand\pindicatorxz{\ensuremath{I_{(l)}^{(i,k)}}}
\newcommand\pindicatory{\ensuremath{I_{(l)}^{(j)}}}

\newcommand\rhop{\ensuremath{\rho_p}} 
\newcommand\Np{\ensuremath{N_p}}          
\newcommand\Vp{\ensuremath{V_p}}
\newcommand\upvec{\ensuremath{\boldsymbol{u}_p}}
\newcommand\up{\ensuremath{u_p}}

\renewcommand\wp{\ensuremath{w_p}}
\newcommand\qpvec{\ensuremath{\boldsymbol{q}_{p}}}
\newcommand\qpx{\ensuremath{q_{p,x}}}
\newcommand\qpz{\ensuremath{q_{p,z}}}

\newcommand\qpxmxz{\ensuremath{\xzavg{\qpx}}}

\newcommand\qpxmxzt{\ensuremath{\xztavg{q_{p,x}}}}

\newcommand\qpxflucx{\ensuremath{q_{p,x}^{\prime\prime}}}
\newcommand\qpzflucx{\ensuremath{q_{p,z}^{\prime\prime}}}
\newcommand\Reb{\ensuremath{Re_b}}    
\newcommand\Ret{\ensuremath{Re_\tau}} 
\newcommand\Dplus{\ensuremath{D^+}}   
\newcommand\Ga{\ensuremath{Ga}}       
\newcommand\shields{\ensuremath{\theta}} 
\newcommand\shieldscrit{\ensuremath{\theta_c}} 
\newcommand\dratio{\ensuremath{\rhop/\rhof}} 
\newcommand\deltanu{\ensuremath{\delta_{\nu}}}
\newcommand\deltax{\ensuremath{\Delta x}}

\newcommand\deltayplus{\ensuremath{\Delta y^+}}

\newcommand\xxi{\ensuremath{x_i}}
\newcommand\yyj{\ensuremath{y_j}}
\newcommand\zzk{\ensuremath{z_k}}

\newcommand\deltaxbin{\ensuremath{\Delta x_{bin}}}
\newcommand\deltaybin{\ensuremath{\Delta y_{bin}}}
\newcommand\deltazbin{\ensuremath{\Delta z_{bin}}}
\newcommand\surfdist{\ensuremath{\Delta}} 
\newcommand\forcerange{\ensuremath{\Delta_c}} 
\newcommand\penetrationl{\ensuremath{\delta_c}} 
\newcommand\normstiff{\ensuremath{k_n}}    
\newcommand\normfriction{\ensuremath{c_n}} 
\newcommand\tangfriction{\ensuremath{c_t}} 
\newcommand\Coulombfriction{\ensuremath{\mu_c}} 
\newcommand\dryrest{\ensuremath{\varepsilon_d}}   
\newcommand\tsep{\ensuremath{\delta t}}

\newcommand\zsep{\ensuremath{\delta z}}
\newcommand\zsepmin{\ensuremath{\delta z_{min}}}

\newcommand\corrhh{\ensuremath{\rho_{hh}}}
\newcommand\corrutau{\ensuremath{\rho_{u,\tau}}}

\newcommand\corruat{\ensuremath{\rho_{ua}^t}}
\newcommand\corruut{\ensuremath{\rho_{uu}^t}}

\newcommand\corrubedt{\ensuremath{\rho_{u,h_b}^t}}

\def\DeltaUplusChanKUinit{3.59}

\def\DeltaUplusChanLRZinit{3.73}

\begin{document}

\maketitle

\newcommand{\KU}{KU2017}
\newcommand{\SKU}{SKU2020}

\begin{abstract}
The role of turbulent large-scale streaks in forming subaqueous sediment ridges on an initially flat sediment bed is investigated numerically.
For this purpose, a series of direct numerical simulations of turbulent open channel flow over a bed of fully-resolved mobile spherical particles has been performed at bulk Reynolds numbers up to $9500$, in computational boxes of different size.
The regular arrangement of quasi-streamwise ridges and troughs at a characteristic spanwise spacing between $1$ and $1.5$ times the mean fluid height is found to be a consequence of the preferential spanwise organization of turbulence in large-scale streamwise velocity streaks at comparable spanwise wavelengths.
Sediment ridges predominantly appear in regions of weaker erosion below large-scale low-speed streaks, while troughs are accordingly found below the corresponding high-speed streaks.
The interaction between the dynamics of the large-scale streaks in the bulk flow and the evolution of sediment ridges on the sediment bed is best described as a `top-down' process, as the arrangement of the sediment bedforms is seen to adapt to changes in the outer large-scale flow organization with a time delay of several bulk time units.
The observed `top-down' interaction of large-scale velocity streaks with the flow in the vicinity of the sediment bed as well as with the sediment ridges agrees fairly well with the conceptual model on causality in canonical channel flows proposed by Jim{\'e}nez (\textit{J. Fluid Mech.}, vol. 842, 2018, P1, \S~5.6).
Mean secondary currents of Prandtl's second kind of comparable intensity and lateral spacing are found over developed sediment ridges and in single-phase smooth-wall channels likewise when averaging the flow field over the streamwise direction and intermediate time intervals of $\mathcal{O}(10)$ bulk time units.
This indicates that the secondary flow commonly observed together with sediment ridges is the statistical footprint of the regularly organized large-scale streaks and their associated Reynolds stress carrying structures.

\end{abstract}


  \section{Introduction}\label{sec:intro}
  Classically, secondary currents have been studied in flows with non-circular cross-section featuring smooth walls  \citep{Nikuradse_1926,Prandtl_1926}, where they are generated in the region close to the side-walls.
It was also Prandtl who proposed the nowadays standard classification of secondary flows into two categories: secondary flows of Prandtl's first kind are due to the skewness of the mean flow axis as in meandering rivers or curved pipes, while secondary flows of the second kind are a pure turbulent phenomenon and originate in anisotropy and non-homogeneity of Reynolds stresses across the flow domain \citep{Bradshaw_1987}. Our current study focusses on the second family of turbulence-induced secondary currents exclusively and in the remainder of the text, we will always refer to this type when using the term secondary flow.

\citet{Hinze_1967,Hinze_1973} later found that secondary flow cells appeared also in flows over bottom-walls covered with laterally alternating smooth and rough regions, centered over the roughness transition. He proposed an interaction between secondary momentum transport and the local production-dissipation balance of the turbulent kinetic energy, through which high-turbulent fluid is carried out of regions with an excess in turbulence production while, simultaneously, low-turbulent fluid is brought into it and vice versa for regions with dominating dissipation.

In hydraulic flows, secondary flow cells mutually interact with the mobile sediment bed,
giving rise to long, essentially streamwise-aligned sediment \textit{ridges}.
Here, the transverse mean flow manifests itself as pairs of quasi-streamwise depth-spanning and counter-rotating flow cells on each side of a ridge, with a transverse spacing of $(1\textup{--}2)\hmean$ \citep{Nezu_Nakagawa_1993}, where $\hmean$ is the mean fluid height.
The interplay between secondary currents and sediment ridges was first claimed in the context of early laboratory studies \citep{Casey_1935,Vanoni_1946,Wolman_Brush_1961} and field observations of dried river beds \citep{Karcz_1967,Culbertson_1967}, but there was disagreement on the actual origin of the process.
\citet{Nezu_Nakagawa_1984} proposed a cascade-like process, according to which the secondary flow cell induced by lateral domain boundaries such as river banks or the side-walls of a laboratory flume give rise to a first ridge in its close proximity, that, in turn, will then trigger a next ridge further apart with the corresponding secondary current and so on.
In a later study, however, \citet{Nezu_Nakagawa_1989} could observe ridges and secondary currents simultaneously rising at different spanwise locations all across the domain.
\citet{Ikeda_1981} followed a different approach and conjectured the underlying mechanism to be
a flow-sediment bed instability that is able to generate sediment ridges and the according secondary flow cells. Although side-wall induced secondary currents may still influence the formation of sediment ridges, they are no necessary condition for their evolution and ridges can accordingly form in absence of side-walls likewise.
These considerations were later confirmed by \citet{Colombini_1993}, who showed by means of a linear stability analysis that a two-dimensional turbulent base flow in an infinitely wide channel is, indeed, linearly unstable with respect to infinitesimal spanwise disturbances of the sediment bed.
For his model, he closed the Reynolds-averaged Navier-Stokes (RANS) equations with a non-linear eddy-viscosity model originally proposed by \citet{Speziale_1987} which allows, in contrast to simpler linear closures, the required anisotropy of the Reynolds stress tensor \citep{Speziale_1982}.
For the set of investigated parameters, a most amplified wavelength of $\lambda_z=1.3\hmean$ could be determined.
Later, \citet{Colombini_Parker_1995} extended the model by including the effect of laterally varying roughness and sand grain distributions as a second source of instability, in agreement with earlier experimental studies \citep{Mueller_Studerus_1979}. In fact, the two different instability mechanisms allow a useful classification of secondary currents in \textit{strip type} (e. g. alternating roughness stripes) and \textit{ridge type} roughness (e. g. topographical variation of the lower wall) or a combination of both \citep{Colombini_Parker_1995,Hwang_Lee_2018}.

Outside the hydraulic community, interest in secondary currents has increased mainly during the past two decades, motivated by observations of secondary flows over complex three-dimensional industrial surfaces such as the replicated surface of a damaged turbine blade investigated by \citet{MejiaAlvarez_Christensen_2010} and \citet{Barros_Christensen_2014}.
Secondary flows seem to represent a robust phenomenon that is ubiquitous in flow configurations that feature a significant laterally heterogeneity. For example, secondary currents have been observed in the flow over
straight riblets \citep{Goldstein_1998},
converging/diverging riblets \citep{Kevin_al_2017,Kevin_al_2019},
spanwise alternating roughness stripes \citep{Mclelland_1999,Wang_Cheng_2005,Wangsawijaya_al_2020},
flow over transverse alternating non-/hydrophobic roughness \citep{Tuerk_2014,Stroh_al_2016},
streamwise aligned artificial ridges of different cross-sectional shape \citep{Wang_Cheng_2006,Vanderwel_Ganapathisubramani_2015,Vanderwel_2019,
Medjnoun_al_2020,Zampiron_al_2020b,Stroh_al_2020b}
or combinations of the above as in \citet{Stroh_al_2020}.
The majority of the listed studies focusses on the general structure of mean secondary motion over the respective heterogeneities, their sense of rotation \citep{Yang_Anderson_2018,Anderson_2019} or the influence of the lateral spacing of the heterogeneities on their intensity and arrangement \citep{Chung_2018,Wangsawijaya_al_2020}.

Less attention has been given to the transient nature of secondary flows. Recently however, fully-resolved direct numerical simulations (DNS) and experiments have allowed to shed some light on instantaneous coherent structures and their relation to the mean secondary flow.
In this spirit, \citet{Uhlmann_al_2007} and \citet{Pinelli_al_2010} were the first to show that the characteristic eight-vortex secondary flow pattern in a square duct is indeed a direct consequence of instantaneous large-scale structures, while the preferential location of the buffer-layer streaks and quasi-streamwise vortices along the four solid side-walls determines the statistics of the mean vorticity distribution.
The close correlation between distinct turbulent coherent structures and the mean secondary flow was further supported by \citet{Uhlmann_al_2010}, who presented a family of traveling wave solutions whose members produce essentially the same secondary flow patterns when averaged over the streamwise direction \citep{Kawahara_al_2012}.
\citet{Sakai_2016} later elaborated that similar relations between coherent structures and the statistical mean secondary flow exist for open ducts just as much.

In canonical laterally homogeneous channels, instantaneous quasi-streamwise vortices appear at all scales in connection with alternating streaks of high and low streamwise velocity \citep{Jimenez_1998,Jimenez_2018}.
The smallest vortices of this kind are those in the buffer layer that form a `self-sustaining process' with the near-wall velocity streaks \citep{Hamilton_Kim_Waleffe_1995,Jeong_Hussain_1997,Schoppa_2002}.
With increasing wall-normal distance, however, energy and enstrophy no longer reside in the same scales: velocity streaks are found to scale self-similarly across the logarithmic layer with the distance to the wall \citep{Jimenez_2013}, whereas vorticity concentrates further away from the wall in structures of essentially the same scale as in the buffer layer \citep{Jimenez_2013b}.
It is thus evident that the role of larger-scale quasi-streamwise rotations that sustain the fully turbulent log-layer streaks cannot be overtaken by a single large-scale vortex, but by the collective effect of a large number of locally isotropic small-scale vortices that globally arrange in large-scale vortex clusters, that are anisotropic enough to generate an average upward motion in their middle \citep{DelAlamo_Jimenez_2006,Jimenez_2018}.
While these collective large-scale rotations are hardly detectable in instantaneous flow visualizations, they have been successfully uncovered in conditional averages \citep{DelAlamo_Jimenez_2006,LozanoDuran_2012} or in low-pass filtered fields \citep{Motoori_Goto_2019}, proving that the self-sustaining process between streaks and quasi-streamwise rotations is a self-similar cascade \citep{Motoori_Goto_2021}, just as the logarithmic layer is \citep{Flores_Jimenez_2010,Kevin_al_2019b}.

Recently, it has been argued by \citet{Kevin_al_2019b} that the mean secondary currents conventionally found over heterogeneous bottom-walls are nothing else than the statistical footprint of these conditional or collective quasi-streamwise rollers. The latter authors argue that the role of the lateral bottom heterogeneities is to laterally lock the large scales structures and quasi-streamwise rollers in the vicinity of the roughness transition, such that they get visible in the long-time average. Structures over homogeneous walls, on the other hand, will appear at any spanwise position with the same probability, explaining the absence of their statistical footprint in form of mean secondary currents when averaged over a sufficiently long time interval. Note that spatial locking should be understood in a sense that the mean lateral position does not vary significantly in time, while the streaks instantaneously meander in response to the quasi-streamwise rollers just as in canonical flows \citep{Kevin_al_2017,Kevin_al_2019b}.
Indeed, such lateral meandering has been detected for a variety of spanwise heterogeneities such as converging/diverging riblets \citep{Kevin_al_2017}, alternating smooth/rough patches \citep{Wangsawijaya_al_2020} and streamwise-aligned artificial ridges \citep{Zampiron_al_2020b}.
In contrast to homogeneous smooth walls, the lateral spacing of bottom wall heterogeneities allows to control the organization of the flow, especially the spacing and location of mean secondary currents.
The above studies show that an optimal lateral spacing of the bottom roughness elements, which enhances the intensity of the secondary currents, corresponds to the typical lateral dimension of the large-scale motions $(1\textup{--}2)\hmean$ \citep{Jimenez_2013,Jimenez_2018}.
For spacings clearly smaller or larger than the outer flow scale, on the other hand, the effect of the heterogeneities is confined to the near-wall region and the regions of roughness transition, respectively, while there is little effect on the outer flow.

In the hydraulic context, sediment ridges similarly form at a lateral spacing of $(1\textup{--}2)\hmean$ \citep{Colombini_1993,Nezu_Nakagawa_1993}, consistent with the lateral dimension of the turbulent large-scale structures with streamwise length $\mathcal{O}(1\textup{--}10)\hmean$ \citep{Tamburrino_Gulliver_1999,Tamburrino_Gulliver_2007,Shvidchenko_Pender_2001,Roy_al_2004}, that have been detected as counterparts to large- and very large-scale motions inherent to canonical turbulent flows \citep{Adrian_Marusic_2012}.
A general relevance of those instantaneous structures for the formation of sediment ridges was formulated, for instance, by \citet{Nezu_2005} based on the smooth-wall duct flow experiments of \citet{Nezu_Rodi_1985}. It was discovered that the mean secondary currents decay towards the duct center, while instantaneous rotating motions remain observable. \citet{Nezu_2005} therefore claimed that the `instantaneous secondary currents may trigger the formation of smooth/rough striping of the bed and the development of sand ribbons'.
\citet{Shvidchenko_Pender_2001} further proposed that passing `macroturbulent structures' with a streamwise spacing of $(4\textup{--}5)\hmean$ induce laterally alternating regions of high and low sediment erosion and transport inside and outside their traveling path, respectively. The lateral variation of sediment transport is due to the fact that sediment erosion is predominantly driven through the action of large-scale high-speed sweeps reaching the bed, rather than by ejections lifting it up in the low-speed regions \citep{Gyr_Schmid_1997,Cameron_al_2020}.

Even though it appears conclusive that organized recurrent large-scale structures are responsible for the formation of sediment ridges, it lacks, to the best of our knowledge, clear experimental and numerical evidence of the above elaborated mechanisms.
While simultaneous measurements of three-dimensional flow structures and individual sediment motion remain challenging \citep{Wang_Cheng_2005}, direct numerical simulations featuring fully-resolved particles have recently proven useful to study complex interactions between mobile sediment and the different scales of wall-bounded turbulence. So, \citet{ChanBraun_al_2011} and \citet{Mazzuoli_Uhlmann_2017} investigated the flow over fixed arrays of fully-resolved spherical roughness elements from the hydraulically smooth to rough regime and quantified the pressure and forces exerted on the particles.
\citet{Kidan_al_2013} showed for moderate Reynolds numbers that the lateral migration of particles in a smooth-walled channel into the near-wall low speed streaks can be attributed to the quasi-streamwise streaks in the buffer-layer.
\citet{Vowinckel_2017b} studied the dynamics of a limited number of particles over a fixed pattern of roughness elements and reported particle clustering in streamwise aligned chains appearing together with mean secondary currents.
Recently, Kidanemariam et al. investigated the evolution of transverse bedforms over a thick mobile sediment bed under laminar and turbulent conditions in a series of simulations with fully-resolved sediment grains \citep{Kidan_Uhlmann_2014a,Kidan_Uhlmann_2017,Scherer_Kidan_Uhlmann_2020}.
To investigate the minimal length of those patterns, \citet{Kidan_Uhlmann_2017} performed a sequence of simulations successively reducing the streamwise box dimension in a similar fashion as the minimal flow units of \citet{Jimenez_Moin_1991}. That way, it was possible to show that there exists a minimal streamwise domain length below which the domain cannot accommodate any unstable pattern wavelength and consequently no ripple-like bedforms evolve.
The correct scaling of this minimal unstable wavelength with the particle diameter $D$ rather than with the fluid height $\hmean$ was shown by \citet{Scherer_Kidan_Uhlmann_2020} for the investigated parameter range, and it was possible to quantify the minimal box length as $L_{x,min}\approx80D$ for any ripple-like pattern to rise.

In most of the simulations conducted by \citet{Kidan_Uhlmann_2017} and \citet{Scherer_Kidan_Uhlmann_2020}, sediment ridges were observed to be the first bedforms to evolve rapidly after the onset of particle motion, while the actually investigated transverse patterns required roughly two orders of magnitude longer to develop and eventually take over the smaller-amplitude ridges.
The current study aims to provide numerical evidence for the ridge formation process by, first, clarifying the role of large-scale turbulent structures in the generation process of sediment ridges and, second, by investigating the relation between coherent structures and mean secondary currents in the vicinity of the bedforms.
In the following \S~\ref{sec:numerics}, we will first give a brief overview of the numerical scheme used in the current simulations, while details on the physical system under consideration will be given in \S~\ref{sec:flow_config}.
\S~\ref{sec:interf_extr} summarizes the applied procedure to extract the fluid-bed interface, which shall form the basis for the analysis of the simulation results in \S~\ref{sec:results}.
The observed `top-down mechanism' between large-scale structures, the near-bed flow organization and sediment ridges will be discussed in \S~\ref{sec:discussion}, before we close with a summary of the relevant outcomes in \S~\ref{sec:conclusion}.

  \section{Numerical method}\label{sec:numerics}
  The numerical scheme used in the current study has been first used for the simulation of sediment transport in a carrying fluid in \citet{Kidan_Uhlmann_2014b}. Its main components are an immersed boundary technique based on the concept proposed by \citet{Uhlmann_2005} and a particle contact model whose features will be discussed below.
The time evolution of the fluid phase is governed by the continuity and Navier-Stokes equations for incompressible Newtonian fluids, extended by an additional force term emanating from the immersed boundary formulation which enforces the no-slip boundary conditions at the phase-boundaries at each particle's surface. This technique allows then to discretize the physical domain by a fixed, uniform staggered finite difference grid, while for the numerical time integration, a mixed explicit-implicit framework is chosen, consisting of a Crank-Nicholson scheme for the diffusive terms and a three-step low storage Runge-Kutta scheme for the convective part of the equations.
Within a fractional step method, the governing equations are first solved disregarding the continuity equation, then projecting the velocity to the space of solenoidal velocity fields.
Back- and forth-transformations of physical information between the global Eulerian fluid grid nodes and the Lagrangian marker points that represent the surface of the particles in the context of the immersed boundary method are performed based on regularized discrete delta functions \citep{Peskin_2002}.

The dynamics of the spherical particles follow the Newton-Euler equations for rigid body motion and are integrated in time alongside the fluid motion, using a sub-stepping procedure with $\mathcal{O}(100)$ particle sub-steps per fluid time step to take care of the different characteristic time scales of fluid and particle motion \citep{Kidan_2015}.
The exchange of linear and angular momentum during particle-wall and inter-particle contacts, from which the latter occur frequently in bedload dominated sedimentary flows, are modeled by a soft-sphere discrete element model that describes the interaction of solid objects by the analogy to simple mechanical mass-spring-damper systems. The individual force and torque components have a finite range of action, that is, two objects are in contact (i.e. they feel the contact force and torque) if and only if the minimal distance between their surfaces $\surfdist$ is below this force range $\forcerange$. In its current form, the collision force and torque include three components, namely a normal elastic, a normal damping and a tangential damping contribution.
The normal elastic component is a linear function of the overlap length
$\penetrationl=\forcerange-\surfdist$ with a constant stiffness coefficient $\normstiff$.
The normal damping component is a linear function of the normal component of the
relative particle velocity between the two particles at the contact point, with a constant proportionality coefficient $\normfriction$.
In a similar fashion, the tangential damping component linearly depends on the relative tangential particle velocity at the contact point, premultiplied by a constant tangential friction coefficient $\tangfriction$. Note that the latter force component has a natural upper traction limit in the Coulomb friction coefficient $\Coulombfriction$. For a more detailed presentation of the described model and an extensive validation study, the interested reader is referred to \citet{Kidan_Uhlmann_2014b}.

The contact model thus comes with a set of five unknown parameters, that are, the four force parameters ($\normstiff$,$\normfriction$,$\tangfriction$,$\Coulombfriction$) plus the force range $\forcerange$. As pointed out by \citet{Kidan_Uhlmann_2014b}, a quantity that is often more accessible than the force parameters is the dry restitution coefficient $\dryrest$ which represents the absolute value of the ratio between the normal components of the relative velocity before and after a dry collision. It is therefore appropriate to use this quantity to eliminate the normal damping coefficient $\normfriction$ from the set of unknowns and to determine it as a function of  $\dryrest$ and $\normstiff$, such that the actually used set of model force parameters reads  ($\normstiff$,$\dryrest$,$\tangfriction$,$\Coulombfriction$).
A parameter sensitivity analysis has been recently presented in \citet{Scherer_Kidan_Uhlmann_2020}, outlining the influence of several of the parameters on the simulation results.
For the simulations carried out in the present work, the normal stiffness coefficient is set in a range $\normstiff=\num{17000}\textup{--}\num{40000}$ times the submerged weight of a single particle divided by the particle diameter $D$, the Coulomb friction limit is chosen in a range  $\Coulombfriction=0.4\textup{--}0.5$ and the friction coefficient is set to  $\tangfriction=\normfriction$.
The force range $\forcerange$ is set equal to the uniform grid spacing $\deltax$. The dry restitution coefficient is $\dryrest=0.3$ except for case \ichanHLRSlonginit{}, where we have used a larger value $\dryrest=0.9$. Note that the variation of $\dryrest$ has no significant effect on the bed formation in the currently investigated bedload-dominated regime, as has been reported in the aforementioned sensitivity analysis.

  \section{Flow configuration and parameter values}\label{sec:flow_config}
  %
%
%
\begin{figure}
    \centering
         \begin{minipage}{.9\linewidth}
           \includegraphics[width=\linewidth]
           {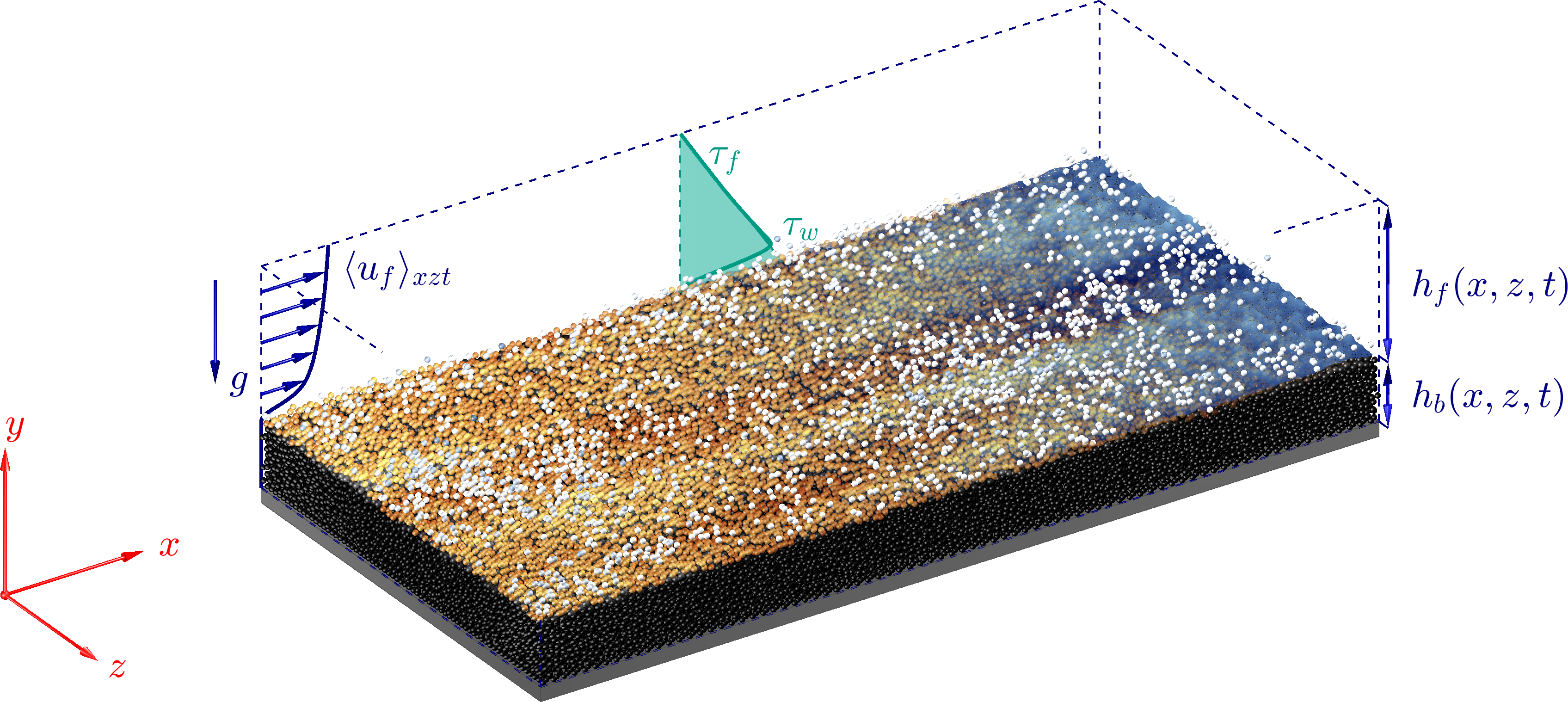}
         \end{minipage}
  	\caption{
    Sketch of the physical system analyzed in the multiphase simulations.
    Mean flow and gravity are pointing in positive $x$- and negative $y$-direction,
    respectively.
    The mean flow profile $\uvecxzt=(\uxzt(y),0,0)^T$ is shown in blue, while the green
    curve represents the wall-normal variation of the mean fluid shear stress $\tau_f(y)$.
    $\hf$ and $\hb$ are the local instantaneous fluid and bed height, respectively.
    Particles are colored depending on their location: bed particles are colored in black, interface particles in orange to yellow with increasing wall distance and transported
    particles are indicated by white color. For the sake of comparison, the bed at the
    downstream end of the domain is overlain by the two-dimensional surface that represents
    the fluid-bed interface as defined in \S~\ref{sec:interf_extr}, the color ranging from dark to bright blue with increasing height.
}
\label{fig:interface_extr}
\end{figure}

%
%
%
\begin{table}
  \begin{center}
%

%
%
\begin{tabular}{l c c c c c c c c}
\multicolumn{1}{c}{Case}&
\multicolumn{1}{c}{\Reb}&
\multicolumn{1}{c}{\Ret}&
\multicolumn{1}{c}{\dratio}&
\multicolumn{1}{c}{\Ga}&
\multicolumn{1}{c}{\Dplus}&
\multicolumn{1}{c}{\HftoD}&
\multicolumn{1}{c}{\HbtoD}&
\multicolumn{1}{c}{\shields}\\[3pt]
  {{\color{col_260} $\solidthick$}} \ichanD{}&3011&259.43&   2.5& 28.37&  9.81& 26.43& 11.97&  0.12 \\[1.5pt]
  {{\color{col_222} $\solidthick$}} \ichanKUinit{}&3011&247.86&   2.5& 28.37&  9.30& 26.66& 11.74&  0.11 \\[1.5pt]
  {{\color{col_1} $\solidthick$}} \ispecHLRSlonginit{}&12100&648.69&   -&   -&   -&   -&   -&   - \\[1.5pt]
  {{\color{col_21} $\solidthick$}} \ichanHLRSlonginit{}&9483&827.81&   2.5& 56.68& 29.12& 28.43&  7.13&  0.26 \\[1.5pt]
  {{\color{col_5} $\solidthick$}} \ispecLRZinit{}&3350&209.83&   -& -& -& -&  -&  -\\[1.5pt]
  {{\color{col_511} $\solidthick$}} \ichanLRZinit{}&3012&249.31&   2.5& 28.37&  9.42& 26.46& 11.94&  0.11 \\[1.5pt]
  \end{tabular}
  \caption{
  Physical parameters of the simulations. $Re_b$, $Re_{\tau}$ and $D^+$ are the bulk Reynolds number, the friction Reynolds number and the particle Reynolds number, respectively. The density ratio $\rho_p/\rho_f$ and the Galileo number $Ga$ are imposed in each simulation, whereas the Shields number $\theta$, the relative submergence $H_f/D$ and the relative sediment bed height $H_b/D$ are computed a posteriori (cf. table~\ref{tab:param_numer}).
  }
\label{tab:param_phys}


\hrulefill \\
\hfill \\

%
%
\begin{tabular}{l c c c c r c}
\multicolumn{1}{c}{Case}&
\multicolumn{1}{c}{$[\Lx \times \Ly \times \Lz]/\hmean$}&
\multicolumn{1}{c}{$[\Lx \times \Ly \times \Lz]/D$}&
\multicolumn{1}{c}{$D/\deltax$}&
\multicolumn{1}{c}{$\min(\deltayplus)$}&
\multicolumn{1}{c}{$N_p$}&
\multicolumn{1}{c}{$\tobs/\tbulk$} \\[3pt]
  {{\color{col_260} $\solidthick$}} \ichanD{}&$  1.94 \times   1.45 \times   2.91$&$  51.2 \times   38.4 \times   76.8$&10&  0.98&  \num{43730}& 678 \\[1.5pt]
  {{\color{col_222} $\solidthick$}} \ichanKUinit{}&$  5.76 \times   1.44 \times   2.88$&$ 153.6 \times   38.4 \times   76.8$&10&  0.93& \num{127070}&  94 \\[1.5pt]
  {{\color{col_1} $\solidthick$}} \ispecHLRSlonginit{}&$  5.33 \times   1.00 \times   2.67$&   - &-&  0.05&      0& 397 \\[1.5pt]
  {{\color{col_21} $\solidthick$}} \ichanHLRSlonginit{}&$  5.00 \times   1.25 \times   2.50$&$ 142.2 \times   35.6 \times   71.1$&36&  0.81&  \num{92292}&  59 \\[1.5pt]
  {{\color{col_5} $\solidthick$}} \ispecLRZinit{}&$ 12.00 \times   1.00 \times  16.00$& -& -&  0.06& 0 &  432 \\[1.5pt]
  {{\color{col_511} $\solidthick$}} \ichanLRZinit{}&$ 11.61 \times   1.45 \times  15.48$&$ 307.2 \times   38.4 \times  409.6$&10&  0.94&\num{1406983}&  84 \\[1.5pt]
\end{tabular}
  \caption{Numerical parameters of the simulations. The computational domain has dimensions $L_i$ in the $i-$th direction and is discretized using a uniform finite difference grid with mesh width $\Delta x=\Delta y=\Delta z$ for the multi-phase simulations, while the smooth-wall simulations were performed using a spectral method featuring a non-uniform distribution of the grid/collocation points in the three spatial directions.
  $N_p$ is the total number of particles in the respective case and $\tobs$ is the total observation time of each simulation starting from the release of the moving particles at $t=0$.
  Time dependent physical and numerical parameters in tables~\ref{tab:param_phys} and \ref{tab:param_numer} ($\Ret$, $\Dplus$, $\hmean$, $\hbedmean$, $\theta$, $\deltayplus$) are computed as an average over the entire simulation period.
  }
  \label{tab:param_numer}
  \end{center}
\end{table}

%
%
%

In the course of the current study, we have conducted four individual simulations of turbulent open channel flow past a mobile sediment bed formed out of spherical sediment particles.
The set of simulations is complemented by two additional reference simulations of single-phase smooth-wall open channel flow performed with a pseudo-spectral code using Fourier and Chebyshev expansions in the periodic and wall-normal directions, respectively \citep{Kim_Moin_Moser_1987}.
The physical system simulated in the two-phase cases is sketched in figure~\ref{fig:interface_extr}. The Cartesian coordinate system has its origin on the bottom wall of the channel, such that the components of an arbitrary spatial position vector $\xvec=(x,y,z)^T$ are measured along the streamwise ($x$), wall-normal ($y$) and spanwise ($z$) direction, respectively.
Accordingly, the fluid velocity vector field at position $\xvec$ has components $\uvec(\xvec)=(\uf,\vf,\wf)^T$. Subscripted letters `$f$' and `$p$' indicate fluid- and particle-related physical quantities.
For the statistical analysis, we introduce the following averaging scheme
\begin{subequations}
 \begin{align}
  \uvec &= \uvect + \uvecfluct \label{eq:t_avg}\\
  \uvec &= \uvecxz + \uvecflucxz, \label{eq:xz_avg}
 \end{align}
\end{subequations}
where \eqref{eq:t_avg} and \eqref{eq:xz_avg} describe the decompositions with respect to time-average and plane-average, respectively. Averaging in the homogeneous directions and/or time is indicated by angular brackets with the respective indices, $\overline{\bullet}$ and $\bullet^{\prime}$ are the corresponding temporal and spatial fluctuations.
We further introduce the following decomposition based upon streamwise averaging only:
\begin{equation}\label{eq:x_avg}
  \uvecx = \uvecxz + \uvecflucx,
\end{equation}
where $\uvecflucx=\xavg{\uvecflucxz}$ are the spatial fluctuations with respect to the streamwise average.

The flow in both, single- and multi-phase systems, is driven by a time-dependent pressure gradient that maintains a constant fluid mass flow rate $\qf$ in the streamwise direction throughout the entire simulation interval. Gravity is acting in the negative $y$-direction with amplitude $g=\absgrav$. The simulation domain is periodically repeated in the wall-parallel $x$- and $z$-directions with fundamental periods $\Lx$ and $\Lz$, respectively. The lower bottom wall and the flat free surface are accounted for as no-slip and free-slip boundary conditions, respectively.
In the wall-normal direction, the computational box has an extent $\Ly$.
Note that in the sediment-laden flows, the domain consists of a particle-dominated subdomain of mean height $\hbedmean$, henceforth denoted as `the sediment bed', and the upper fluid-dominated region of mean height $\hmean=\Ly-\hbedmean$. A rigorous definition of the two distinct sub-domains and their mean heights will be given in \S~\ref{sec:interf_extr}. In the sediment-laden simulations, the fluid-bed interface takes the role of a virtual wall, and hence we will frequently refer to a shifted wall-normal coordinate $\yrel = y - \hbedmean$.

The mean wall-shear stress $\tauw$ is evaluated at the wall-normal location of the virtual wall $\yrel=0$ by interpolating the pure fluid stress
\begin{equation*}
  \tauf(\yrel)/\rhof=\fnu \mathrm{d}\uvecxzt/\mathrm{d}\yrel - \xztavg{\ufluct\vfluct}
\end{equation*}
from the bulk where it varies essentially linearly down to this location.
Two additional characteristic length scales of the system are the particle diameter $D$ and the viscous length $\deltanu=\fnu/\utau$, respectively, where $\utau=\sqrt{\tauw/\rhof}$ is the friction velocity. Following the general conventions, we shall denote quantities that are non-dimensionalized with $\utau$ and/or $\fnu$ with a superscript $(\bullet)^+$ and refer to them as scaled in inner or wall units.

From these three length scales, we derive the friction Reynolds number $\Ret=\hmean^+=\hmean/\deltanu$, the particle Reynolds number $\Dplus=D/\deltanu$ and the relative submergence $\HftoD$, respectively. Let us further introduce the bulk Reynolds number as $\Reb=\qf/\fnu=\ubulk\hmean/\fnu$, where $\ubulk=\qf/\hmean$ is the bulk velocity.
Note that in virtue of the increased parameter space that comes with the additional degrees of freedom of the mobile particles, it requires two additional non-dimensional numbers to fully describe the two-phase system. Here, we choose the density ratio of the two phases $\rhop/\rhof=2.5$, where the value of $2.5$ is close to that of glass beads or sand grains in water, and the Galileo number $Ga=\ugrav D/\fnu$, where $\ugrav=\sqrt{(\rhop/\rhof - 1)\absgrav D)}$ is the gravitational velocity.
The squared ratio of the Galileo and the particle Reynolds number $\shields=(\Dplus/\Ga)^2=(\utau/\ugrav)^2$, the Shields number, can be understood as a ratio between destabilizing and stabilizing effects and is thus a measure for the ability of a flow to erode sediment.
A conventionally applied critical value of $\shieldscrit= 0.03\textup{--}0.05$ marks the onset of sediment erosion in a turbulent stream, that slightly depends on the Galileo number \citep{Soulsby_Whitehouse_1997,Wong_Parker_2006,Franklin_Charru_2011}.

Tables~\ref{tab:param_phys} and \ref{tab:param_numer} summarize the relevant parameters of the simulations. As a short-hand notation, we will identify each case with a name combining the size of the underlying computational domain and its inherent friction Reynolds number in the remainder of this work. The smooth-wall single-phase flows are indicated by a respective subscript.
The short (S), medium (M) and large (L) domains feature streamwise and lateral periods of
approximately $\Lx/\hmean\in\{2,5,12\}$ and $\Lz/\hmean\in\{3,16\}$, respectively, and friction Reynolds numbers $250\lesssim \Ret \lesssim 850$.
Let us remark that for the smooth-wall reference simulations, the values of the friction Reynolds number are set to $\Ret=210$ and $\Ret=650$. These values match those of cases \ichanLRZinit{} and \ichanHLRSlonginit{}, respectively, but when the bed is still at rest. As the bed evolves, the values of $\Ret$ increase as a result of the sediment bed modulation.

The box dimensions $\Lx^+$ and $\Lz^+$ are sufficiently larger than those of \citet{Jimenez_Moin_1991}'s minimum flow unit and as such accommodate the full near-wall regeneration cycle. A comparison of the outer-scaled box width $\Lz/\hmean$ with the findings of \citet{Flores_Jimenez_2010} for closed channels, on the other hand, indicates that our narrow domains are just wide enough to host a single full regeneration cycle of the largest log-layer streaks, for which the authors have estimated a minimal box width $\Lz/\hmean \approx 3$. Surely, the comparison to our open channel is limited in the vicinity of the free-surface, but should give a good estimate for the rest of the domain \citep{Bauer_Sakai_Uhlmann_2021}.

The preparation of each individual simulation follows a careful procedure described in detail by  \citet{Kidan_Uhlmann_2014a}, in which first a pseudo-randomly arranged sediment bed is created through settling of a large amount of particles under the action of gravity in a quiescent fluid. The number of particles per simulation has been set such that all simulations feature a comparable relative submergence $\HftoD\approx26\textup{--}28$, while the Galileo number is chosen a priori such that the a resulting Shields number is sufficiently larger than its critical value to allow for particle erosion.
First, a turbulent flow field is developed by fixing all particles in space.
Then, the particles are released again at a time which we define as $t=0$.
Let us remark that a small fraction of the particles, namely the layer closest to the bottom wall, remain fixed in space even for times $t>0$ to create a rough boundary.
In the following, we will use the bulk time unit $\tbulk=\hmean/\ubulk$ as a reference time.

As has been stated earlier, ridges appear shortly after particles are released, but eventually, transverse bedforms appear and dominate the evolution process after $100\textup{--}200 \tbulk$. For the investigation of ridge evolution at the current parameter point, consequently, there exist basically two options: either the observation time $\tobs$ in arbitrary long domains is limited to the initial simulation phase in which transverse bedform instabilities are still small enough to be of minor importance, or the streamwise domain length $\Lx/D$ is reduced below the critical value of ${\Lx}_{,c}/D \approx 80$ determined by \citet{Scherer_Kidan_Uhlmann_2020} such that transverse bedform instabilities are effectively suppressed for arbitrary simulation times.

Again recalling that the main concern of the current study is the role of large-scale structures for the generation of sediment ridges, we prefer the former concept that allows the accommodation of longer streaks and reduces the domain size effects, accepting the limited observation interval as it is still sufficient to study all relevant processes due to the fast formation of the sediment ridges within a few bulk time units.
In the single-phase simulations \ispecHLRSlonginit{} and \ispecLRZinit{}, on the other hand, we have exploited the opportunity to gather statistics over clearly longer time intervals than in their multi-phase counterparts.
For comparisons with longer time series over ridges, we have additionally performed simulation~\ichanD{} that features a streamwise box length $\Lx=51.2D$ to suppress the rise of transverse bed features.
This concept is very similar to the streamwise-minimal channels of \citet{Toh_Itano_2005} in that sense that the large-scale streaks can be considered as infinitely long due to the missing spatial de-correlation, while the self-sustained regeneration cycle in the buffer layer \citep{Hamilton_Kim_Waleffe_1995} is captured by the domain without any restrictions.

Eventually, it should be stressed that case \ichanKUinit{} is found at the same parameter point as case $H6$ of \citet{Kidan_Uhlmann_2017}, but represents an own independent simulation conducted in the context of this study. Due to technical reasons, data in the first roughly $5\tbulk$ of the simulation are not available for the study.
Furthermore, to the best of the authors' knowledge, the newly presented simulations \ichanLRZinit{} and \ichanHLRSlonginit{} represent the largest number of fully-resolved particles simulated and the highest ever obtained Reynolds number in DNS studies with fully-resolved particles, respectively.
The resolution which is required to correctly resolve all flow scales at comparably high Reynolds numbers naturally comes with immense computational costs, summing up to a total amount of approximately 9 million CPU hours consumed for the investigated simulation interval of around 60 bulk time units in case \ichanHLRSlonginit{}, including a number of around 16.8 billion grid nodes.

  \section{Definition of the fluid-bed interface}\label{sec:interf_extr}
  The determination of the two-dimensional interface that separates the domain in two distinct regions, that are, the sediment bed and the fluid-dominated region above it follows the procedure described in \citet{Scherer_Kidan_Uhlmann_2020}. For the sake of completeness, we will briefly recapitulate the concept in the following. For a more detailed presentation of the methodology, the interested reader is referred to the original study.

First, let us decompose the vertical domain length $\Ly$ for each point of the ($x$,$z$)-plane in two contributions as
\begin{equation*}
  \Ly = \hb(x,z,t) + \hf(x,z,t),
\end{equation*}
where $\hb$ and $\hf$ are the local bed height and thickness of the fluid dominated region, respectively. In the special coordinate system that we have chosen, the local \textit{fluid-bed interface} that separates the two distinct sub-domains is found at $y=\hb(x,z,t)$.
Implicitly contained in these considerations is that the sediment bed contour can be formulated in a continuous way, while the sediment bed is naturally only represented by a set of discrete particles. Several methods are known to approximate the presence of the sediment bed in a continuous way, for instance, by identifying the interface with the wall-normal location at which a threshold for the solid volume fraction is attained, as in \citet{Kidan_Uhlmann_2014a,Kidan_Uhlmann_2017}.

The methodology proposed in \citet{Scherer_Kidan_Uhlmann_2020} follows another approach that first identifies a set of `interface particles' which form the uppermost sediment layer of the bed by an algorithm that will be outlined below.
In a second step, a continuous two-dimensional manifold is obtained through triangulation between the interface particles and interpolation to a regular grid.

The sorting-algorithm bases on the conventional morphological classification of sediment as either belonging to the quasi-stationary bed or being transported within the bedload or suspended load layer \citep{Bagnold_1956,VanRijn_1984}.
Potential candidates for `bed particles' are only those that feature a negligible kinetic energy $(|\upvec|/\ugrav)^2$ and a non-vanishing wall-normal contact force $F_{c,y}/F_{w}$, which a particle necessarily feels while being part of a densely packed bed. Here, $F_{w} = (\rhop-\rhof)\absgrav \pi D^3/6$ is the submerged weight of a single spherical particle.
All particles that fulfill these requirements are classified as bed particles. From this set, we determine in a second step the uppermost sediment layer (the interface particles) geometrically using the $\alpha$-shape algorithm of \citet{Edelsbrunner_Muecke_1994}. While conceptually similar to a conventional convex hull around a set of points, the $\alpha$-shape allows non-convexity for length scales over some threshold radius $\alpha$ (here taken as $1.1$ times the particle diameter) while it is strictly convex for length scales smaller than this threshold. Under these restrictions, an enclosing surface can be generated by means of triangulation. Nodes that relate to interface particle centers are those that bound a triangle with an outward pointing normal with positive wall-normal component. The information about the local bed height is eventually transferred to a regular equidistant Eulerian grid in the ($x$,$z$)-plane with sampling width of $1D$ by means of linear interpolation.
An exemplary visualization of bed, interface and mobile particles together with the generated fluid-bed interface is supplemented to figure~\ref{fig:interface_extr}.

  \section{Results}\label{sec:results}
  Classically, sediment ridges have been described as long, streamwise
aligned patterns \citep{Casey_1935,Vanoni_1946}, that are
`parallel to each other, of little relief,
and of a uniform transverse spacing' \citep{Allen_1968}.
%
%
%
\begin{figure}
       \centering
    \includegraphics
    {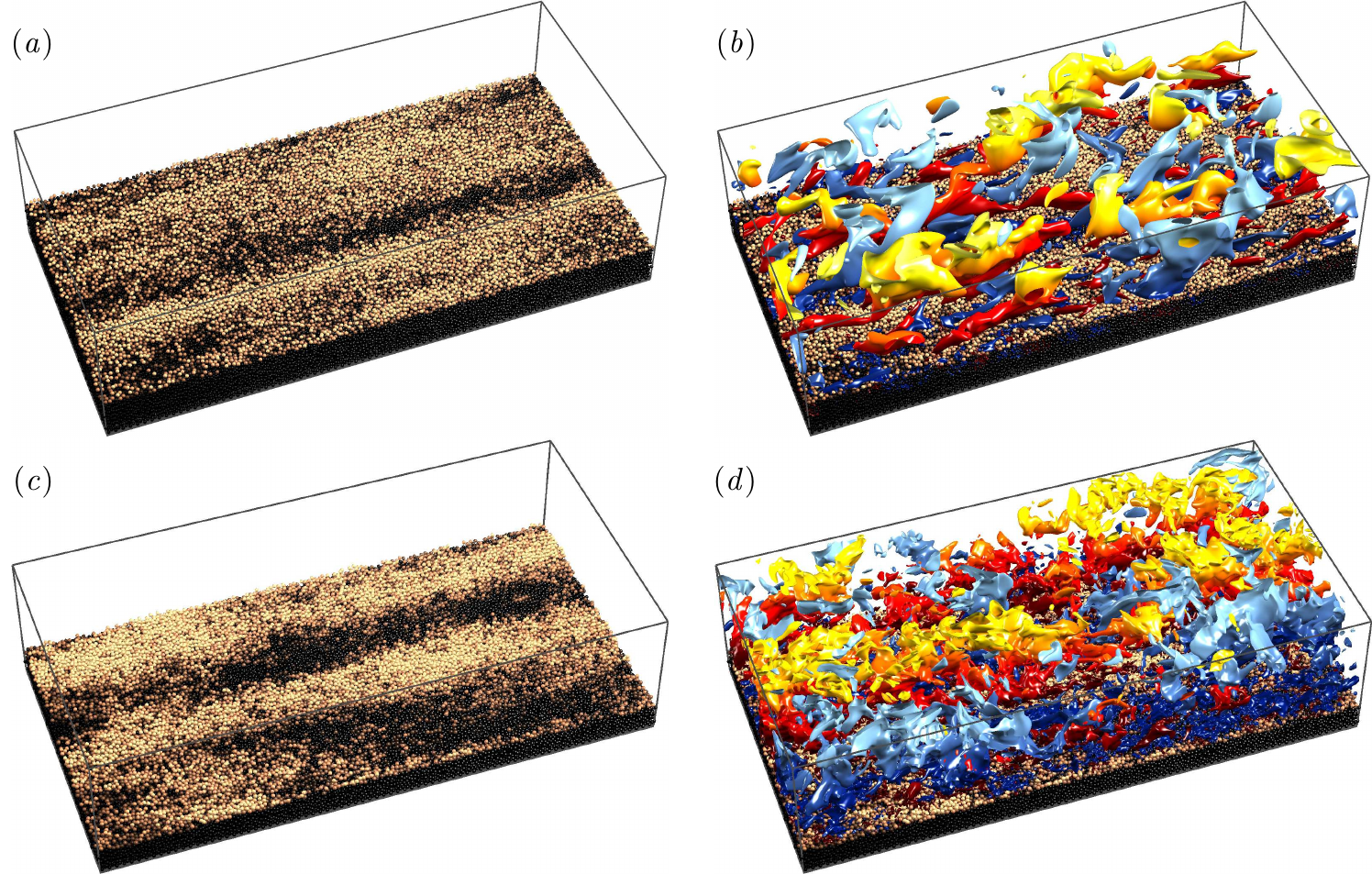}
  	\caption{Instantaneous visualization of the evolved sediment ridges
    (\textit{a},\textit{c}) compared with the instantaneous distribution of three-dimensional Reynolds stress carrying $Q^-$ structures
    (\textit{b},\textit{d}) characterized by connected regions fulfilling
    $|-\uflucxz(\xvec,t)\vflucxz(\xvec,t)| > H \, u_{f,rms}(y)v_{f,rms}(y)$ with $H=1.75$ \citep{LozanoDuran_2012}. Ejection structures are indicated by reddish colors, while sweeps are colored in blue, with brighter colors indicating a larger distance to the bottom wall.
    Particles are colored depending on their wall-normal location, ranging form dark to light brown with increasing coordinate $y$. For the sake of clarity, only bed and interface particles are shown (cf. \S~\ref{sec:interf_extr}).
    In each panel, flow is from bottom left to top right.
    (\textit{a},\textit{b}) \ichanKUinit{} ($t/\tbulk=40$),
    (\textit{c},\textit{d}) \ichanHLRSlonginit{} ($t/\tbulk=59$).
}
\label{fig:3d_snaps_bed_flow}
\end{figure}

%
%
%
The instantaneous snapshots of the sediment bed provided in
figure~\ref{fig:3d_snaps_bed_flow} clearly reveal that the
ridges observed in the current simulations share all
these features. The sediment bed is covered by
laterally alternating ridges and troughs of small amplitude
($1D-2D$) that are essentially parallel to each other and to the mean flow direction.
The bedforms are seen to span over the entire streamwise domain for all cases,
which is in complete agreement with experimental observations that ridges can easily reach streamwise extensions of $\mathcal{O}(10\hmean)$ \citep{Wolman_Brush_1961}.
Perhaps even more important is the fact that these observations confirm for the first time that sediment ridges can form due to mechanisms that are completely independent of side-wall induced secondary currents \citep{Ikeda_1981,Colombini_1993}.

To give a first impression on the comparable lateral organization of bed and large-scale structures, we have supplemented to figure~\ref{fig:3d_snaps_bed_flow} instantaneous visualizations of the Reynolds stress carrying ejection ($\uflucxz<0$,$\vflucxz>0$) and sweep structures ($\uflucxz>0$,$\vflucxz<0$) (collectively termed as $Q^{-}$'s) introduced by \citet{LozanoDuran_2012} as a generalization of the classical quadrant analysis \citep{Wallace_1972} to three dimensional objects.
According to the former study, coherent ejection and sweep structures are connected sub-domains for which $|-\uflucxz(\xvec,t)\vflucxz(\xvec,t)| > H \, u_{f,rms}(y) v_{f,rms}(y)$ with $H=1.75$ and $u_{f,rms}=\sqrt{\uuprime}$.

It is remarkable that regions of intense ejections seem to reasonably well correlate with the lateral positions of preferential sediment deposition, i.e. ridges, while, vice versa, intense sweep events seem more likely to occur above the troughs where erosion is dominant. Our observations support those of \citet{Gyr_Schmid_1997} that sediment erosion is mainly due to local sweep events that are naturally directed towards the bed \citep{Jimenez_2018}.
Note that, by definition, large-scale ejections live in the low-speed streaks whereas sweeps populate the high-speed streaks of the logarithmic layer, thus we can equivalently conclude that ridges are found below the large low-speed streaks and troughs accordingly next to the high-speed streaks, which has been also verified by comparable visualizations (plots not shown).
To avoid confusion, the term `streak' will always refer to the velocity structures of the logarithmic layer in the remainder of this work if not otherwise stated, whereas the nearest-wall structures will be always explicitly termed as buffer layer structures.

\subsection{Sediment ridge evolution}
In the following, we provide more rigorous quantitative analysis of the relation between bedforms and the turbulent flow.
%
%
%
\begin{figure}
  \centering
  \includegraphics
  {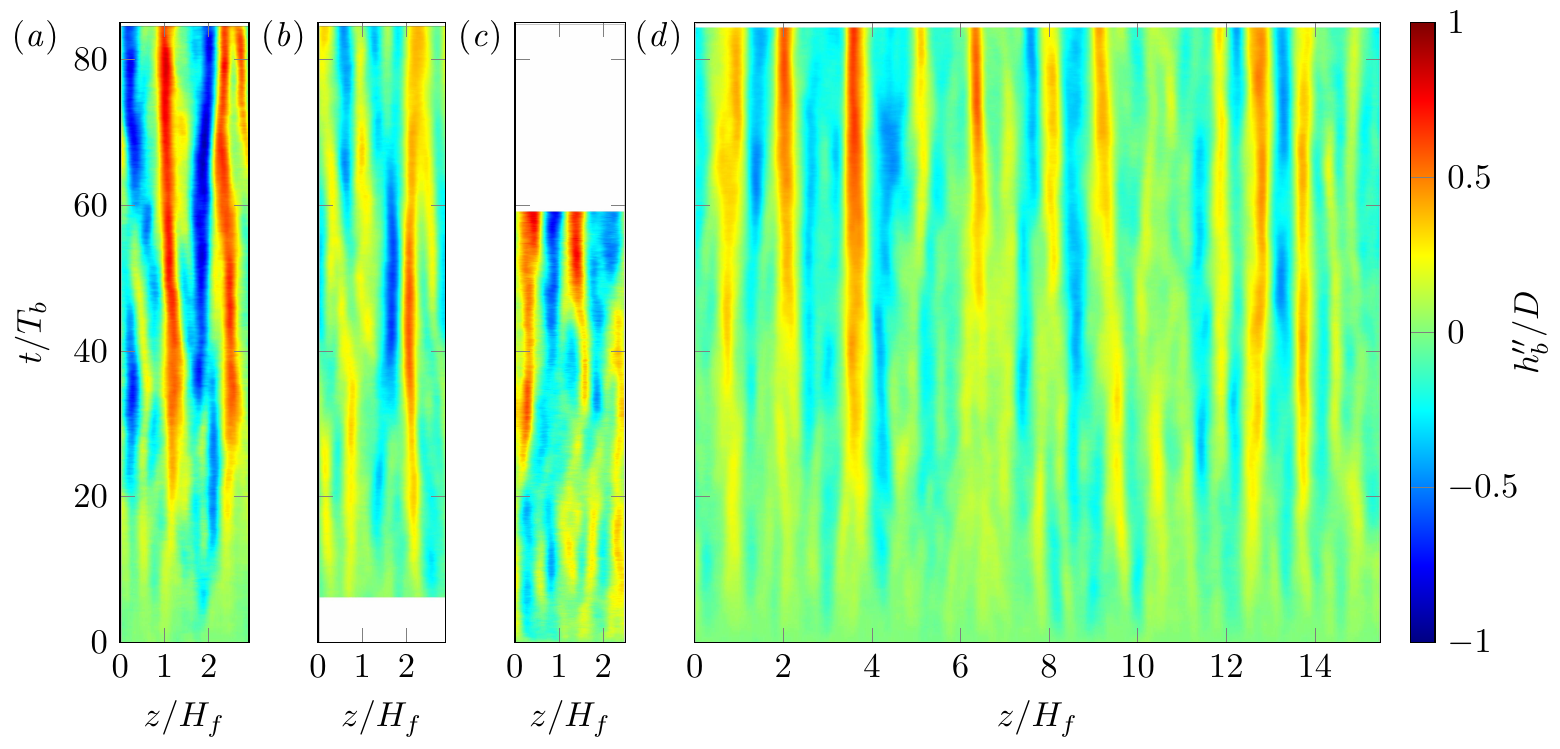}
\caption{Space-time plot of the streamwise-averaged sediment bed height fluctuations
$\hbflucx(z,t)/D$. Blue and red regions refer to
troughs and crests of the streamwise-averaged fluid-bed interface profiles, respectively.
Cases
(\textit{a}) \ichanD{},
(\textit{b}) \ichanKUinit{},
(\textit{c}) \ichanHLRSlonginit{},
(\textit{d}) \ichanLRZinit{}.
}
\label{fig:spacetime_interf_xavg}
\end{figure}

%
%
As mentioned in \S~\ref{sec:flow_config}, all simulations are initiated with an initially macroscopically flat sediment bed and ridges form solely under the action of the turbulent structures after a few bulk time units. Figure~\ref{fig:spacetime_interf_xavg} shows the time evolution of the fluctuations of the streamwise-averaged sediment bed $\hfflucx(z,t)$. Note that
the sediment ridges in the time period of interest are essentially parallel to and without relevant perturbations in the streamwise direction such that they can be considered as essentially
statistically invariant in the mean flow direction.

First, it is concluded that ridges form at different spanwise locations more or less simultaneously and hence independently from each other right after the release of the particles.
It is interesting to see that in the early phase, say the first $20$ bulk time units of each simulation, the lateral spacing between adjacent crests and troughs is clearly lower than the conventionally found values of $(1\textup{--}2)\hmean$ \citep{Wolman_Brush_1961,Ikeda_1981,Colombini_1993}.
Advancing in time, however, a coalescence of the bedforms is observed that causes a reduction of the number of individual ridges.
After approximately $40$ bulk time units, merging or splitting events between ridges are less frequent but still observable and the remaining bedforms now feature a lateral spacing approximately in the expected range and a more pronounced amplitude throughout all simulations. Note that an exact match between the observed/estimated spacings and our results is not to be expected, as the time scales at which those former are observed differ by several orders of magnitude and represent quasi-asymptotic states whereas the currently observed state is highly transient. Further recall that the majority of the available experimental datasets have been conducted under the influence of lateral side-walls in low to moderate aspect ratio flumes and underlie the side-wall induced secondary currents.
The larger and further developed ridges appear to be rather immobile in the lateral direction over the observed time period, i.e. the mean spanwise position of these ridges is quite stable. This
and the regular spacing of sediment ridges seem to be no effect of laterally narrow domain sizes, as ridge crests follow the same straight vertical space-time lines in the
sufficiently wide channel of case \ichanLRZinit{}.

We further conclude that the narrow boxes with a lateral domain period $\Lz/\hmean \approx 3$
are capable of accommodating one or two ridges, whereas the large domain of \ichanLRZinit{}
exhibits nine to ten ridge units. We can therefore consider the small to medium domains as close to minimal in the context of the number of available ridges, which shall be favourable for the subsequent analysis in a sense that individual ridges and their relation to turbulent coherent structures can be investigated excluding possible merging or splitting effects between individual bedforms. The large domain simulation \ichanLRZinit{}, on the other hand, contains a sufficient number of individual ridges to allow statements on the collective behaviour of the bedforms.

%
%
%
\begin{figure}
  \centering
  \includegraphics
  {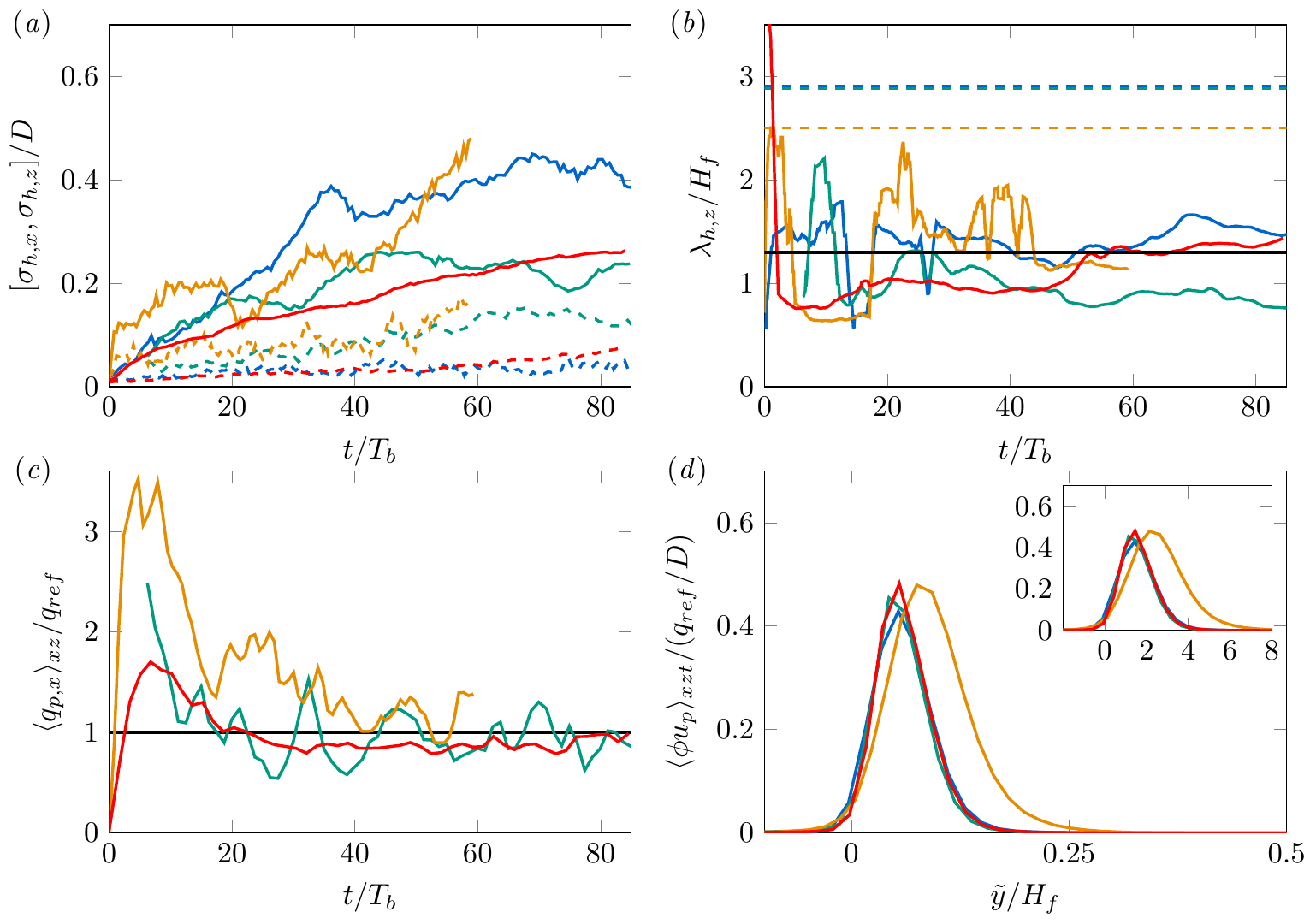}
  \caption{Time evolution of fluid-bed interface dynamics.
    (\textit{a}) Root mean square of the fluctuation of the streamwise-averaged $\sigma_{h,z}/D$ ($\solidthick$) and spanwise-averaged fluid-bed interface $\sigma_{h,x}/D$ (\dashed).
    (\textit{b}) Mean lateral ridge wavelength $\lambda_{h,z}/\hmean$.
    The black solid line marks the most-amplified wavelength determined in the linear stability analysis of \citet{Colombini_1993}, while horizontal dashed lines of matching color mark the lateral domain size $\Lz/\hmean$.
    (\textit{c}) Streamwise particle flux $\qpxmxz/q_{ref}$ as a function of time.
    The reference particle flux $q_{ref}$ is computed based on \citet{Wong_Parker_2006}'s version of the classical formula of \citet{MPM_1948}, that is, $q_{ref}/(\ugrav\, D)=4.93(\shields(t)-\shieldscrit)^{1.6}$, with the critical Shields number $\shieldscrit=0.034$ \citep{Soulsby_Whitehouse_1997}.
    (\textit{d}) Mean particle flux density $\xztavg{\phi\up}/(q_{ref}/D)$ as a function of the bed/wall distance. The inset shows the same quantity in the near-bed region, with the wall-distance scaled with the particle diameter $D$.
    \ichanD{} ({{\color{col_260} $\solidthick$}}),
    \ichanKUinit{} ({{\color{col_221} $\solidthick$}}),
    \ichanHLRSlonginit{} ({{\color{col_21} $\solidthick$}}),
    \ichanLRZinit{} ({{\color{col_511} $\solidthick$}}).
    }
\label{fig:global_bed_stats}
\end{figure}

%
%
%
Figure~\ref{fig:global_bed_stats} provides the time evolution of the mean bedform geometry and the particle transport rate.
Panels~\ref{fig:global_bed_stats}(\textit{a,b}) show the development of the mean pattern height expressed by the root mean square of the sediment bed height fluctuations $\sigmaz$ and that of the lateral spacing in terms of the mean bedform wavelength $\lambdahz$ (cf. appendix~\ref{subsec:bedform_dim_definition} for definitions).
In accordance with the discussed space-time plots, it is seen that the bedform amplitude globally increases with time during the initial approximately $40$ bulk time units.
In the first approximately $10$ bulk time units, low amplitude disturbances first form, before they merge in the subsequent phase between $t=10\tbulk$ and $t=20\tbulk$, leading to a net reduction of the number of ridges. The bedforms continue growing in amplitude predominantly during the time interval between $t=20\tbulk$ and $t=40\tbulk$.
The lower growth rate in the first $5$ bulk time units of the three low Reynolds number cases \ichanD{}, \ichanKUinit{} and \ichanLRZinit{} is a consequence of the lower Shields numbers
when compared to the high Reynolds number case \ichanHLRSlonginit{}, in which the high
Shields number causes a stronger erosion and thus a faster pattern growth.
For the sake of comparison, panel~\ref{fig:global_bed_stats}(\textit{a}) additionally provides the evolution of the spanwise-averaged root mean square of the sediment bed height fluctuations $\sigmax$, that is a measure for the amplitude of possibly evolving transverse ripple-like bed features. It is thereby verified that transverse bed features are indeed not relevant in this initial time frame, consistent with the instantaneous bed snapshots in figure~\ref{fig:3d_snaps_bed_flow}.
The variation of the number of individual patterns is revealed by the oscillations of the mean wavelength $\lambdahz$ during the first $40$ bulk time units. For $t>40\tbulk$, then, $\lambdahz$ settles without further strong oscillations in all cases, attaining final values of $1.47\hmean$ (\ichanD{}), $1.14\hmean$ (\ichanHLRSlonginit{}) and
$1.44\hmean$ (\ichanLRZinit{}) that are of comparable size to the dominant wavelength $\lambdahz=1.3\hmean$ obtained by means of linear stability analysis \citep{Colombini_1993}.
The clearly lower attained final value of $0.76\hmean$ in case \ichanKUinit{} is in agreement with the corresponding space-time plot in figure~\ref{fig:spacetime_interf_xavg} which shows that the bed in the final phase indeed features three ridge patterns.

Panel~\ref{fig:global_bed_stats}(\textit{c}) contains the temporal evolution of the streamwise particle flux $\qpxmxz(t)$ (cf. equation~\eqref{eq:def_qpx} in appendix~\ref{sec:append_B}) showing that the particle flow rate asymptotically approaches a quasi-steady state that is well estimated based on the empirical formula of \citet{MPM_1948} in the modified form of \citet{Wong_Parker_2006}, \textit{viz.}
\begin{equation}\label{eq:MPM_WP}
  q_{ref}/(\ugrav\, D)=4.93(\shields(t)-\shieldscrit)^{1.6},
\end{equation}
where we have chosen a critical Shields number $\shieldscrit=0.034$ \citep{Soulsby_Whitehouse_1997}.
This further supports earlier findings in the context of transverse patterns by
\citet{Kidan_Uhlmann_2017} that relation~\eqref{eq:MPM_WP} reasonably well describes the mean particle flow in the presence of sediment bedforms, even though it was actually developed for flow over a flat sediment bed.
It should be stressed that the apparent 'overshoot' of $\qpxmxz(t)/q_{ref}$ during the initial 20 bulk time units is not a feature of the particle flux evolution $\qpxmxz(t)$ which increases in an essentially monotonic way in this time window, before it settles to a quasi-stationary plateau \citep[cf., for instance,][]{Kidan_Scherer_Uhlmann_2021}.
Instead, it is observed that relation \eqref{eq:MPM_WP} underpredicts the actual particle flux in this initial transient regime owing to the different growth rates of the particle flux and the non-dimensional wall-shear stress, while it provides a good approximation in the quasi-steady interval in which both quantities vary only weakly.

All discussed simulations belong to the bedload-dominated regime, that is, particle transport focusses in a layer of thickness $\mathcal{O}(D)$ above the bed in which particles are transported by rolling, sliding and small jumps (saltation) without loosing the contact with the bed for longer time intervals \citep{VanRijn_1984}.
This is verified in panel~\ref{fig:global_bed_stats}(\textit{d}) which shows the wall-normal profile of the mean particle flux density $\xztavg{\phi\up}(y)$
(cf. equation~\eqref{eq:def_qpx_density}), highlighting that most of the transported particles indeed remain in the near bed region.
It becomes evident from the wall-normal expansion of the flux density that the higher Shields number $\shields$ in case \ichanHLRSlonginit{} leads to a thickening of the bedload layer by around $1D$ compared to the remaining cases. We conclude that in this latter case, a layer of $0.25\hmean$ thickness above the interface is characterized by a non-negligible particle flux.

\subsection{Turbulent mean flow}
%
%
\begin{figure}
    \centering
    \includegraphics
    {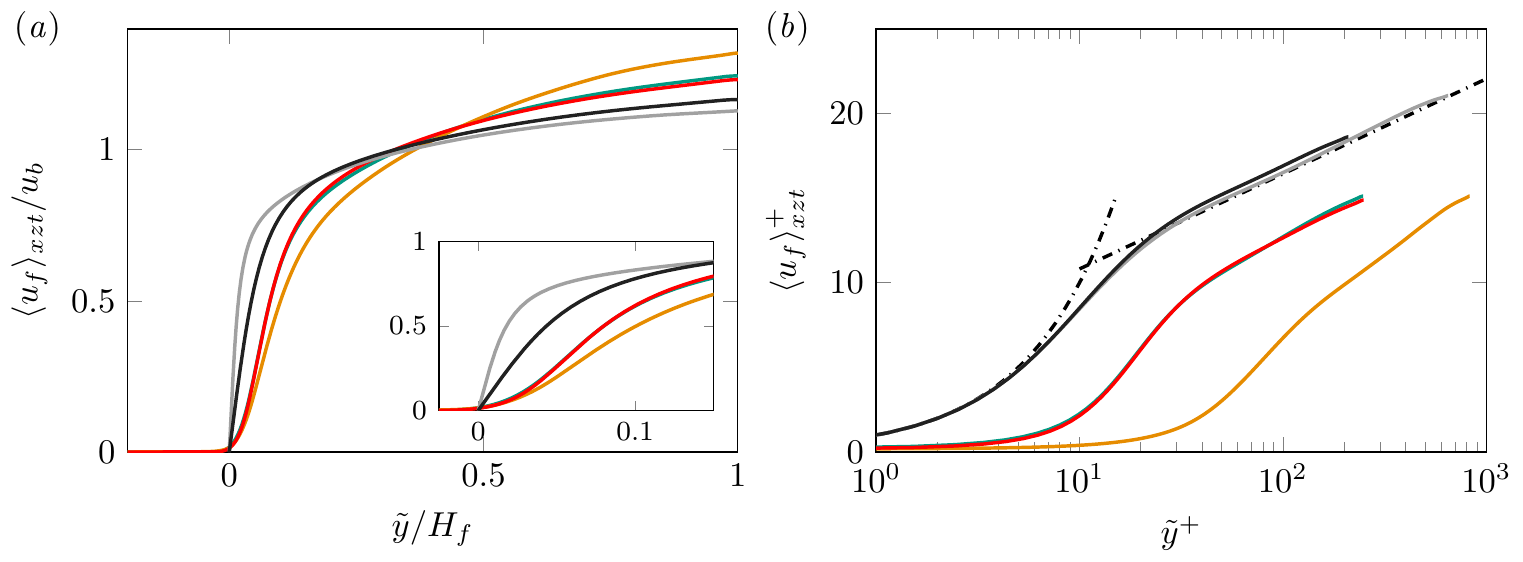}
    \caption{
    Wall-normal profiles of the double-averaged mean velocity (\textit{a}) $\uxzt/\ubulk$ as a function of $\yrelh$ (linear scaling) and (\textit{b}) $\uxzt^+$ as a function of $\yrelplus$ (semi-logarithmic scaling).
    The inset in (\textit{a}) shows a close-up of the same data in the vicinity of the (virtual) wall.
    In (\textit{b}), the dashed-dotted line shows the standard predictions for the linear viscous sublayer and logarithmic layer, respectively, $\uxzt^+=\yrelplus$ and $\uxzt^+=1/\kappa \ln(\yrelplus)+B$ ($\kappa=0.41$, $B=5.2$).
    \ichanKUinit{} ({{\color{col_221} $\solidthick$}}),
    \ichanHLRSlonginit{} ({{\color{col_21} $\solidthick$}}),
    \ichanLRZinit{} ({{\color{col_511} $\solidthick$}}),
    \ispecHLRSlonginit{} ({{\color{col_1} $\solidthick$}}),
    \ispecLRZinit{} ({{\color{col_5} $\solidthick$}}).
     }
\label{fig:um_profiles}
\end{figure}

%
%
%
The presence of the mobile sediment severely modifies the mean fluid flow characteristics.
Figure~\ref{fig:um_profiles} shows the mean fluid velocity profile $\uvecxzt(y)$ compared to that of the single-phase cases.
First, the sediment-laden cases feature inflection points in the near-wall region as a consequence of the underlying porous beds, which are susceptible to a Kelvin-Helmholtz instability \citep{Jimenez_al_2001}.
Generally, the near wall region is strongly modified in the particle-laden simulations in the sense that the mean shear $\mathrm{d}\uvecxzt/\mathrm{d}y$ reduces to negligible value when approaching the mean fluid-bed interface at $\yrel=0$, whereas in the smooth-wall case the mean shear peaks at the wall and is responsible for the entire wall-shear stress. In the multi-phase cases, on the other hand, the shear at the location of the virtual wall is clearly dominated by the contributions stemming from particle-fluid interactions \citep{Kidan_Uhlmann_2017}.
The mean velocity $\uvecxzt$ increases significantly slower with increasing distance to the bed in all particulate cases, leading to a velocity deficit compared to the smooth-wall cases, an essential feature arising in turbulent flows over rough-walls due to an enhanced friction \citep{Flores_Jimenez_2006,ChanBraun_al_2011}.
Farther away from the wall, in particular for $\yrelh>0.5$, the opposite is true, i.e. the particle-laden flows feature a higher mean velocity than their smooth wall counterparts. This is expected as the reduced mass flow rate in the vicinity of the sediment bed as a consequence of the fluid-particle interactions has to be compensated by a higher flow rate in the outer flow to maintain the constant bulk flow rate $\qf$.

As a consequence, the slope of the velocity profile in the semi-logarithmic visualization deviates from that in the logarithmic layer of the smooth wall channels. For the low Reynolds number simulations, the deviation is rather weak due to the comparably lower particle flux, such that the
effect of the sediment bed can be mostly compensated by the conventional shift of the log-layer velocity profile by an offset $\uroughshift$ known from analysis of rough wall turbulence \citep{Jimenez_2004}
\begin{equation}\label{eq:logprofile_shift}
  \uxzt^+=\dfrac{1}{\kappa} \ln(\yrelplus) + B - \uroughshift.
\end{equation}
Using a standard choice of von K\'{a}rm\'{a}n's constant $\kappa=0.41$ and the coefficient $B=5.2$, we obtain $\uroughshift=\DeltaUplusChanKUinit$ (\ichanKUinit{}) and $\uroughshift=\DeltaUplusChanLRZinit$ (\ichanLRZinit{}) as values for the roughness function.
For the high Reynolds number case \ichanHLRSlonginit{}, on the other hand, it is observed that the slope in the semi-logarithmic representation differs markedly compared to the smooth-wall reference solutions, such that a simple vertical shift will not lead to a satisfactory match of the profiles.
%

%
%
\begin{figure}
  \centering
  \includegraphics[width=\linewidth]
  {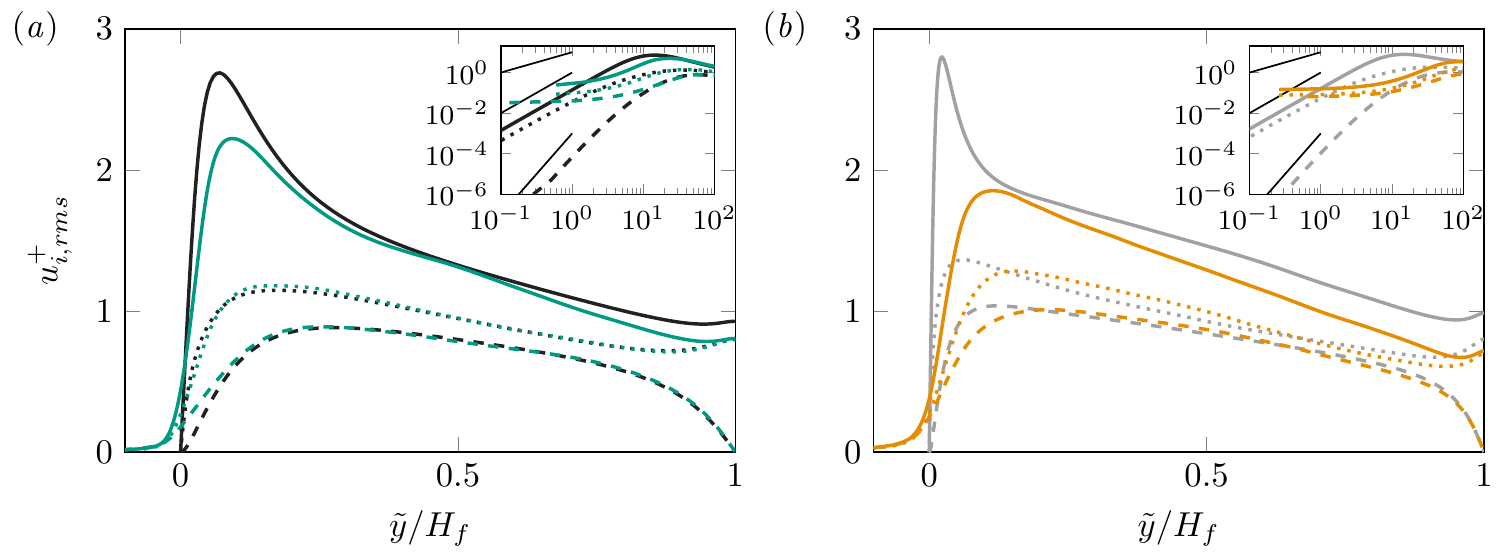}
  \caption{
    Wall-normal profiles of the mean turbulence intensities
    $u_{i,rms}^+=\xztavg{\overline{u}_{fi} \overline{u}_{fi}}^{1/2}/\utau$ ($i=x,y,z$) of cases
    (\textit{a}) \ichanKUinit{} ({{\color{col_221} $\solidthick$}}) and
                 \ispecLRZinit{} ({{\color{col_5} $\solidthick$}}),
    (\textit{b}) \ichanHLRSlonginit{} ({{\color{col_21} $\solidthick$}}) and
                 \ispecHLRSlonginit{} ({{\color{col_1} $\solidthick$}}).
    The small insets show close-ups of the same data close to the (virtual) origin as a function of the wall-normal distance $\yrelplus$ in double-logarithmic scaling. Solid lines indicate logarithmic decay as $\sim \yrel$, $\sim \yrel^2$ and $\sim \yrel^4$, respectively.
    $u_{rms}$ (\solidthick),
    $v_{rms}$ (\dashed),
    $w_{rms}$ (\dotted).
  }
\label{fig:uuvvww_turbint}
\end{figure}

%
%
%
The modifications of the mean velocity profile have direct implications for the intensity and distribution of the Reynolds stresses, which are depicted in figure~\ref{fig:uuvvww_turbint} exemplary for cases \ichanKUinit{} and \ichanHLRSlonginit{}, respectively, together with the corresponding single-phase smooth-wall data. Recall that the presented statistics in the multi-phase cases cannot be taken as fully converged as the observation interval is quite short and in addition highly transient. While we can assume to have captured enough statistics for the short-living small scales in the near-wall region, deviations between single- and multi-phase simulations close to the free surface are expected to be a consequence of the restricted averaging interval $\tobs$ with respect to the lifetime of the largest scales.

In all simulations, the typical buffer-layer peak of the streamwise normal stress $\uuprime$ is visibly reduced when compared to the smooth-wall cases. This is not unexpected, as the mean shear which represents the source of energy for the fluctuating field has been observed to be lower in the vicinity of the bed \citep{ChanBraun_al_2011}. For the low Reynolds number cases \ichanKUinit{} and \ichanLRZinit{} (plot of the latter case not shown), the modulation w.r.t. the smooth-wall reference case is moderate in the sense that the peak is, even though damped, clearly detectable and found at essentially the same wall-normal position. In the high Reynolds number case \ichanHLRSlonginit{}, on the contrary, the differences are more pronounced: the intense particle motion in the bedload layer of particles with significant size results in the complete absence of the characteristic buffer-layer peak of $\uuprime$.
Note that the essentially same phenomenon appears in flows past fully-rough walls where the roughness height scaled in wall-units is also large enough to destroy the self-sustained near-wall cycle \citep{Jimenez_2004,Flores_Jimenez_2006,Mazzuoli_Uhlmann_2017}.
In the low Reynolds number cases, on the other hand, particles are smaller by a factor of three in terms of wall-units ($D\approx 10\deltanu$), which has been observed to be sufficiently small for particles to be transported by the buffer-layer structures rather then to destroy the near-wall cycle \citep{Kidan_al_2013}.
Deviations for the remaining normal stresses $\vvprime$ and $\wwprime$ are also observable, but are clearly weaker and visually restricted to a layer of thickness $\yrel\approx0.11\hmean$ ($\yrel\approx3D$) above the bed in the low Reynolds number cases and $\yrel\approx0.17\hmean$ ($\yrel\approx4.5D$) in the high Reynolds number case, respectively.
The insets in figure~\ref{fig:uuvvww_turbint} show the Reynolds stress decay in the near-bed region in double-logarithmic scaling. It is seen that all Reynolds stresses are higher for the multi-phase simulations as there is no strict boundary condition that enforces them to vanish at $\yrel=0$ and thus they are observed to settle at small but finite levels when entering the bed whereas their counterparts over smooth walls decay at the theoretically predicted decay rates when approaching the physical wall \citep[cf.][p.284]{Pope_2000}.

Our conclusions from the afore-discussed global flow statistics highlight the fact that a general mechanism for the creation of sediment ridges cannot be due to the dynamics of the usual buffer-layer structures, as these readily disappear once particles are released in case \ichanHLRSlonginit{}.
Also, natural flows over fully-rough beds with $\Dplus\gg1$ lack a viscous near-wall cycle just as much. It is thus conclusive to focus in the following on the dynamics of the outer flow structures as the main candidates to drive ridge formation. This idea is further strengthened by our observation that ridges feature a very similar lateral spacing of $\lambdahz/\hmean\approx1\textup{--}2$ in the current small to intermediate Reynolds number simulations and in the high Reynolds number experiments.

%
%
%
\begin{figure}
  \centering
  \includegraphics
  {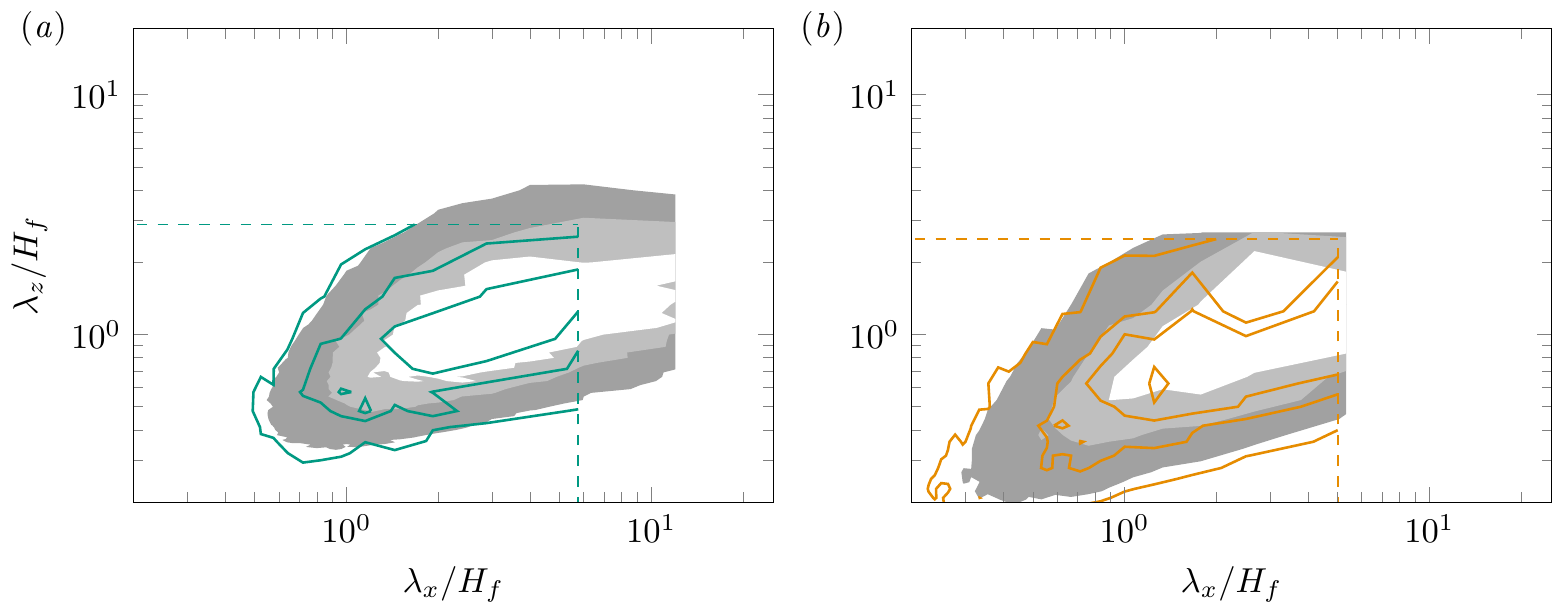}
  \caption{
    Time-averaged premultiplied streamwise energy spectra
    $\kx \kz \uuft(\lamx,y,\lamz)\,\hmean^2/\utau^2$ at the reference wall normal-location
    $\yrelh = 0.5$ 
    for cases
    (\textit{a}) \ichanKUinit,
    (\textit{b}) \ichanHLRSlonginit.
    Colored isolines are 0.2(0.2)0.6 times the maximum value of the respective energy spectra, while gray shaded areas represent the same quantity evaluated for the smooth-wall reference simulations (\textit{a}) \ispecLRZinit{} and (\textit{b}) \ispecHLRSlonginit{}, respectively. The streamwise and spanwise domain periods $\Lx$ and $\Lz$ of the particle-laden simulations are highlighted by dashed lines in the respective colors.
  }
\label{fig:psuuxz}
\end{figure}

%
%
%
In the following, we first compare whether the mobile sediment bed has a detectable influence on the distribution of the kinetic energy among the different scales away from the wall. To this end, the premultiplied streamwise energy spectra of cases \ichanKUinit{} and \ichanHLRSlonginit{} are analyzed in figure~\ref{fig:psuuxz} in the center of the clear-fluid region ($\yrelh=0.5$)
for smooth-wall and particle-laden simulations.
The instantaneous streamwise energy spectra at a given wall-normal location $y$ is introduced as
\begin{equation}
\uuf(\kx,y,\kz,t)=\ufour\uconj,
\end{equation}
where $\ufour(\kx,y,\kz,t)=\mathcal{F}(\uflucxz(\xvec,t))$ is the Fourier transform of the fluctuating field in the two wall-parallel directions, with the asterisk indicating complex conjugation.
$\kx=2\pi/\lamx$ and $\kz=2\pi/\lamz$ are the streamwise and spanwise wavenumber-wavelength pairs. Velocity spectra for the remaining velocity components are defined accordingly.

Apart from some expected fluctuations due to the limited statistics, one may conclude a good agreement between the spectral patterns in both flow configurations.
As expected, the medium domains cover only part of the full spectra, but this part is again observed to reasonably well collapse with that of the smooth-wall reference simulations.
The good match between spectra in single- and multi-phase simulations is an indication that the evolving sediment bed does not significantly alter the streamwise energy spectra away from the bed and that the large streaks over ridges are in fact essentially the same structures as those observed over smooth walls. While it is established for flow over fixed roughness elements that the roughness effect remains restricted to a layer of several multiples of their height \citep{Jimenez_2004,Jimenez_2018,Mazzuoli_Uhlmann_2017} and does not strongly affect the organization of the large scales outside the roughness, a similar statement was not evident for the current case.

%
%
%
\begin{figure}
  \centering
  \includegraphics
  {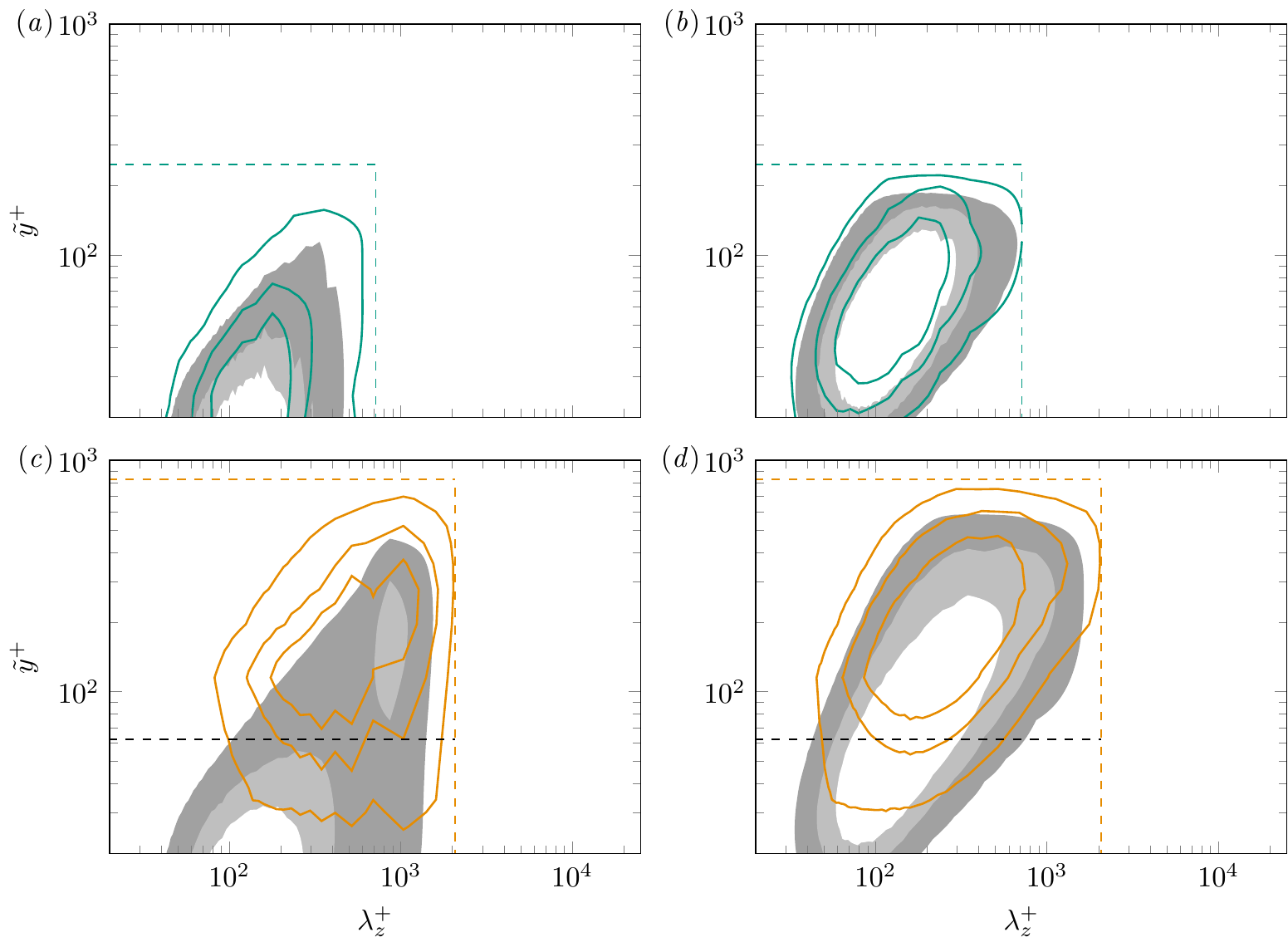}
  \caption{
    Time-averaged and streamwise-integrated premultiplied energy spectra
    $\kz \int \uufvec(\lamx,y,\lamz) \mathrm{d}\kx \,\hmean^2/\utau^2$ as a function of the inner-scaled spanwise wavelength $\lamz^+$ and wall distance $\yrelplus$, respectively.
    (\textit{a,c}) streamwise $\uuft$ and
    (\textit{b,d}) wall-normal $\vvft$ energy spectra
    for cases
    (\textit{a,b}) \ichanKUinit,
    (\textit{c,d}) \ichanHLRSlonginit.
    Colored isolines are 0.2(0.2)0.6 times the maximum value of the respective energy spectra, while gray-shaded areas indicate the same quantities determined for the smooth-wall reference simulations
    (\textit{a,b}) \ispecLRZinit{} and
    (\textit{c,d}) \ispecHLRSlonginit{},
    respectively.
    The mean fluid height $\hmean^+=\Ret$ and the spanwise domain period $\Lz^+$ of the particle simulations are marked by colored dashed lines. The dashed black line refers to the wall-normal distance at which the mean particle flux density $\xztavg{\phi\up}$ attains its maximum (cf. figure~\ref{fig:global_bed_stats}(\textit{d})).
  }
\label{fig:psuuzy_inner}
\end{figure}

\begin{figure}
  \centering
  \includegraphics
  {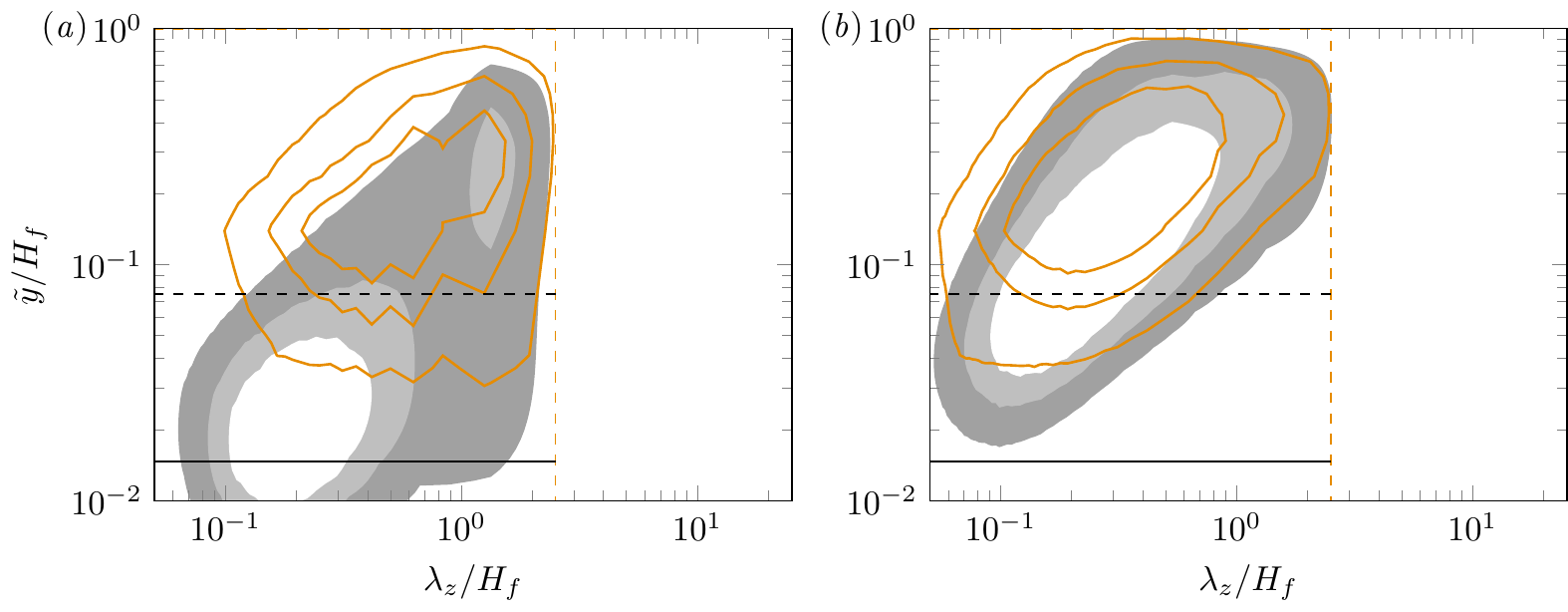}
  \caption{
    Time-averaged and streamwise-integrated premultiplied energy spectra for cases \ichanHLRSlonginit{} and \ispecHLRSlonginit{}.
    Data is identical to that in figure~\ref{fig:psuuzy_inner}(\textit{c,d}), but this time the lateral wavelength $\lamz/\hmean$ and the wall-normal distance $\yrelh$ are scaled in outer units. An additional reference height is added in form of the crest height of the mean fluid-bed interface averaged over $t/\tbulk\,\in\,[40,85]$ ($\solidthick$).
    (\textit{a}) streamwise $\uuft$ and
    (\textit{b}) wall-normal spectra $\vvft$.
  }
\label{fig:psuuzy_outer}
\end{figure}

%
%
%
The wall-normal variation of the streamwise (left column) and wall-normal energy spectra (right column) is clearly identifiable in figures~\ref{fig:psuuzy_inner} and \ref{fig:psuuzy_outer}, where the premultiplied streamwise-integrated energy spectra are presented as functions of the lateral wavelength $\lamz$ and the wall-normal distance $\yrel$ in inner and outer scales, respectively. Note that we have scaled each spectrum by its maximum value over all wall-normal locations
to highlight how the energy distribution over the wall-normal direction is modified through the mobile sediment.
Let us first conclude that the medium domains are just wide enough to accommodate the major fraction of the wall-normal energy spectra and hence are at the limit of being minimal near the free surface \citep{Jimenez_2013b}, which further supports our observations in \S~\ref{sec:flow_config}.
Also, it is consistent with our claim that the relative particle size $\Dplus$ together with the high particle transport rate in the bedload layer of case \ichanHLRSlonginit{} cause the immediate breakdown of the buffer layer regeneration cycle. We observe that the near-wall peak in the streamwise spectra at $\lamz^+\approx100$ associated with the buffer-layer streaks is still present for the low Reynolds number case \ichanKUinit{} and kinetic energy is concentrated around this region whereas it is absent in case \ichanHLRSlonginit{}, in which the kinetic energy peaks further away from the wall outside the buffer layer. As indicated by the dashed line, most of the energy agglomerates in this latter case in regions above the bedload layer with less intense particle transport. A direct comparison with the smooth-wall reference case shows that this region is connected to the already existing second energetic peak at $\lamz/\hmean\approx1\textup{--}1.5$ ($\lamz^+\approx1000$) that reflects the large-scale motions at higher Reynolds numbers.

Naturally, the wall-normal energy spectra in smooth-wall channels has to concentrate in regions away from both, the lower wall and the upper free surface, as the boundary conditions prevent intense wall-normal motion in both regions.
Again, the spectra of the low Reynolds number case \ichanKUinit{} almost collapse with that of the smooth-wall dataset with slight deviations only due to the slight difference in $\Ret$, highlighting that the distribution of energy over the scales and the wall-normal layers is essentially the same as over the smooth wall.
In the high Reynolds number case \ichanHLRSlonginit{}, the wall-normal kinetic energy is significantly reduced throughout the bedload layer and the region of intense vertical motion is found above the latter. However, the general ellipsoidal shape and the inclination angle are maintained. As can be seen in the outer-scaled figure~\ref{fig:psuuzy_outer}(\textit{b}), the spectra contours almost collapse in the vicinity of the free surface and for large wavelengths, whereas the intense energy region in the bulk is compressed, restricted by the bedload layer that damps the wall-normal energy even before the bed is reached, quite similar to the streamwise spectra.

\subsection{Large-scale flow organization}
%
%
%
\begin{figure}
  \centering
  \includegraphics
  {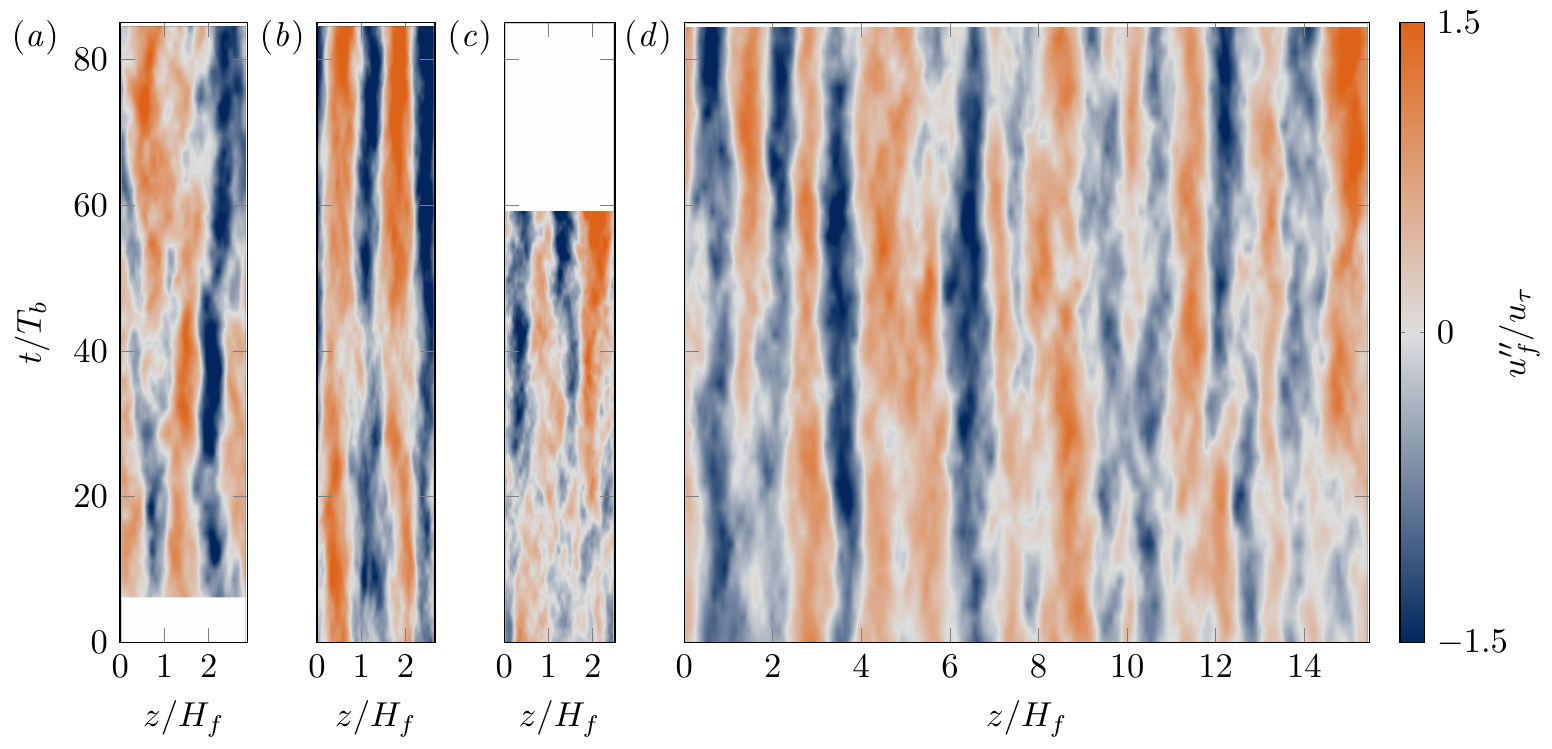}
\caption{Space-time plot of the streamwise-averaged fluctuations of the streamwise velocity component $\uflucx/\utau$ extracted at $\yrelh=0.5$. Blue and red regions refer to streamwise-averaged low and high-momentum regions, respectively.
Cases
(\textit{a}) \ichanKUinit{},
(\textit{b}) \ispecHLRSlonginit{},
(\textit{c}) \ichanHLRSlonginit{},
(\textit{d}) \ichanLRZinit{}.
}
\label{fig:spacetime_ufluc_outer_xavg}
\end{figure}

%
%
%

Let us now turn to the dynamics of the large-scale flow structures in physical space.
Figure~\ref{fig:spacetime_ufluc_outer_xavg} shows the space-time evolution of the streamwise velocity fluctuations w.r.t. the streamwise average, $\uflucx(z,t)$. Note that streamwise averaging filters out lateral meandering of structures while retaining the footprint of the large streamwise elongated structures.
In fact, the streamwise-averaged low-momentum and high-momentum regions are found to develop comparably straight in the space-time plane, indicating that there is only weak lateral migration of these zones.
It is noteworthy that there is no qualitative difference in this point between ridge-featuring and smooth-wall cases during the current limited time interval, showing that even laterally homogeneous smooth channels are able to maintain a substantial spanwise modulation of the mean flow profile over an intermediate time scale as a consequence of the long living large-scale streaks \citep{Jimenez_2013b}, which is expected to disappear for sufficiently long time averaging intervals leading to statistical homogeneity. We shall discuss the role of these intermediate time-scale dynamics in the creation of ridges below.
Note that a comparison between the medium domain simulations and case \ichanLRZinit{} in figure \ref{fig:spacetime_ufluc_outer_xavg}(\textit{d}) verifies that the observations are not a consequence of the limited box width $\Lz/\hmean\approx3$, but that comparable regular organization of the streaks over intermediate time intervals similarly occur in domains as wide as $\Lz/\hmean\approx16$. Also, it shows that the streaks are of significant length, as they are clearly detectable even in streamwise averages over long streamwise extents $\Lx/\hmean\approx12$.

Intermittently, the amplitude of the high and low-momentum regions is seen to reduce as, for instance, around $t=40\tbulk$ in case \ispecHLRSlonginit{} (panel ~\ref{fig:spacetime_ufluc_outer_xavg}(\textit{b})) or in a period $10\lesssim t/\tbulk \lesssim 20$ in case \ichanHLRSlonginit{}. It appears likely that at least part of these `events' are related to regularly occurring bursting events in which the large streaks bend and eventually break, just as their viscous counterparts in the buffer layer do \citep{Flores_Jimenez_2010}.
Interestingly, there are events after which streaks recover at practically the same lateral position, while in other situations as after $t=45 \tbulk$ in case \ichanKUinit{} (panel~\ref{fig:spacetime_ufluc_outer_xavg}(\textit{a})), high and low-momentum regions reorganize and partly even change the number of streaks, as it is here the case. We will investigate the implications of such `events' on sediment ridges in figure~\ref{fig:spacetime_interf_qpx_ufluc_xavg} below.

Throughout the previous sections, it has become clear that the largest streaks over ridges are no other structures than those found in smooth-wall turbulence, and we have in fact observed these structures throughout the entire ridge evolution interval.
For homogeneous roughness, \citet{Flores_Jimenez_DelAlamo_2007} similarly report that the self-similar vortex clusters and the associated velocity streaks are not significantly modified outside the roughness layer.
We have, however, not explicitly discussed the question of cause and effect: based on the results seen so far, it seems reasonable that the spanwise heterogeneity of the flow in form of the well-organized low and high-speed streaks generates a laterally varying bed shear stress that leads accordingly to spanwise heterogeneous erosion from the very beginning, hence implying that the streaks trigger the first ridges and not vice versa. Note that this does not rule out the possibility of a feed-back of developed ridges on the mobility or meandering tendency of the streaks in later stages of their lifetime.


Large-scale streaks are entirely turbulent structures and as such not as smooth as their viscous counterparts in the buffer layer \citep{Jimenez_2013b}.
Also, large-scale instantaneous quasi-streamwise rotations are almost impossible to detect by the classical gradient-based vortex detection methods such as the $\lambda_2$-criterion of \citet{Jeong_Hussain_1995} as the velocity gradients are rather weak, as opposed to the near-wall quasi-streamwise vortices.
Hence, for the understanding of the large-scale processes it is often customary to study fields in which small-scale motions are filtered out, either indirectly as in the context of ensemble averaging \citep[see, for instance,][]{DelAlamo_Jimenez_2006,LozanoDuran_2012} and in some sense in our own streamwise-averaging procedure, or directly by means of instantaneous flow field filtering, for instance, with a Gaussian filter as in \citet{Motoori_Goto_2019,Motoori_Goto_2021}.
The advantage of direct filtering is that there is no information loss for all scales larger than the filter widths. In the current study, we therefore extend our analysis by applying a Gaussian cut-off filter to the flow field. The low-pass filtered field of the $i$-th velocity component is obtained by the following convolution of the field with an anisotropic Gaussian kernel \citep{LozanoDuran_al_2016}:
\begin{equation}\label{eq:Gaussfilter3d}
\ufifilt(\xvec) = \iiint_{V} G \cdot \ufi(\xvec-\xvec^{\prime})
                  \exp \left(
                  - \left( \dfrac{\pi x^{\prime}}{\deltaxfilt} \right)^2
                  - \left( \dfrac{\pi y^{\prime}}{\deltayfilt} \right)^2
                  - \left( \dfrac{\pi z^{\prime}}{\deltazfilt} \right)^2
                  \right)
                  \mathrm{d}x^{\prime} \mathrm{d}y^{\prime} \mathrm{d}z^{\prime}.
\end{equation}
Therein, $\Delta_i$ ($i=x,y,z$) is the filter cut-off width in the three Cartesian directions and $\xvec^{\prime}$ is the inner-convolution coordinate.
The boundary conditions are treated similarly as in \citet{LozanoDuran_al_2016}, i.e. the flow field is periodically extended in the wall-parallel directions, while it is mirrored vertically at the bottom wall, reversing the sign of the wall-normal velocity component to take care of the fundamental symmetries. The free surface can be treated in a completely analogous way as the symmetries along this plane resulting from the free-slip boundary condition do not differ from that at a smooth wall. $V$ is the filter volume and the constant coefficient $G$ is normalized to ensure that the integral of the kernel over $V$ is unity.
Note that the filtered flow field in a layer of thickness $\deltayfilt/\hmean$ will be affected by the velocity field inside the bed. In the following, however, we will study only filtering results in regions sufficiently away from this wall-normal region or for single-phase flow simulations, in which the above effect is naturally absent.
If not otherwise stated, we conventionally use filter widths
{$[\deltaxfilt,\deltayfilt,\deltazfilt]/\hmean=[0.6,0.2,0.3]$}
that obey the typical aspect ratio of the vortex clusters as reported by \citet{DelAlamo_Jimenez_2006}.
In the following, we will also use a two-dimensional analogue of equation~\eqref{eq:Gaussfilter3d} in wall-parallel $(x,z)$-planes whenever we are mainly interested in the structure's wall-parallel extensions, using filter widths
{$[\deltaxfilt,\deltazfilt]/\hmean=[0.6,0.3]$}.

%
%
%
\begin{figure}
  \centering
  \includegraphics
  {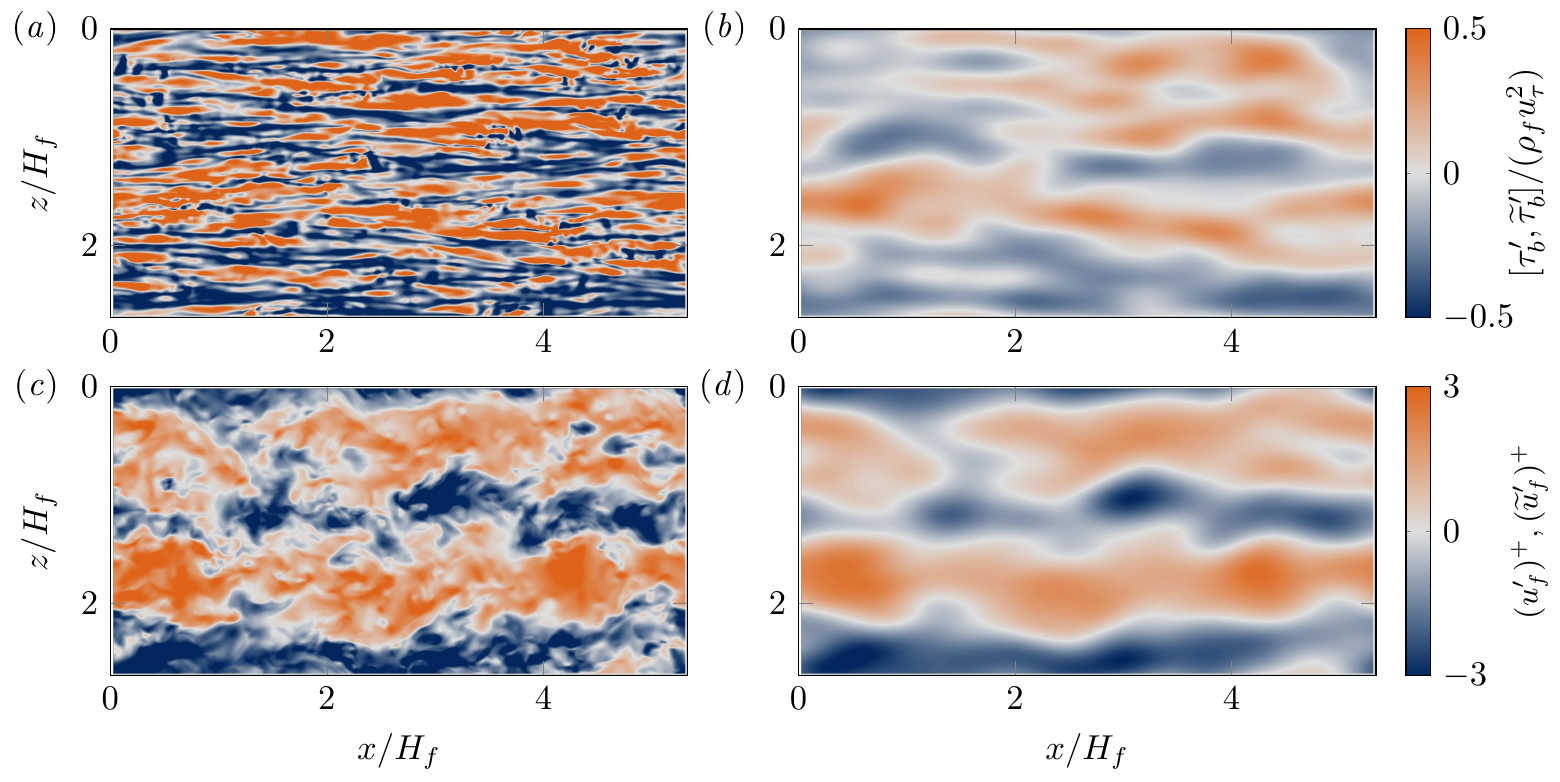}
  	\caption{
      (\textit{a},\textit{b}) Instantaneous wall-shear stress fluctuations
      $\taubflucxz(x,z,t)/(\rhof \utau^2)$ for the smooth-wall simulation \ispecHLRSlonginit{}
      (\textit{a}) without applied filtering and
      (\textit{b}) after a 2D Gaussian cut-off filter has been applied in the two homogeneous   directions with filter widths $\deltaxfilt=3y_{ref}$ and $\deltazfilt=1.5y_{ref}$,   respectively. The reference value has been set to the height $y_{ref}=0.3\hmean$.
      (\textit{c},\textit{d}) Corresponding streamwise velocity fluctuation field $(\uflucxz)^+$ extracted at $\yrelh=0.3$
      (\textit{c}) non-filtered and
      (\textit{d}) filtered with the same filter size as in (\textit{b}).
      The time at which all snapshots have been extracted corresponds to the second marker point in   figure~\ref{fig:uuf_time_evo_modes}(\textit{b}), $t/\tbulk=83$.
    }
\label{fig:tauw_snaps}
\end{figure}

%
%
%
In the previous paragraphs, it has been discussed that the lateral organization of large-scale high and low-speed velocity streaks and the associated ejection and sweep events may cause a laterally varying instantaneous shear stress along the bed or the lower wall, respectively, as recently discussed in the hydraulic context by \citet{Lemmin_Bagherimiyab_2018}.
Underlying this hypothesis is today's understanding that the near-wall region is not exclusively populated by short scales of order $\mathcal{O}(100\deltanu)$, but that it also features large scales of order $\mathcal{O}(\hmean)$ that scale in outer units and that are correlated with the structures away from the wall \citep{DelAlamo_Jimenez_2003}.
Even though the near-wall regeneration cycle itself is capable of functioning autonomously \citep{Jimenez_Pinelli_1999}, large-scale influences have nonetheless been observed to modulate regions even close to the wall \citep{Jimenez_DelAlamo_Flores_2004}. \citet{Toh_Itano_2005} stated that there is a frequent interplay between small near-wall structures and large structures away from it, the former necessarily agglomerating below the large-scale ejections due to continuity \citep{Jimenez_2018}.

In figure~\ref{fig:tauw_snaps}, the organization of buffer-layer structures with dimensions $\mathcal{O}(100\deltanu)$ into larger essentially $\hmean$-scaled streaks and the corresponding outer large-scale streaks at $\yrelh=0.3$ can be observed exemplary for the smooth-wall case~\ispecHLRSlonginit{} in visualizations of unfiltered and corresponding spatially-filtered fields.
Note that qualitatively similar results are found for the flow organization over the sediment beds just before the particle release.
To avoid confusion with the mean wall shear $\tauw$, let us introduce the instantaneous bottom-shear stress $\taub(x,z,t)=\fnu \mathrm{d}\uf/\mathrm{d}\yrel|_{\yrel=0}$ for the following analysis.
In figure~\ref{fig:tauw_snaps}, the high-speed buffer-layer streaks whose position can be inferred by their induced zones of locally increased shear are seen to cluster in two streamwise elongated  streak-like structures, from which one spans the entire box length in a slightly meandering way, while the other is roughly half as long and laterally shifted by an offset somewhat larger than $\hmean$.
These two large-scale structures are found to lay essentially below the corresponding high-speed streaks at the upper end of the logarithmic layer, $\yrelh=0.3$, and are of comparable lateral extension, even though the streamwise extension differs in particular for the patchy structure with center around $z/\hmean=0.5$.
In order to quantify the correlation of the boundary shear with the flow field at each height of the channel in \ispecHLRSlonginit{} in the streamwise-averaged framework, let us define the instantaneous two-point correlation coefficient as follows:
\begin{equation}\label{eq:corrutau}
	 \corrutau(y,t)=
	       \zavg{\taubflucx(z,t) \, \uflucx(y,z,t)}/
	       \left[
          \zavg{(\taubflucx(z,t))^2} \, \zavg{(\uflucx(y,z,t))^2}
          \right]^{1/2}.
\end{equation}
It turns out that, in accordance with the previous findings, regions of high and low momentum over the entire logarithmic layer are highly correlated with the corresponding wall shear regions along the bottom wall, with a mean value of the correlation in the channel center of
$\tavg{\corrutau}(\yrelh=0.5) \approx 0.5$.
Note that the restricted periodic domain length makes it hard to visually detect a streamwise phase shift between the wall-shear object and the log-layer streaks which is however expected to exist as a consequence of the varying mean shear and thus propagation speed between the two wall-normal locations.

The observed correlation between log-layer streaks and the organization of the bottom shear stress is of direct relevance for the current simulations, as it provides a mechanism that couples the erosion rate along the bed with the large-scale structures.
While it appears inevitable that large scales drive the particle motion in case \ichanHLRSlonginit{} due to the absence of the near-wall cycle, the observed link between large-scale structures and the local wall shear might explain why the same ridge spacing is found for the low Reynolds number cases in which small scales are, in general, still able to transport particles \citep{Kidan_al_2013}.

\subsection{Streak-ridge interaction}
%
%
%
\begin{figure}
  \centering
  \includegraphics
  {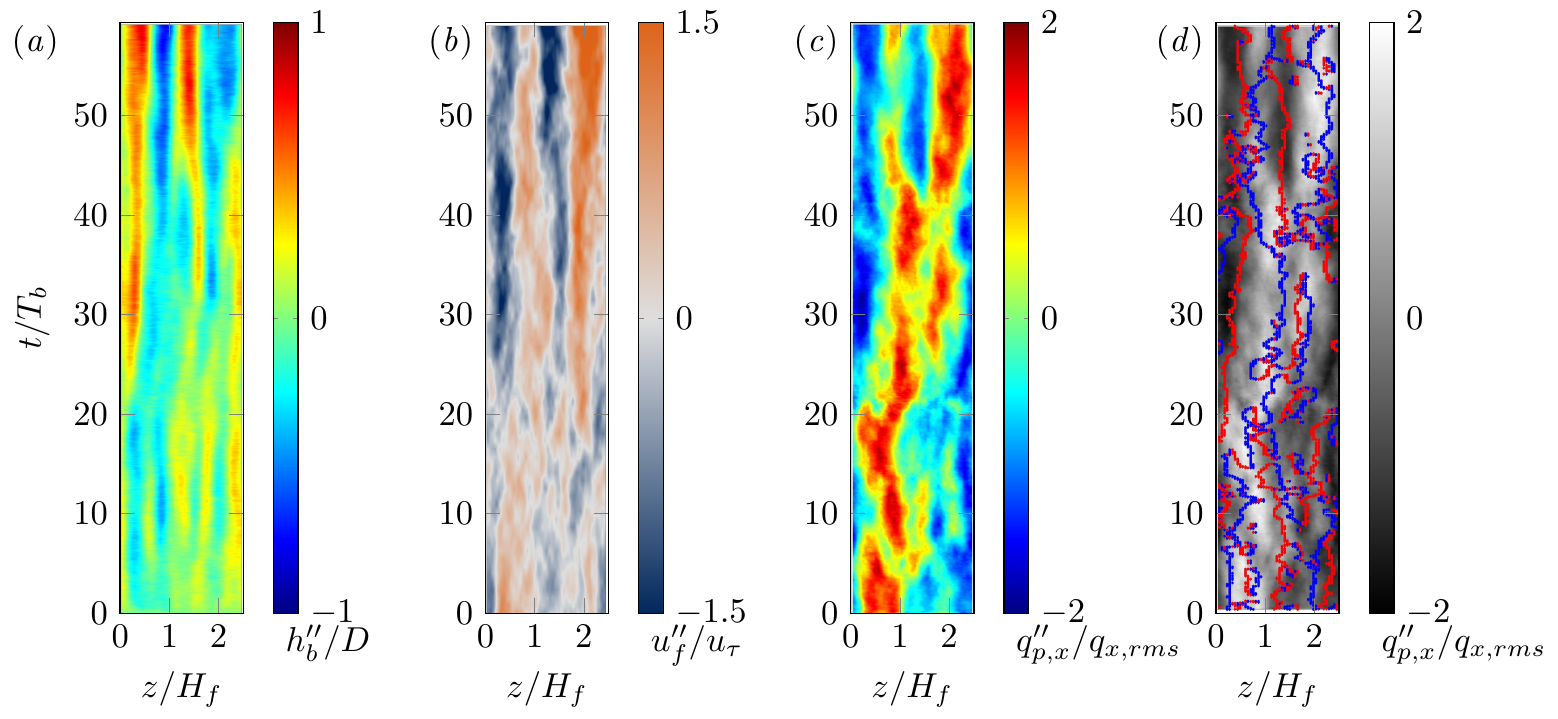}
\caption{
Space-time evolution in case \ichanHLRSlonginit{} of
(\textit{a}) the sediment-bed height fluctuation of the streamwise-averaged bed (identical to figure~\ref{fig:spacetime_interf_xavg}(\textit{c}));
(\textit{b}) the streamwise-averaged streamwise velocity fluctuation $\uflucx/\utau$ at $\yrelh=0.5$ (identical to figure~\ref{fig:spacetime_ufluc_outer_xavg}(\textit{c}));
(\textit{c}) The streamwise-averaged particle flux fluctuation $\qpxflucx/q_{x,rms}$, with
             $q_{x,rms}(t)=(\zavg{\qpxflucx\qpxflucx})^{1/2}$.
(\textit{d}) Same data as in (\textit{c}) is shown in the background as gray map, supplemented with red (blue) dots marking regions of lateral bed growth (decrease), that are, regions with vanishing lateral particle flux $\qpzflucx=0$ and negative (positive) lateral gradient $\partial_z \qpxflucx(z,t)$.
}
\label{fig:spacetime_interf_qpx_ufluc_xavg}
\end{figure}

%
%
%
The above arguments imply that particle erosion and hence bed topography are linked with the large-scale structures via their near-bed effects and the local wall shear. Indeed, it appears that
the number of adjacent alternating high and low-speed regions agrees reasonably well with the spanwise wavelength $\lamz/\hmean \approx1.1\textup{--}1.3$ of the outer energetic peak determined from the streamwise energy spectra, which, in turn, is of striking similarity to the observed mean pattern wavelength $\lambdahz/\hmean$.
That, indeed, regions of high and low streamwise velocity are spatially correlated with the bed contour is visualized in figure~\ref{fig:spacetime_interf_qpx_ufluc_xavg}, exemplary for case \ichanHLRSlonginit{}. Here, the space-time evolution of the sediment bed fluctuation $\hbflucx(z,t)$ and the respective fluctuation of the streamwise velocity component at $\yrelh=0.5$ are repeated for convenience, supplemented by similarly presented data for the streamwise and spanwise particle flux components $\qpxflucx$ and $\qpzflucx$, respectively.
The observed dependences can be summarized as follows:
regions of high momentum which are the streamwise-averaged analogue to the large-scale streaks lead to spanwise alternating zones of locally increased erosion due to turbulent sweep structures that are located in the high-speed streaks. The wall-shear stress is locally increased and with it the local Shields number. Higher shear stress and erosion directly imply an increased streamwise particle flux $\qpxflucx$ in these regions.
The effect is further supported by the action of the lateral particle transport $\qpzflucx$, which is predominantly directed in such sense that sediment is transported from the regions of high streamwise particle transport ($\qpxflucx>0$) to areas with lower streamwise particle flux, i.e. $\qpxflucx<0$, as indicated by the red and blue lines in figure~\ref{fig:spacetime_interf_qpx_ufluc_xavg}(\textit{d}).
The consequence is a local decrease of the sediment bed height in form of a local trough.
Corresponding opposite relations are found for low-momentum regions, ridges and regions of low streamwise particle transport $\qpxflucx<0$.

The lateral particle flux is found to be two orders of magnitude smaller than the corresponding streamwise particle flux $\qpxflucx$ in all simulations. It is therefore conceived that the initial sediment ridges readily appearing after the onset of particle motions are mostly a consequence of the laterally varying particle erosion in the streamwise direction, resulting in the comparably stronger streamwise particle flux. Lateral particle transport from the troughs to the crests which is due to the weaker lateral fluid motion then further supports the growth of these initial sediment ridges once the particle fluxes have reached a quasi-stationary level (cf. figure~\ref{fig:global_bed_stats}(\textit{c})).

%
%
%
\begin{figure}
    \centering
    \includegraphics
    {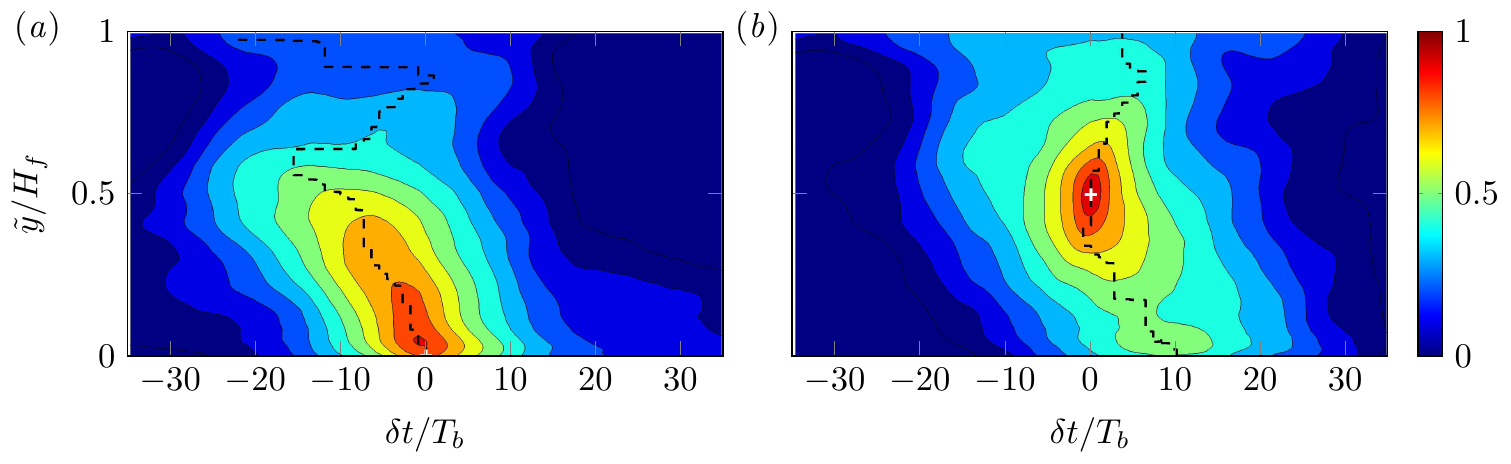}
  	\caption{Two-point cross-time correlations between $\uflucx(y,z,t+\tsep)$ and:
    (\textit{a}) the sediment bed height fluctuations, $-\corrubedt(y,\tsep)$;
    (\textit{b}) the velocity fluctuations at the reference height $\uflucx(\yrelh=0.5,z,t)$, $\corruut(y,\tsep)$.
    The reference time and reference wall-normal position are indicated by a white cross.
    Red (blue) regions represent strong (weak) correlation of the compared quantities.
    Black dashed lines connect the maximum correlation values at each wall-normal distance $\yrelh$.
    Contours separating the colored areas indicate values of 0(0.1)1 in all panels.
    Data is for case \ichanHLRSlonginit{}.
}
\label{fig:corr_u2d_bed_qpx}
\end{figure}

%
%
%
The visually identified correlation between large-scale structures and the bed contour in the multi-phase simulations can be quantified with the aid of a two-point cross-correlation as defined in equation~\eqref{eq:corrutau}, replacing the wall-shear stress $\taubflucx$ by the sediment bed height fluctuation $\hbflucx$.
In agreement with the previous observations, it turns out that over most of the simulation time, the bed contour shows a high instantaneous correlation with the flow field over most of the channel height reaching values of more than 0.6 at the bulk flow centerline $\yrelh=0.5$ (figure omitted).
This holds, however, not true in the phase between $t=10\tbulk$ and $t=30\tbulk$, in which the bed contour is found to be essentially uncorrelated with the flow structures away from the bed.
The missing correlation between the large-scale structures and the location of sediment ridges is also observed in figures~\ref{fig:spacetime_interf_qpx_ufluc_xavg}(\textit{a,b}): between $t=10\tbulk$ and $t=20\tbulk$, the amplitude of the streamwise-averaged velocity fluctuations reduces and the previously and subsequently observed intense regions temporarily `diffuse' into several narrower regions. We have argued earlier that the streaks are believed to break up in this phase. Two large-scale streaks recover between $t=20\tbulk$ and $t=25\tbulk$, but at somewhat different lateral positions. Interestingly, the bed contour is seen to adapt to the new spanwise organization of the streaks, laterally migrating in time towards these spanwise positions, however with a noticeable time delay.

This time lag between the bed evolution and the streak organization is instructive as it indicates how information is propagating across the channel. In order to determine the time delay between the dynamics of the large-scale motions and the deformation of the sediment bed contour, we introduce the two-time cross-correlation between the streamwise-averaged flow field and an arbitrary physical quantity $a$ as
\begin{equation}\label{eq:corruat}
  \corruat(y,\tsep)=
                  \zavg{\uflucx(y,z,t+\tsep) \, a^{\prime\prime}(z,t)}/
                  \left[
                  \zavg{(\uflucx(y,z,t+\tsep))^2} \, \zavg{(a^{\prime\prime}(z,t))^2}
                  \right]^{1/2}.
\end{equation}
Panel~\ref{fig:corr_u2d_bed_qpx}(\textit{a}) shows the cross-time correlation between the flow field and the bed contour, i.e. $a^{\prime\prime}=\hbflucx(z,t)$, while panel~\ref{fig:corr_u2d_bed_qpx}(\textit{b}) provides the cross-time correlation between the flow field at varying wall-normal positions and that at a reference height chosen as $\yrelh=0.5$, i.e.
$a^{\prime\prime}=\uflucx(\yrelh=0.5,z,t)$.
Note that the correlation between bed and flow field is presented premultiplied with a global factor $-1$ to take account of the fact that a locally higher velocity implies a local decrease of the sediment bed height.
In both panels, figure~\ref{fig:corr_u2d_bed_qpx}(\textit{a}) and  \ref{fig:corr_u2d_bed_qpx}(\textit{b}), a time lag is clearly identifiable.
Figure~\ref{fig:corr_u2d_bed_qpx}(\textit{a}) verifies our earlier observations that ridges align themselves with the location of the large-scale streaks with a time lag of around $10\tbulk$ (equivalent to approximately $0.9\hmean/\utau$). A very similar time delay is observed in figure~\ref{fig:corr_u2d_bed_qpx}(\textit{b}) between the dynamics of the flow field in the channel center and that close to the bed.
The observed time lag implies that the generation of the initial sediment ridges is indeed predominantly dictated by the organization of the large-scale streaks in the channel center which interact in a `top-down mechanism' with the fluid layers closer to the bed.

%
%
%
\begin{figure}
  \centering
  \includegraphics
  {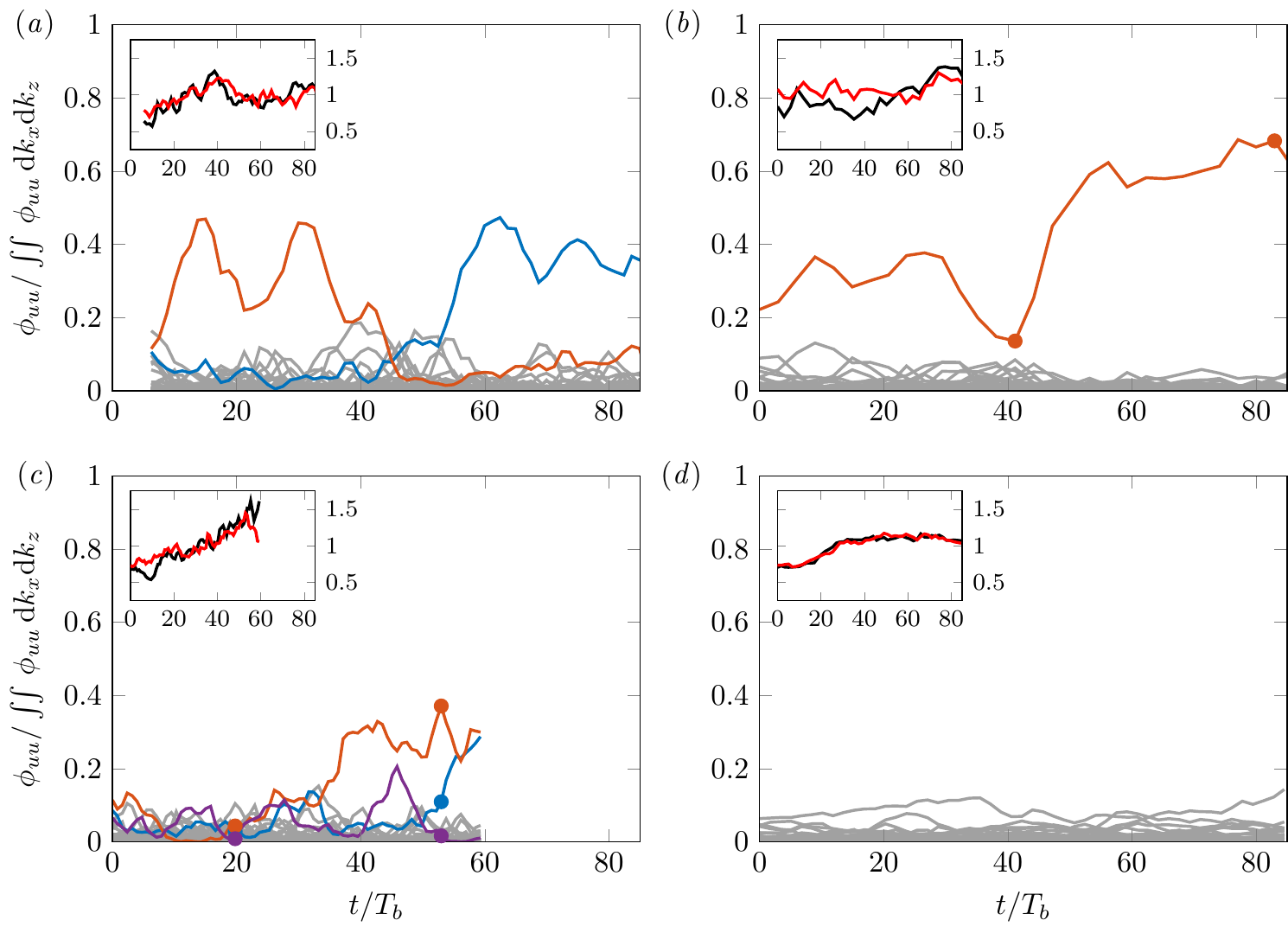}
  \caption{
    Time evolution of the energy distribution between individual modes of the streamwise
     velocity spectrum $\uuf(\kx,\kz,t)$ evaluated at a wall-normal position $\yrelh\approx 0.5$.
     Note that the spectrum at each time has been normalized such that the contributions of all
     individual modes sum up to unity.
     Note that the most dominant modes (those modes that carry more than $20\%$ of the total energy at least once during the observation time) are highlighted using the following color scheme:
    ({{\color{col1_matlab} $\solidthick$}}) ($0$,$1$)-mode,
    ({{\color{col2_matlab} $\solidthick$}}) ($0$,$2$)-mode,
    ({{\color{col4_matlab} $\solidthick$}}) ($1$,$2$)-mode.
    The smaller insets show the time evolution of the total streamwise and wall-normal fluctuation energy at the same wall-normal position,
    ({{\color{black} $\solidthick$}}) $\xzavg{\uflucxz\uflucxz}$ and
    ({{\color{red} $\solidthick$}}) $\xzavg{\vflucxz\vflucxz}$, normalized by their time-averaged mean.
    Cases
    (\textit{a}) \ichanKUinit{},
    (\textit{b}) \ispecHLRSlonginit{},
    (\textit{c}) \ichanHLRSlonginit{},
    (\textit{d}) \ichanLRZinit{}.
}
\label{fig:uuf_time_evo_modes}
\end{figure}

\begin{figure}
  \centering
  \includegraphics
  {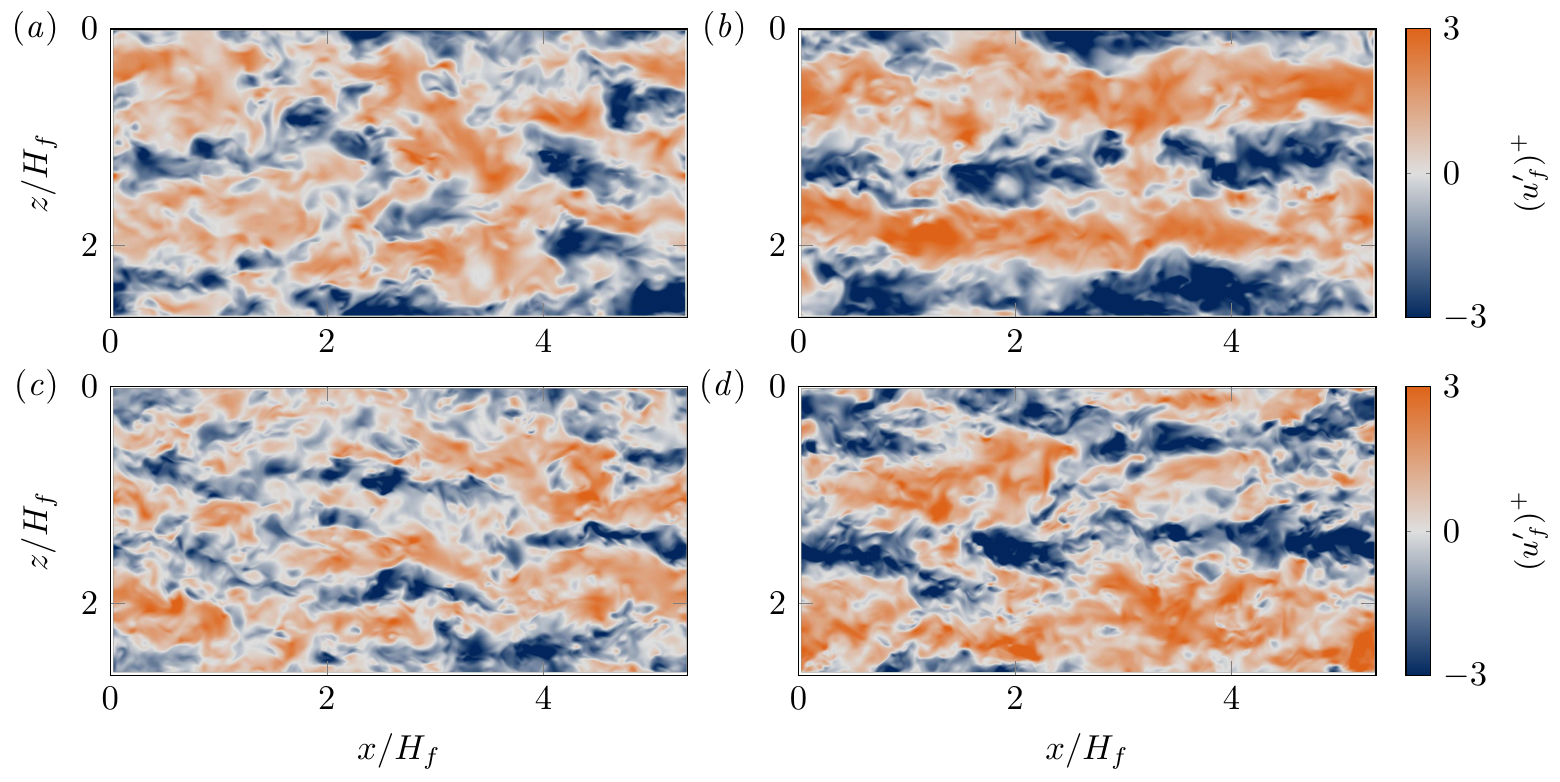}
  	\caption{Wall-parallel plane of the streamwise velocity fluctuation field $(\uflucxz)^+$ extracted at $\yrelh\approx 0.5$ for cases
    (\textit{a,b}) \ispecHLRSlonginit{}
    (\textit{c,d}) \ichanHLRSlonginit{}
    at two different times. The flow fields refer to the instances marked by filled circles in figure~\ref{fig:uuf_time_evo_modes}(\textit{b}) and (\textit{c}), respectively.
}
\label{fig:uprime_snaps}
\end{figure}

%
%
%

It also indicates that the existence or absence of a dominant large-scale velocity streaks in the core of the channel is of direct relevance for the stability and position of the underlying sediment ridges and troughs, respectively.
In the following, we shall analyze such events during which a clear signature of high- and low-momentum regions is missing, avoiding streamwise averaging for the moment to show
that the modulation of $\uflucx$ in such phases is indeed due to a break-up of the streaks.
Figure~\ref{fig:uuf_time_evo_modes} shows the time evolution of the most-energetic Fourier modes of the streamwise velocity fluctuations, $\uuf(\kx,y,\kz,t)$, in wall-parallel planes at $\yrelh=0.5$. For the sake of readability, let us refer to modes with wavelength pairs $\lamx=\Lx/i$ and $\lamz=\Lz/k$ ($i,k \, \in \mathbb{N}_0$) as $(i,k)$-mode in the remainder. Note further that the shown curves are normalized by the total streamwise kinetic energy contained in the fluctuating field at that time and wall-normal position, respectively, such that all contributions sum up to unity.
This is particularly noticeable regarding the transient nature of the currently studied systems: the insets show the absolute value of the streamwise $\xzavg{\uflucxz\uflucxz}$ (black) and wall-normal energy $\xzavg{\vflucxz\vflucxz}$ (red) contained in the same plane, indicating that in all sediment-laden cases the turbulent kinetic energy globally increases with time (at least initially) as a consequence of the increasing friction along the bed due to particle transport and bedform development.

In the three medium-sized boxes, the range of wavelengths is limited by the box size and only a few large-scale modes can be accommodated, such that most of the time a large fraction of the total kinetic energy is contained in these few dominant modes that are highlighted in different colors.
Here, we have classified those modes as dominant that contain more than $20\%$ of the total perturbation energy at least once during the observation interval.
From the four dominant highly energetic mode pairs that are of relevance in the medium boxes, three feature the streamwise zero-mode, i.e. the corresponding structures are of infinite streamwise length and do not appear in the premultiplied energy spectra.

These modes are in fact the large alternating high and low-speed streaks identified earlier, spanning the entire streamwise domain and are thus clearly detectable in the streamwise average.
In the large box \ichanLRZinit{}, on the other hand, there are no specially exposed modes, but energy is rather uniformly distributed among all modes.
Comparing the dynamics of the individual modes in the medium boxes with the space-time plots in figure~\ref{fig:spacetime_ufluc_outer_xavg}, the times during which the clear signature of high and low-momentum regions disappears correspond to instances in which there is temporarily no clear dominant mode in the spectra, in particular the energy in the infinitely long modes is markedly reduced.
In case \ichanKUinit{} (panel~\ref{fig:uuf_time_evo_modes}(\textit{a})), the earlier discussed change from two to only one pair of high and low-speed streaks is clearly visible around $t\approx50\tbulk$, accompanied by a reduction in the total energies $\xzavg{\uflucxz\uflucxz}$ and $\xzavg{\vflucxz\vflucxz}$, suggesting that at this time, the streaks indeed break up.
Panel~\ref{fig:uuf_time_evo_modes}(\textit{b}) shows a comparable behavior for the smooth-wall case \ispecHLRSlonginit{} with the difference that after the breakdown of the infinitely long two-streak mode, the same mode recovers instead of being replaced by a single pair of streaks as in the previous case.
The two states without and with clear dominant mode indicated by the two symbols in figure~\ref{fig:uuf_time_evo_modes}(\textit{b}) are visualized in physical space in terms of fluctuating streamwise velocity fields in figure~\ref{fig:uprime_snaps}(\textit{a,b}), which also demonstrates that there is no strong lateral meandering.
In the high Reynolds number sediment-laden case \ichanHLRSlonginit{} provided in panel~\ref{fig:uuf_time_evo_modes}(\textit{c}), though, the streamwise zero-modes carry only slightly more energy than the remaining modes at the beginning of the simulation, but shortly after the particle release (between $t=10\tbulk$ and $t=20\tbulk$) even this slight plus in energy is lost and they cannot be differentiated from the remaining modes any more. Again, this is in line with the space-time plots and the flow field visualization in figure~\ref{fig:uprime_snaps}(\textit{c}) underlines the absence of a clear dominating large structure.
Later, however, the mode that represents two pairs of elongated streaks increases first mildly, then stronger until it settles at $t\approx40\tbulk$ and dominates the spectra for the rest of the simulation. The dominance of this mode is also seen in figure~\ref{fig:uprime_snaps}(\textit{d}).

\subsection{Streaks and mean secondary flow}

Up to this point, we have seen that quasi-streamwise large-scale streaks characterize the appearance of the flow field in particular in the medium boxes where a considerable fraction of the energy resides in the infinitely long streamwise modes.
Some authors have hypothesized that sediment ridges may `lock' the spanwise position of these streaks such that they leave a footprint in the streamwise time-average as up- and downwelling of the mean velocity profile together with mean secondary currents \citep{Nezu_2005}.

%
%
%
\begin{figure}
    \centering
    \includegraphics
    {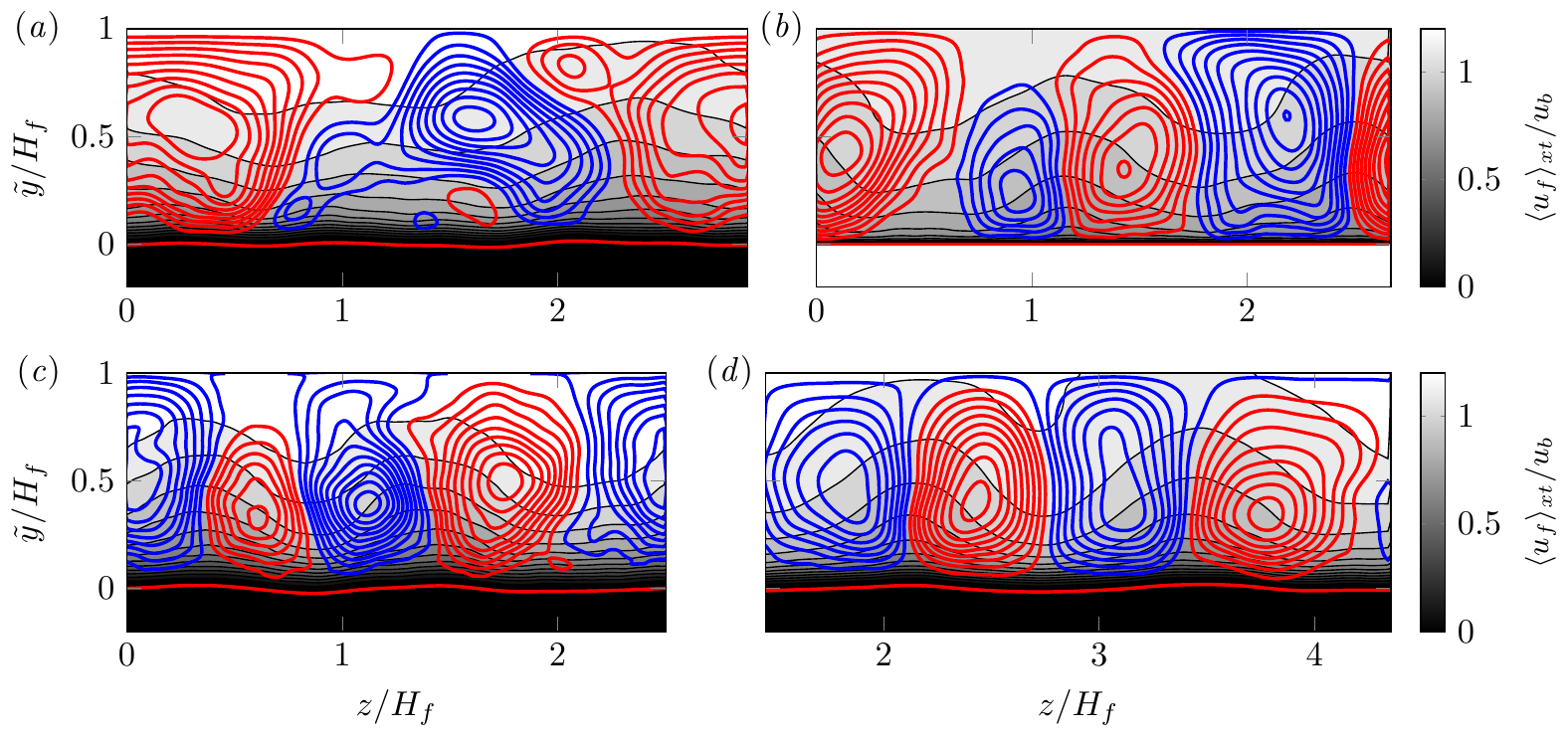}
  	\caption{
    Mean primary and secondary flow pattern for cases
    (\textit{a}) \ichanKUinit{},
    (\textit{b}) \ispecHLRSlonginit{},
    (\textit{c}) \ichanHLRSlonginit{},
    (\textit{d}) \ichanLRZinit{}.
    Note that for the sake of comparison, in panel (\textit{d}) we have arbitrarily chosen a sub-domain of the entire cross-section of case~\ichanLRZinit{} with a lateral width similar to $\Lz/\hmean$ in the medium domain cases.
    Isolines of the primary flow $\uxt/\ubulk$ are shown in intervals 0(0.1)1.2, while
    the secondary flow pattern $(\vxt,\wxt)^T/\ubulk$ is shown in terms of the secondary mean flow stream function $\xtavg{\psi}$.
    Clockwise (counter-clockwise) secondary flow rotation is indicated by red (blue) contours.
    The time-averaged fluid-bed interface profile is indicated by a red curve.
    The time window over which data has been accumulated can be seen in figure~\ref{fig:secflow_timeevo}.
    For the shown contours, the range $[\displaystyle\min_{y,z}\xtavg{\psi},\displaystyle\max_{y,z}\xtavg{\psi}]$ is divided in 20 equally spaced subintervals.
}
\label{fig:secflow_mean_yz}
\end{figure}

\begin{figure}
    \centering
    \includegraphics
    {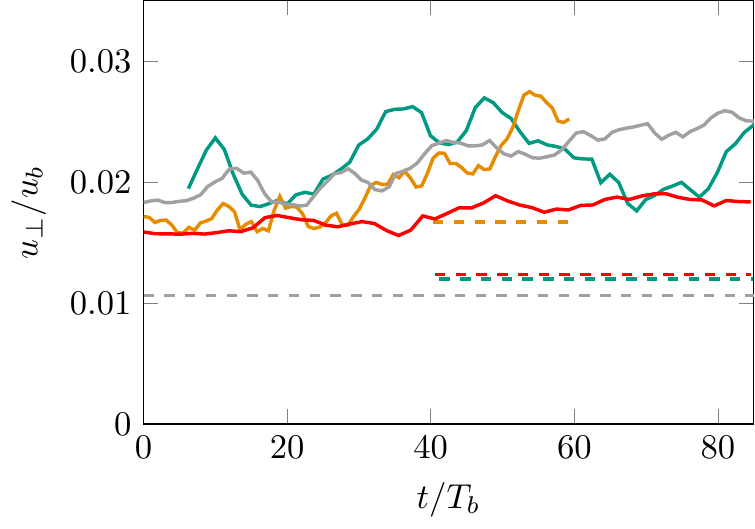}
	  \caption{
    Instantaneous secondary flow intensity $\usec/\ubulk$ as a function of time (solid lines) and the corresponding value determined for the time-averaged fields in  figure~\ref{fig:secflow_mean_yz} (dashed lines) following the definition in equation~\eqref{eq:usect}. The length of the dashed lines indicates the time window over which the fields shown in the respective panels of figure~\ref{fig:secflow_mean_yz} have been averaged:
    \ichanKUinit{} ({{\color{col_221} $\solidthick$}}),
    \ispecHLRSlonginit{} ({{\color{col_1} $\solidthick$}}),
    \ichanHLRSlonginit{} ({{\color{col_21} $\solidthick$}}),
    \ichanLRZinit{} ({{\color{col_511} $\solidthick$}}).
}
\label{fig:secflow_timeevo}
\end{figure}

%
%
%
Having shown that, at least over intermediate time intervals, the large-scale streaks indeed propagate little in the lateral direction and observing that instantaneous spatial meandering appears to be rather weak, the assumption that mean secondary currents are the statistical footprints of large-scale streaks appears conclusive.
Indeed, depth-spanning vortices are seen to populate the cross-section of all four periodic channels in figure~\ref{fig:secflow_mean_yz}, which provides mean flow field data averaged over short time intervals (between $20$ and $80$ bulk time units) and over the streamwise direction.
The secondary mean flow is expressed in terms of the stream function $\xtavg{\psi}(y,z)$, which is defined by the requirement that
$\nabla_{2D} \xtavg{\psi}(y,z)= (\partial_y \xtavg{\psi},\partial_z \xtavg{\psi})^T = (-\wxt,\vxt)^T$.
For the particle-laden simulations, the conventionally discussed configuration of primary flow upwelling over ridge crests and downwelling over the troughs is observed with accordingly oriented secondary currents \citep{Nezu_Nakagawa_1993}.
A comparison between the cases over mobile sediment beds and the smooth-wall simulation~\ispecHLRSlonginit{} in panel~\ref{fig:secflow_mean_yz}(\textit{b}) shows that over adequately chosen intermediate time intervals of the order of the large-scale streaks' lifetime, even statistically spanwise homogeneous simulations do generate a mean secondary flow pattern that is similar to the secondary currents next to developed ridges in panels~\ref{fig:secflow_mean_yz}(\textit{a,c,d}).

In particular, the secondary flow cells in all cases share roughly the same spanwise spacing, which is equivalent to the preferential spanwise wavelength of the instantaneous streaks, $\lamz/\hmean\approx1\textup{--}1.5$. An exception is case~\ichanKUinit{} for which in the observed time interval intermittently only one pair of low- and high-speed streaks exists, as has been seen earlier.
Local mean up- and downflows are the statistical footprints of ejections and sweeps, occurring in the respective low- and high-speed streaks.

Also, we do not observe any significant quantitative difference in the amplitude of the secondary currents from case to case. Figure~\ref{fig:secflow_timeevo} shows the time-dependent secondary flow intensity $\usec(t)$ defined from the cross-plane averaged kinetic energy contained in the cross-flow field $(\xavg{\vf},\xavg{\wf})^T$, \textit{viz.}
\begin{equation}
\usec(t) = \left[ \dfrac{1}{\Lz \, \xzavg{h_f}(t)}
                  \displaystyle \int_{0}^{\Lz}
                  \displaystyle \int_{\xzavg{h_b}(t)}^{\Ly}
                  \left( \xavg{\vf}^2 + \xavg{\wf}^2 \right)
                  \mathrm{d}y \mathrm{d}z\right]^{1/2}.
                  \label{eq:usect}
\end{equation}
Note that a comparable formulation has been originally presented by \citet{Sakai_2016} for quantification of side-wall induced mean secondary flow in open ducts and has been adapted here to the multi-phase configuration.
It is seen that $\usec(t)$ oscillates between $1.5\%$ and $2.5\%$ of the bulk velocity for all cases, with slightly lower values in case~\ichanLRZinit{} as a consequence of the larger averaging ensemble in a sense that the wider domain allows accommodation of a larger number of secondary flows cells (by a factor of three to four).
Note that the secondary flow intensity of the time-averaged cross-flow $(\xtavg{\vf},\xtavg{\wf})^T$ in figure~\ref{fig:secflow_mean_yz} is, as expected, systematically  lower than their instantaneous counterparts.
Interestingly, the latter values are of comparable size to the usually reported strength of side-wall induced mean secondary flows in open ducts. For instance, \citet{Sakai_2016} determined the intensity of secondary flow within a distance $z=1\hmean$ from the side-walls of a smooth-wall open duct in a range $1.1\%$ to $1.3\%$ of the bulk velocity.

Note that the observed secondary flow pattern disappears for conventional time and streamwise averaging over sufficiently long time intervals in the smooth-wall case, and it is only recovered in the context of conditional averaging over individual streaks \citep{LozanoDuran_2012,Kevin_al_2019b}.
Long time prognosis of the situation over ridge-covered beds is only possible for the shortest simulation \ichanD{} (cf. figure~\ref{fig:append_spacetime_interf_qpx_ufluc_xavg} in appendix~\ref{sec:append_A}) where the observation time is not restricted through the development of transverse bedforms. Here, observations over longer time intervals of roughly $600\tbulk$ suggest that the observed ridges undergo continuous variations in form of merging, splitting or short lateral propagations, probably reacting to changes in the structure of the large-scale streak organization as seen before in the initial phase. The global trend, however, shows that the lateral positions at which ridge crests are found do not vary strongly in this narrow computational domain.
In figure~\ref{fig:append_spacetime_interf_qpx_ufluc_xavg} it can be seen that the dominant ridge recovers at essentially the same lateral position even after an intermediate disappearance during the period between $t=400\tbulk$ and $t=500\tbulk$.

  \section{Discussion}\label{sec:discussion}
  The previous analysis has revealed that large-scale velocity streaks in turbulent open channel flow are able to generate the characteristic pattern of spanwise alternating sediment troughs and ridges on an initially flat sediment bed.
The observed generation mechanism can be summarized as follows:
in the initial phase during which the bed is macroscopically flat and particles are still immobile, the streamwise velocity field is organized in streamwise elongated streaks of laterally alternating high and low momentum just as in canonical smooth-wall channel flows. The largest representatives of this self-similar streak family are of dimensions $\mathcal{O}(\hmean)$.
At the onset of particle motion, particles are predominantly eroded in regions below the large high-speed streaks, where high-momentum fluid is brought towards the bed in form of sweep structures that have been identified as the main driver of sediment erosion \citep{Gyr_Schmid_1997}.
The laterally non-homogeneous erosion of sediment, in turn, gives rise to a wave-like deformation of the initially flat sediment bed, with ridges found below large-scale low-speed streaks and troughs accordingly associated with regions of relatively higher streamwise velocity. This goes hand in hand with a lateral variation of the streamwise sediment flux.
The described formation mechanism implies that during the initial formation phase investigated in the current work, bedform evolution is predominantly controlled by the large-scale velocity structures in the channel core and their associated Reynolds stress carrying structures.
This hypothesis has been supported by the two-time cross-correlations in figure~\ref{fig:corr_u2d_bed_qpx}, which reveal a time lag between the lateral organization of large-scale streaks at $\yrelh=0.5$ and that of the flow structures in the vicinity of the bed on the one hand and that of the sediment ridges on the other hand. For the current case~\ichanHLRSlonginit{}, the time lag has been estimated as approximately $10\tbulk$ or, equivalently in terms of the eddy turnover time, $0.9\hmean/\utau$.

%
The observed mechanism conceptually differs from the instability studied in the classical linear stability analysis of \citet{Colombini_1993} in which cause and effect are reversed compared to the above discussed process:
a strictly one-dimensional turbulent base profile is disturbed by an infinitesimal sinusoidal perturbation of the sediment bed across the spanwise direction that triggers the evolution of secondary flow cells. The fluid motion is captured by the RANS equations that have been closed with a nonlinear eddy-viscosity model \citep{Speziale_1987} to allow for the necessary anisotropy of the Reynolds stress tensor.
The fluid-related equations of the model are quasi-stationary and one-way coupled to those capturing the sediment bed dynamics such that the computation can be performed sequentially:
all quantities feature a sinusoidal lateral perturbation that is `geometrically induced' by the disturbances of the bed. The most-amplified wavelength of the system is then directly obtained from the perturbed and linearized equations of fluid motion only and, as such, does not depend on the properties of the sediment bed and the particle flux. The sediment bed continuity equation \citep{Exner_1925} is afterwards considered to determine the growth rate of the initial perturbations a posteriori, balancing the destabilizing lateral bed-shear stress exerted on the sediment bed by the perturbed flow field and a constant stabilizing gravitational term that is a function of the Shields number only.
The latter is thus in particular independent of the wavenumber and, consequently, its effect reduces to a modification of the growth rate amplitude while the chosen bedform wavelength is entirely controlled by the bed-shear stress which features the most-amplified wavelength determined from the linearized fluid equations.

We conclude that in the linear stability analysis, the initially deformed lower domain boundary is indispensable to trigger the spanwise disturbance of the flow field and, that way, to cause the secondary flow instability in the context of the time-independent RANS equations.
By contrast, considering the full time-dependent Navier-Stokes equations as in the current study, lateral finite-amplitude modulations of the turbulent mean flow profile naturally exist in form of turbulent large-scale streaks, whose statistical footprints in the streamwise and time average are the secondary flow cells.
In particular, streaks and the secondary currents exist over intermediate time intervals in open channels over flat smooth walls and macroscopically flat sediment beds likewise, highlighting that the currently observed mechanism does not require an initial spanwise disturbance of the sediment bed to trigger the lateral modulation of the flow field as in the linear stability model.

Despite the conceptual differences between the model of \citet{Colombini_1993} and the interaction of streaks and sediment ridges observed in the current study, interestingly, the organization of sediment ridges and secondary currents show remarkable similarities.
The characteristic spacing of sediment ridges turn out to be in a very similar range $\lambdahz/\hmean\approx1\textup{--}1.5$, which is comparable to values $\lambdahz/\hmean\approx1\textup{--}3$ found in experiments \citep{Wolman_Brush_1961,Ikeda_1981,Mclelland_1999}.
Also, a comparison of the secondary flow cells that are obtained in Colombini's analysis (recomputed, as plots are not provided in the article) with those presented in figure~\ref{fig:secflow_mean_yz} shows a qualitatively similar structure concerning shape, wall-normal location of the center of rotation and lateral spacing of the secondary currents. In particular, in both situations the lateral wavelength of the secondary currents and of the sediment ridges results from the momentum balance in the flow and is essentially independent of the bed properties.
Therefore, we believe that a more detailed analysis of the implications of the present results for the linear stability of a sediment bed should be performed in the future.

%
The observed behaviour of the large-scale streaks and the interaction of flow structures at different wall distances agree fairly well with the conceptual model of \citet[][\S~5.6 and references therein]{Jimenez_2018} in canonical smooth-wall bounded flows according to which the main role of the lower domain boundary is to provide a mean shear that feeds turbulence with energy. The model turns away from the idea that turbulent structures are almost exclusively produced in the buffer layer where the mean shear peaks, and it assumes that structures can form at all wall distances.
This is in contrast to hairpin-vortex based models \citep{Adrian_al_2000,Adrian_2007}, in which
near-wall generated structures are assumed to migrate outwards to form larger structures.
It is therefore inherent to this class of models that the preferential `direction of causality' is directed away from the wall.
In the conceptual model of Jim{\'e}nez, on the other hand, the lack of a single wall-parallel layer in which structures are preferentially generated reposes the question how scales at different wall-normal distances interact and whether there exists a net preferential direction of this interaction.
In \citet{Jimenez_2018}, this question is discussed after reanalyzing the data obtained by \citet{Flores_Jimenez_2010} and it is concluded that for the minimal simulation boxes considered therein, there is such a preferential direction in the logarithmic layer bringing information from outer regions towards the wall (i.e. `top-down'), rather than the other way around.

Support for the `top-down'-concept comes, amongst others, from direct numerical simulations which show that logarithmic and outer layer coherent structures are relatively persistent in situations in which the near-wall self-sustaining regeneration cycle is effectively suppressed, either artificially \citep{Mizuno_Jimenez_2013} or through the presence of significant roughness \citep{Flores_Jimenez_DelAlamo_2007}.
Our simulations support these findings, highlighting that even under severe perturbations of the near-bed flow through the presence of mobile particles and a consequently destroyed buffer layer regeneration cycle, the regular organization of large-scale velocity streaks in the outer layer remains essentially unaffected.
The large-scale structures in our simulations are seen to span the entire depth of the channel reaching down to the bed, and the discussed time lag of flow organization in the near-bed region compared to that of the streaks in the channel bulk agrees with the preferential direction of information propagation towards the lower boundary found by \citet{Jimenez_2018}.
Also, individual streaks are seen to intermittently break up (see, for instance, case~\ichanHLRSlonginit{} between $t=10\tbulk$ and $t=20\tbulk$) in a kind of bursting process comparable to that observed in \citet{Flores_Jimenez_2010}.
After such break-up events, the large-scale streaks are seen to first regenerate in the channel bulk essentially uncorrelated to the flow in other regions of the channel, while the flow near the bed later organizes in similar structures with the discussed time lag (cf. figure~\ref{fig:corr_u2d_bed_qpx}).
This further strengthens the idea that coherent structures can indeed form at all distances from the bed/wall autonomously through the action of the local shear.

  \section{Conclusion}\label{sec:conclusion}
  In the current study, we have investigated the role of turbulent large-scale velocity streaks in the formation cycle of sediment ridges in open channels and their connection to mean secondary currents. To this end, four direct numerical simulations with fully-resolved spherical particles representing the immersed mobile sediment have been performed, supplemented with two reference simulations of single-phase smooth-wall open channel flow in matching domains and for comparable values of the friction Reynolds number $\Ret$. Simulations were performed in three different domain sizes with streamwise and spanwise periods $\Lx/\hmean=2\textup{-}12$ and $\Lz/\hmean=2.5\textup{-}16$, respectively. The variation of the domain size allows to investigate pairs of ridges and large-scale streaks either isolated or in interacting groups.
The friction Reynolds number has been varied in a range $\Ret=250\textup{-}850$, from which  case~\ichanHLRSlonginit{} features with $\Ret=850$, to the best of the authors' knowledge, the highest Reynolds number ever reached in a direct numerical simulation with fully-resolved particles.

The current simulations verify assumptions according to which sediment ridges can form out of an instability process between a mobile sediment bed and a turbulent stream that is devoid of side-wall induced secondary currents \citep{Ikeda_1981}.
The observed mean spanwise spacing ${\lambdahz/\hmean=1\textup{--}1.5}$ of sediment ridges in the current simulations is comparable to values $\lambdahz/\hmean\approx1\textup{--}3$ found in experimental studies \citep{Wolman_Brush_1961,Ikeda_1981,Mclelland_1999}.
The currently observed lateral sediment ridge wavelength and the overall structure of the secondary flow cells show similarly good agreement with predictions obtained by means of linear stability analysis \citep{Colombini_1993}, even though the exact physical mechanism underlying the therein proposed instability conceptually differs from the currently observed formation process.
The qualitatively similar results therefore motivate a more detailed investigation of the instability found by \citet{Colombini_1993} in comparison with the results obtained in the current simulations in a future study.

The typical lateral spacing of sediment ridges in the present work is found to be a consequence of the preferential spanwise organization of turbulence in large-scale streamwise velocity streaks in the core of the fluid-dominated region at comparable spanwise wavelengths. Interestingly, the premultiplied energy spectra over developed sediment ridges do not significantly differ from their counterparts in single-phase smooth-wall channels in regions sufficiently away from the bed, in good agreement with data from rough-wall channel flows in which only a certain roughness layer in the vicinity of the wall is modified while large-scale structures remain essentially unaffected \citep{Flores_Jimenez_2006,Flores_Jimenez_DelAlamo_2007}.

Even though the bed in none of the current simulations can be classified as fully rough, the development of a bedload layer of thickness $\mathcal{O}(D)$ above the bed in which intense particle transport takes places is seen to destroy the near-wall self-sustaining regeneration cycle just as in the context of fully-rough boundaries \citep{Jimenez_2004,Cameron_al_2008}. Condition for this to occur is that particles are sufficiently large compared to the buffer-layer streak spacing, as in the current case~\ichanHLRSlonginit{} where $\Dplus\approx 30$.
Irrespective of the existence or absence of the characteristic buffer-layer cycle, large-scale streaks are seen to be correlated to those large-scale structures into which the wall/bed shear stress organizes, an effect that has been discussed earlier in the context of canonical closed channel flows \citep{DelAlamo_Jimenez_2003,Jimenez_DelAlamo_Flores_2004}.

The organization of the bed shear in laterally varying regions of high and low stresses causes a spanwise varying ability of the flow to erode sediment. It is consequently observed that at the onset of particle motion, significantly more particles are eroded in regions below large-scale high-speed streaks rapidly leading to the formation of a local trough, while local sediment ridges are preferentially located below low-speed streaks.
The streamwise particle flux is accordingly higher in high-speed regions, where turbulent sweeps reside that are mainly responsible for particle erosion \citep{Gyr_Schmid_1997}.

An important implication of the formation process found in this study is that the initial appearance of sediment ridges is effectively controlled by large-scale streaks and their associated Reynolds stress carrying structures. 
The preferred direction of information propagation has indeed been found to be oriented from the bulk flow towards the wall, by evaluating temporal cross-correlation data between the bulk flow and the bed. Sediment ridges and troughs adapt with a time delay to changes in the organization of the large-scale streaks, indicating that the coherent structures at different distances
from the wall interact in a `top-down' fashion.
This agrees with the observations of \citet{Jimenez_2018} in minimal flow simulations of canonical closed channels that the `preferential direction of causality' is directed towards the wall in the sense that Reynolds stresses near the wall are correlated to those away from the wall at earlier times.

The formation of new large-scale structures in the bulk of the channel has been observed to occur after intermittent break-ups of `older' streaks, in a process that is consistent with the log-layer streak bursting reported by \citet{Flores_Jimenez_2010}. Our simulations give thus further numerical evidence that coherent structures can form at significant distances to the lower domain boundary independently only due to the action of the local shear, in accordance with the model outlined in \citet[][\S5.6 and references therein]{Jimenez_2018}.

Finally, we have discussed how the instantaneous coherent structures correlate with the commonly observed mean secondary flow pattern in the presence of sediment ridges. Here we have shown that the
characteristic upwelling and downwelling of the mean primary flow over ridges and troughs are the statistical footprints of the streamwise elongated streaks when averaged over the streamwise direction and short time intervals of $\mathcal{O}(10\tbulk)$, while upward and downward secondary motions are the collective effect of individual sweeps and ejections adjacent to the large-scale streaks. In that sense, the observed secondary flow cells are closely linked to the conditional quasi-streamwise rotations found between conditionally averaged streaks \citep{DelAlamo_Jimenez_2006,LozanoDuran_2012}.
This point is further strengthened by the observation that over the limited time interval under consideration, a very similar secondary flow pattern with comparable intensity and lateral spacing is obtained also for the smooth-wall single-phase simulations.
For sufficiently long averaging time intervals, the mean secondary flow pattern is only maintained for spanwise heterogeneous flow configurations \citep[e.g. duct flow with side-walls,][]{Pinelli_al_2010}, while it is absent for smooth-wall channel flows.
For fixed spanwise heterogeneities, such as alternating roughness stripes on the lower domain boundary, this effect is commonly explained by a `locking' of the mean spanwise position of instantaneous large-scale flow structures due to the spanwise heterogeneities \citep{Kevin_al_2019b}.

For self-formed sediment ridges, the situation is less clear as they are mobile themselves, and in the current study their position is seen to be relatively sensitive to rearrangements of the large-scale flow structures. In the current simulations, we cannot give a conclusive answer on the long-time evolution of sediment ridges as the observation time is limited in most of our cases due to the appearance of larger transverse ripple-like bedforms after $\mathcal{O}(100)$ bulk time units.
Experimentally observed sediment ridges, on the other hand, are usually described to be time-persistent and to not significantly propagate laterally \citep{Nezu_Nakagawa_1993}. However, it should be kept in mind that the experimentally studied bedforms are usually subject to a stabilizing effect of lateral side-walls in laboratory flumes, which are absent in the current channel configuration.
To this end, it would be highly desirable to determine the effect of lateral domain boundaries on the formation and stability of sediment ridges in the future, which will then provide direct comparability between experiments and numerical simulation.

  \vspace*{2ex}
  \section*{Declaration of Interests}
  The authors report no conflict of interest.

  \vspace*{2ex}
  \section*{ORCID}
  M. Scherer,
  \href{https://orcid.org/0000-0002-6301-4704}{https://orcid.org/0000-0002-6301-4704};\\
  M. Uhlmann,
  \href{https://orcid.org/0000-0001-7960-092X}{https://orcid.org/0000-0001-7960-092X};\\
  A. G. Kidanemariam,
  \href{https://orcid.org/0000-0003-4276-671X}{https://orcid.org/0000-0003-4276-671X};\\
  M. Krayer,
  \href{https://orcid.org/0000-0002-6251-9698}{https://orcid.org/0000-0002-6251-9698}.

  \section*{Acknowledgements} \label{sec:acknow}
  The current work was supported by the German Research Foundation (DFG) through grants UH242/2-1 and UH242/12-1. MK has received funding from Baden-W{\"u}rttemberg Stiftung through the program `High Performance Computing II' (project `MOAT').
Part of the work was performed on the supercomputer ForHLR II at the Steinbuch Centre for Computing funded by the Ministry of Science, Research and the Arts Baden-W{\"u}rttemberg and by the Federal Ministry of Education and Research.
The remaining simulations have been carried out on SuperMUC-NG at the Leibniz Supercomputing Centre at the Bavarian Academy of Science and Humanities and on Hazel Hen/Hawk at the High-Performance Computing Center at the University Stuttgart.
The computer resources, technical expertise and assistance provided by the staff at these computing centres are gratefully acknowledged.

  \begin{appendices}

    \section{long-time evolution of case~\ichanD{}}\label{sec:append_A}
    %
%
%
\begin{figure}
\centering
  \begin{minipage}{0.35\textwidth}
  \raggedright{(\textit{a})}\\
   \begin{tikzpicture}
     \begin{axis}[width=3cm,height=7cm,enlargelimits=false,axis on top,
       point meta min=-1,
       point meta max=1,
       xlabel={$z/\hmean$},
       ylabel={$t/\tbulk$},
       colormap/jet,
       colorbar,
       colorbar style={
          anchor=north west,
          width=0.25cm,
          ytick={-1.0,0,1.0},
          ylabel={$\hbflucx/D$},
          every axis y
          label/.style={at={(ticklabel cs:0.5)},rotate=90,anchor=center},
          }]
          \addplot graphics[xmin=0,xmax=2.91,ymin=0,ymax=678]
       {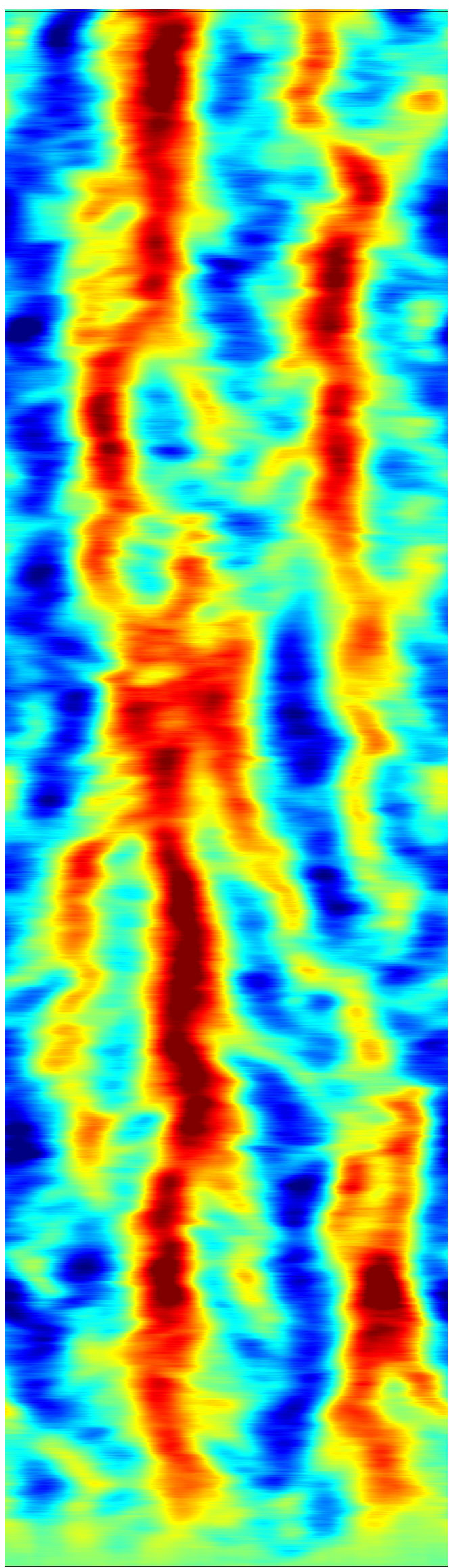};
     \end{axis}
   \end{tikzpicture}
  \end{minipage}
  %
  %
  \begin{minipage}{0.35\textwidth}
  \raggedright{(\textit{b})}\\
  \begin{tikzpicture}
    \begin{axis}[width=3cm,height=7cm,enlargelimits=false,axis on top,
      point meta min=-2,
      point meta max=2,
      xlabel={$z/\hmean$},
      colormap/jet,
      colorbar,
      colorbar style={
         anchor=north west,
         width=0.25cm,
         ytick={-2.0,0,2.0},
         ylabel={$\qpxflucx/q_{x,rms}$},
         every axis y
         label/.style={at={(ticklabel cs:0.5)},rotate=90,anchor=center},
         }]
      \addplot graphics[xmin=0,xmax=2.91,ymin=0,ymax=678]
      {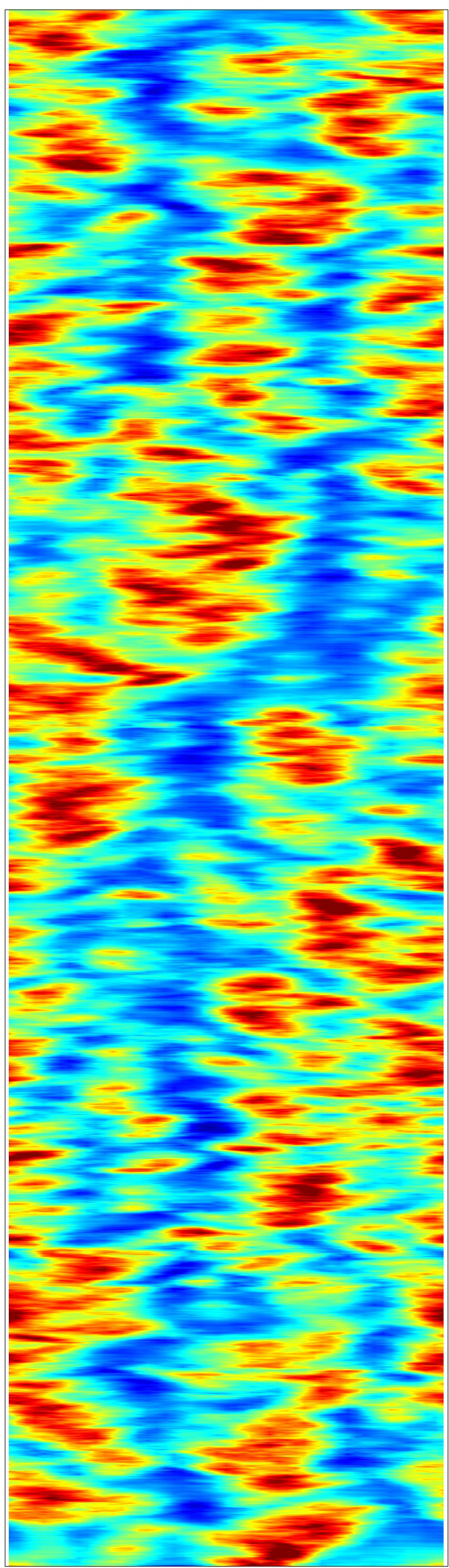};
    \end{axis}
  \end{tikzpicture}
  \end{minipage}
\caption{
Space-time evolution in case \ichanD{} of
(\textit{a}) the sediment-bed height fluctuation of the streamwise-averaged bed and
(\textit{b}) the streamwise-averaged particle flux fluctuation $\qpxflucx/q_{x,rms}$, with
             $q_{x,rms}(t)=(\zavg{\qpxflucx\qpxflucx})^{1/2}$.
}
\label{fig:append_spacetime_interf_qpx_ufluc_xavg}
\end{figure}

%
%
%
Figure~\ref{fig:append_spacetime_interf_qpx_ufluc_xavg} provides the space-time evolution
of the streamwise-averaged sediment bed height fluctuation $\hbflucx(z,t)$ and streamwise particle flux fluctuation $\qpxflucx(z,t)$ for case~\ichanD{}, respectively, for the entire observation interval $\tobs$.
As such, figure~\ref{fig:spacetime_interf_xavg}(\textit{a}) represents a close-up of the initial $85\tbulk$ of figure~\ref{fig:append_spacetime_interf_qpx_ufluc_xavg}(\textit{a}).

    \section{Quantification of sediment bed related quantities}\label{sec:append_B}
%
%
\subsection{Quantification of bedform dimensions}\label{subsec:bedform_dim_definition}
In an abstract sense, we can interpret the evolution of sediment bedforms as wave-like deformations of the defined two-dimensional surface that represents the fluid-bed interface. In this context, it is natural to quantify the wall-parallel and wall-normal bedform dimensions as wave amplitude and wavelengths, respectively. To distinguish between the effects of streamwise and spanwise bedforms, we investigate the length scales for the streamwise- and spanwise-averaged interface $\hbx(z,t)$ and $\hbz(x,t)$ separately.
As a measure for the ridge height, we choose a statistical approach based on the root mean square of the sediment bed height fluctuations \citep{Langlois_Valance_2007,Kidan_Uhlmann_2017}, \textit{viz.}
\begin{equation}
\sigmaz(t) = \sqrt{\zavg{\hbflucx(z,t) \cdot \hbflucx(z,t)}}.
\end{equation}
$\sigmax$ is accordingly defined based on the fluctuations of the spanwise-averaged interface.
Alternative approaches to determine bedform dimensions are reviewed in \citet{Coleman_Nikora_2011}.
The spanwise mean wavelength of the sediment bed, $\lambdahz$, is defined below based on the instantaneous two-point correlation coefficient
\begin{equation}\label{eq:2pointcorr_1D}
\corrhh(\zsep,t)=
\zavg{\hbflucx(z,t) \cdot \hbflucx(z+\zsep,t)}/\sigmaz^2(t),
\end{equation}
where $\zsep$ denotes the spanwise separation between two positions.
For a given time $t$, we identify the mean wavelength as twice the separation length for which $\corrhh$ attains its first and global minimum,
\begin{equation}\label{eq:wlength_1D}
  \begin{array}{rcl}
   \corrhh(\zsepmin,t) &\leq& \corrhh(\zsep,t) \quad \forall \; \zsep \;  \in \; [0,\Lz/2] \\[1ex]
   \lambdahz(t) &=& 2 \zsepmin.
  \end{array}
\end{equation}

%
%
\subsection{Quantification of particle transport}\label{subsec:qp_definition}
In order to transform the discretely spaced information of Lagrangian particle properties to the Eulerian observation framework, we apply a binning technique similar to that of \citet{Kidan_2015}, but generalized to the two-dimensional case.
Here we choose to discretize the two periodic directions in bins of width $\deltaxbin=\deltazbin\approx1.5D$ spanning over the entire wall-normal box length \Ly{}, resulting in a number of $N_{x,bin}$ and $N_{z,bin}$ bins in the streamwise and spanwise direction, respectively.
The local particle flux in the $(i,k)$-th bin ($1\leq i \leq N_{x,bin}$, $1 \leq k \leq N_{z,bin}$) is then defined as the volumetric particle flow rate averaged over that bin, that is, the sum of the particle velocity of all particles centered in the bin times the particle volume divided by the bin base area $\deltaxbin\deltazbin$, \textit{viz.}
\begin{subequations}
  \begin{align}
    \qpx (\xxi,\zzk,t) &= \dfrac{\Vp}{\deltaxbin \deltazbin} \displaystyle \sum_{l=1}^{\Np}
                          \up^{(l)}(t) \;\pindicatorxz(t) \label{eq:def_qpx}\\[1.5ex]
    \qpz (\xxi,\zzk,t) &= \dfrac{\Vp}{\deltaxbin \deltazbin} \displaystyle \sum_{l=1}^{\Np}
                          \wp^{(l)}(t) \;\pindicatorxz(t) \label{eq:def_qpz}.
  \end{align}
\end{subequations}
where $\qpx$, $\qpz$ are the streamwise and spanwise components of the particle flux vector $\qpvec$.
$\Vp=(\pi/6)D^3$ denotes the volume of a single spherical particle and $\up^{(l)}$, $\wp^{(l)}$ are the streamwise and spanwise component of the particle velocity vector $\upvec^{(l)}$ of particle $l$, respectively. The bin centers are located at $\xxi=((i-1)+i)\deltaxbin/2$ and $\zzk=((k-1)+k)\deltazbin/2$.
$\pindicatorxz(t)$ is an indicator function that is unity if the $l$-th particle is centered in the
$(i,k)$-th bin, i.e.
\begin{equation}
  \pindicatorxz(t)=
  \begin{cases}
    1 \quad \; \parbox[t]{.6\textwidth}{if $
                         (i-1)\deltaxbin \leq \xp^{(l)}(t) \leq i\deltaxbin \\
                         \;\wedge\; (k-1)\deltazbin \leq \zp^{(l)}(t) \leq k\deltazbin $}\\[3ex]
    0 \quad \text{else.}
  \end{cases}
\end{equation}
Note that the particle flux is an integral measure and as such does not reflect the inhomogeneous distribution of particle transport intensity over the wall-normal direction. Let us therefore additionally introduce the streamwise particle flux density as
\begin{equation}\label{eq:def_qpx_density}
  \xzavg{\phi\up} (\yyj,t) = \frac{\Vp}{\Lx \Lz \deltaybin} \sum_{l=1}^{\Np} \up^{(l)}(t) \; \pindicatory(t),
\end{equation}
with the wall-normal bin centers
$\yyj=((j-1)+j)\deltaybin/2$ ($1 \leq j \leq N_{y,bin}$) and the indicator function $\pindicatory(t)$ defined as
\begin{equation}
  \pindicatory(t)=
  \begin{cases}
    1 \quad \; \parbox[t]{.6\textwidth}{if $
                         (j-1)\deltaybin \leq \yp^{(l)}(t) \leq j\deltaybin$}\\[3ex]
    0 \quad \text{else}.
  \end{cases}
\end{equation}
In a similar way as in the previous definitions, we have subdivided the wall-normal box length into $N_{y,bin}$ bins with however smaller bin height $\deltaybin\approx0.5D$, this time spanning over the entire box length $\Lx$ and width $\Lz$. By definition, the following relation holds \citep{Lobkovsky_al_2008,Chiodi_al_2014}:
\begin{equation}
  \qpxmxzt = \int_{y=0}^{\Ly} \xztavg{\phi\up} \mathrm{d}y.
\end{equation}

  \end{appendices}

  \phantomsection
  \setlength{\bibsep}{.4ex}
  \addcontentsline{toc}{section}{References}
  \bibliography{literature.bib}


\end{document}